\documentclass[%
 reprint,
 superscriptaddress,
 amsmath,amssymb,
 aps,
 prb,
]{revtex4-2}
\usepackage{soul}
\usepackage{graphicx}
\usepackage{dcolumn}
\usepackage[all]{xy}
\usepackage{tikz-cd}
\usepackage{bm}
\usepackage{xcolor}
\usepackage{multirow}
\usepackage{comment}
\usepackage{hyperref}
\hypersetup{
    colorlinks=true,
    linkcolor=blue,
    citecolor=blue
}

\definecolor{amethyst}{rgb}{0.6, 0.4, 0.8}

\begin{document}

\title{Symmetry indicators in commensurate magnetic flux}

\author{Yuan Fang}
\affiliation{Department of Physics and Astronomy, Stony Brook University, Stony Brook, New York 11974, USA}
\author{Jennifer Cano}
\affiliation{Department of Physics and Astronomy, Stony Brook University, Stony Brook, New York 11974, USA}
\affiliation{Center for Computational Quantum Physics, Flatiron Institute, New York, New York 10010, USA}
\date{\today}
\begin{abstract}

We derive a framework to apply topological quantum chemistry in systems subject to magnetic flux.
We start by deriving the action of spatial symmetry operators in a uniform magnetic field, which extends Zak's magnetic translation groups to all crystal symmetry groups.
Ultimately, the magnetic symmetries form a projective representation of the crystal symmetry group.
As a consequence, band representations acquire an extra gauge invariant phase compared to the non-magnetic theory.
Thus, the theory of symmetry indicators is distinct from the non-magnetic case. 
We give examples of new symmetry indicators that appear at $\pi$ flux.
Finally, we apply our results to an obstructed atomic insulator with corner states in a magnetic field.
The symmetry indicators reveal a topological-to-trivial phase transition at finite flux, which is confirmed by a Hofstadter butterfly calculation.
The bulk phase transition provides a new probe of higher order topology in certain obstructed atomic insulators.

\end{abstract}
\maketitle

\section{\label{sec:intro} Introduction}

Threading magnetic flux through a two-dimensional crystal changes the single particle band spectrum into a Hofstadter butterfly spectrum that exhibits a fractal structure with an infinitude of mini gaps~\cite{hofstadter1976energy}. The Hofstadter butterfly is the lattice counterpart of Landau levels in the continuum.
While the Landau levels of a continuum model are often easier to compute than the Hofstadter butterfly of the corresponding lattice model, diagnosing band topology in the presence of magnetic flux requires the lattice because topological invariants are defined over the entire Brillouin zone.
The topology of Hofstadter bands has been a subject of intense recent study ~\cite{herzog2020hofstadter,wang2020classification,zuo2021topological,asaga2020boundary,otaki2019higher,zhang2022fractional}.

In the absence of magnetic flux, the topology of a band structure can be classified by the theory of topological quantum chemistry (TQC)  \cite{bradlyn2017topological,vergniory2017graph,elcoro2017double,cano2018building,bradlyn2018band,vergniory2019complete,cano2021band,elcoro2021magnetic}. 
A practical diagnosis comes from studying the space group representations of bands at high symmetry momenta, which are known as symmetry indicators \cite{po2017symmetry}.
However, in its present form, TQC cannot be directly applied to systems in a magnetic field because it does not account for the Aharonov-Bohm phase.


\par


\begin{figure}
    \centering
    \includegraphics[width=\linewidth]{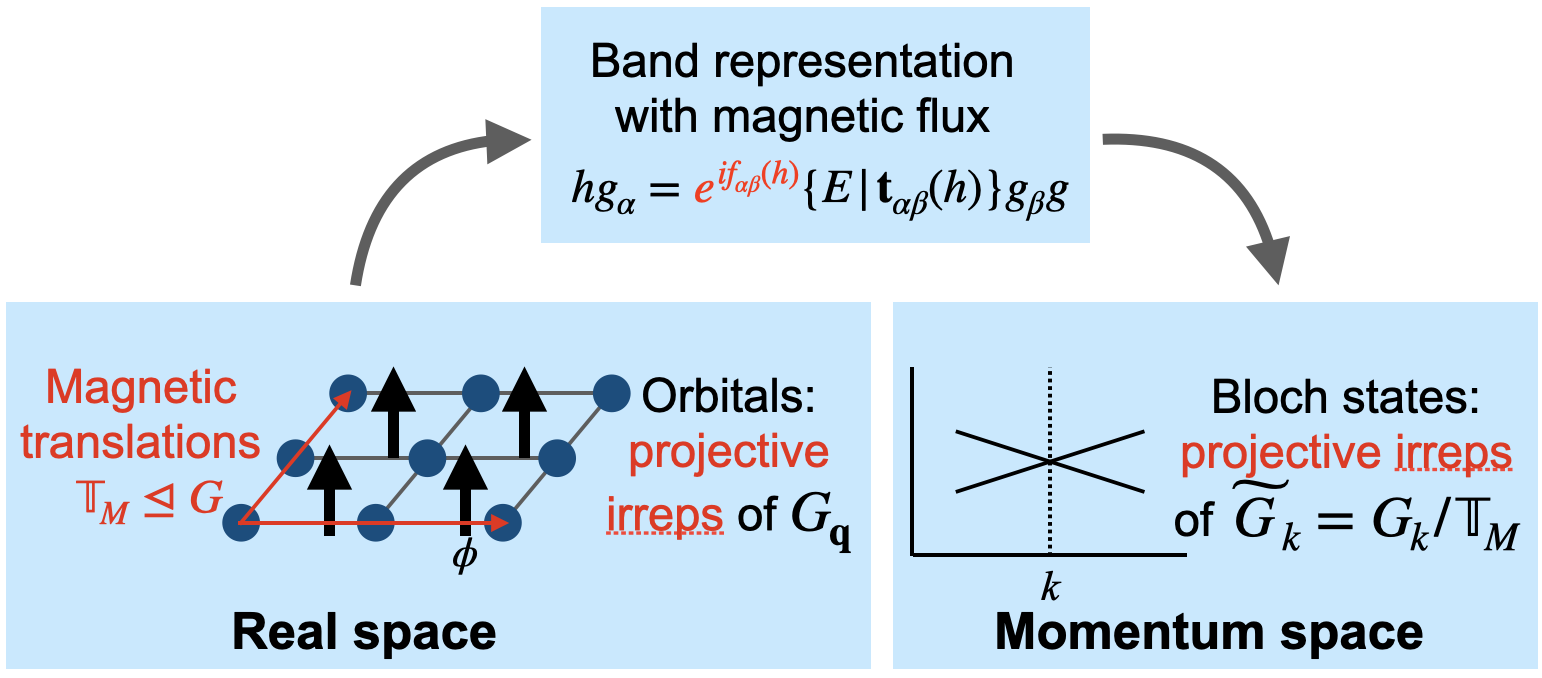}
    \caption{Framework for TQC in the presence of magnetic flux. A representation of the site-symmetry group induces a band representation of the entire space group, which is subduced to a representation of the little co-group at each high-symmetry momentum, i.e., the symmetry indicator. The new element introduced by the magnetic field is that the symmetry operators form a projective representation of the space group. The red font indicates differences between TQC with and without magnetic flux.}
    \label{fig:workflow}
\end{figure}

In the present manuscript we derive a framework to generalize TQC and the classification of symmetry indicators to band structures in the presence of a rational magnetic flux per unit cell.
The workflow is shown in Fig.~\ref{fig:workflow}.
We find that the two key ingredients in the theory of TQC -- the irreducible representations of bands at high symmetry points in momentum space and the induced representations of localized orbitals in real space -- are modified from their non-magnetic counterparts due to the presence of magnetic flux. 
The essential reason for this modification is that the commutation relations between crystal symmetries change in the presence of magnetic flux due to the Aharonov-Bohm phase.
As a result, the symmetry operators form  non-trivial projective representations of the space group. 
The earliest example of this is Zak's magnetic translation group~\cite{zak1964magnetic,zak1964magnetic2}.
Our theory builds on Zak's theory by including crystalline symmetries.

Our theory of TQC in commensurate magnetic flux is distinct from ``magnetic TQC''~\cite{elcoro2021magnetic}.
While magnetic TQC classifies topological band structures according to the representations of magnetic space groups, which describe the symmetry of magnetically ordered crystals, 
magnetic TQC does not yet accommodate magnetic flux through each unit cell, as it deals with zero flux configurations of orbitals.

The mathematical formalism utilized in this manuscript is also distinct from (magnetic) TQC.
Specifically, (magnetic) TQC uses the usual {linear} (co-) representations of point groups, which were tabulated prior to those works.
In contrast, our present study requires projective representations.
The projective representations of point groups are not systematically tabulated.
We derive an algorithm to construct magnetic symmetry operators using irreducible projective representations of point groups.
Our algorithm provides a framework that encompasses Zak's magnetic translation groups~\cite{zak1964magnetic,zak1964magnetic2}, as well as the Benalcazar-Bernevig-Hughes model~\cite{benalcazar2017electric,benalcazar2017quantized}. 
We then use the magnetic symmetry operators to derive symmetry indicators for topological phases on lattices with a particular symmetry group and subject to commensurate flux.
We give several explicit examples that have not appeared in previous literature.

Our manuscript proceeds as follows.
In Sec.~\ref{sec:mag_sym}, we derive the space group symmetry operators at rational flux in both real and momentum space.
The results are at the crux of the theory of TQC in a magnetic field that we derive in Sec.~\ref{sec:TQC}.
We then use the theory to compute symmetry indicators for magnetic layer groups, $p2$, $p4$, $p3$, $p6$, and $p4/m'$, at $\pi$ flux per unit cell.
The strong indicator in $p2$ recovers an earlier formula for the Chern number in Ref.~\onlinecite{matsugatani2018universal}, which is a stronger version of the formula in Ref.~\onlinecite{fang2012bulk}.
In group $p4/m'$, our theory gives rise to a new strong $\mathbb Z_2$ indicator, which is simply the filling per unit cell mod $2$. This $\mathbb Z_2$ non-trivial phase is protected by translation symmetry: the non-trivial phase does not permit exponentially localized and symmetric Wannier functions, but such Wannier functions exist when translation symmetries within the magnetic unit cell are broken.
The Chern number indicators for $p3$, $p4$, and $p6$ do not appear in earlier work.
\par

In Sec.~\ref{sec:Model}, we study a tight binding model introduced in Ref.~\onlinecite{wieder2020strong} that realizes an obstructed atomic limit (OAL) on the square lattice at zero flux. 
The Hofstadter butterfly spectrum shows that the system undergoes a gap-closing phase transition at finite flux after which the corner states that were present in the OAL phase disappear.
By applying our theory of TQC in a magnetic field to this model, 
we show that the gap closing corresponds to a phase transition from an OAL to a trivial phase that can be diagnosed by symmetry indicators.
\par

\section{\label{sec:mag_sym} Magnetic symmetries}
In quantum mechanics the coupling of a magnetic field to a charged particle is described by replacing the momentum $\mathbf P$ of the particle with the canonical momentum $\mathbf p=\mathbf P+\mathbf A$ in the Hamiltonian (without loss of generality, we have used natural units and assumed positive unit charge). 
To account for the Aharonov-Bohm phase, terms in the single-particle tight binding Hamiltonian are modified by the usual Peierls substitution:
\begin{equation}
    c^{\dagger}_{\mathbf r_2} c_{\mathbf r_1} \mapsto e^{i\int_{\mathbf r_1}^{\mathbf r_2} \mathbf A(\mathbf r)\cdot d\mathbf r}c^{\dagger}_{\mathbf r_2} c_{\mathbf r_1},
    \label{eq:peierls}
\end{equation}
where the path of the integral is the straight line connecting $\mathbf{r}_1$ and $\mathbf{r}_2$.

However, if the zero-field Hamiltonian is invariant under a crystal symmetry $\hat{g}: c_{\mathbf{r}} \mapsto c_{\hat{g}\mathbf{r}}$,
the Hamiltonian modified by the Peierls substitution in Eq.~(\ref{eq:peierls}) is not necessarily invariant under $\hat{g}$, even if the physical system is unchanged by the symmetry. Consequently, the operator $\hat{g}$ must be modified from its zero-field form by a gauge transformation that accounts for the Aharonov-Bohm phase. 
Specifically, the magnetic field requires $\hat{g}$ be replaced by $g\equiv \tilde G_g \hat g$, where $\tilde{G}_g=e^{i\sum_{\mathbf x}\lambda_g(\mathbf x)c^\dagger_{\mathbf x} c_{\mathbf x}}$ is a gauge transformation that acts on the electron annihilation operators by~\cite{herzog2020hofstadter}:
\begin{align}
    \tilde{G}_g c_{\mathbf{r}}  \tilde{G}_g^{-1}&= e^{-i\lambda_g(\mathbf{r})} c_{\mathbf{r}}, \label{eq:defG1}\\
    \tilde{G}_g c^\dagger_{\mathbf{r}}  \tilde{G}_g^{-1}&= e^{i\lambda_g(\mathbf{r})} c^\dagger_{\mathbf{r}},
    \label{eq:defG2}
\end{align}
where $\lambda_g$ is a scalar function defined for each symmetry $\hat{g}$ that we will derive momentarily. 
Acting on terms in the Hamiltonian in the form of Eq.~(\ref{eq:peierls}), $\tilde{G}_g$ has the effect of mapping $\mathbf{A}(\mathbf{r}) \mapsto \mathbf{A}(\mathbf{r}) + \nabla \lambda_g(\mathbf{r})$.

Similar gauge transformations were introduced by Zak for the magnetic translation operators in Refs.~\cite{zak1964magnetic,zak1964magnetic2}.
More recently, the magnetic operators for rotations about the origin and for time-reversal symmetry were considered in Refs.~\cite{de2011exponentially,matsugatani2018universal,herzog2020hofstadter}.
Here, we develop a general theory for any symmetry group in the presence of a magnetic field, thereby extending previous works to include more general rotations and glide reflection symmetries. Doing so allows us to apply the theory of symmetry indicators to diagnose topological phases in the presence of a magnetic field.
\par

We now derive the gauge transformation $\lambda_g$ in Eq.~(\ref{eq:defG1}): we require that if a single-particle Hamiltonian in zero field is invariant under a symmetry $\hat{g}$, then in the presence of a magnetic field that preserves $\hat{g}$, the Hamiltonian modified by the  Peierls substitution in Eq.~(\ref{eq:peierls}) is invariant under the combined symmetry operation $g \equiv \tilde{G}_g \hat{g}$, i.e., we require
\begin{equation}
	g : e^{i\int_{\mathbf r_1}^{\mathbf r_2} \mathbf A(\mathbf r)\cdot d\mathbf r}c^{\dagger}_{\mathbf r_2} c_{\mathbf r_1} \mapsto e^{i\int_{g\mathbf r_1}^{g\mathbf r_2} \mathbf A(\mathbf r')\cdot d\mathbf r'}c^{\dagger}_{g\mathbf r_2} c_{g\mathbf r_1}
	\label{eqn:covariance}
\end{equation}
Acting on the left-hand-side by $g= \tilde{G}_g \hat{g}$, using the definition of $\tilde{G}_g$ in Eqs.~(\ref{eq:defG1}) and (\ref{eq:defG2}), and equating with the right-hand-side yields
\begin{equation}
	e^{i\int_{\mathbf r_1}^{\mathbf r_2} \mathbf A(\mathbf r)\cdot d\mathbf r +i\int_{g\mathbf{r}_1}^{g\mathbf{r}_2} \nabla \lambda(\mathbf{r}')\cdot d\mathbf{r}' }  = e^{i\int_{g\mathbf r_1}^{g\mathbf r_2} \mathbf A(\mathbf r')\cdot d\mathbf r'}
	\label{eqn:covariance2}
\end{equation}
A few lines of algebra (detailed in Appendix~\ref{app:Eqlambda}) show that Eq.~(\ref{eqn:covariance2}) is satisfied when $\lambda_g(\mathbf r)$ satisfies

\begin{equation}
\label{eqn:lambda}
    \nabla \lambda_g(\mathbf r) = \mathbf A(\mathbf r)-R_g \mathbf A (g^{-1} \mathbf r),
\end{equation}
where $R_g$ is the point group part of $g$. Eq.~(\ref{eqn:lambda}) applies equally well to uniform or non-uniform magnetic fields. For simplicity, we restrict ourselves to the uniform field case for the remainder of this manuscript.
Eq.~(\ref{eqn:lambda}) determines each $\lambda_g$ up to a constant. We choose the constant such that for translation~\cite{herzog2020hofstadter}
\begin{equation}
\label{eqn:lambdaT}
    \lambda_{T(\mathbf a)}(\mathbf r) = \int_{\mathbf r-\mathbf a}^{\mathbf r} \mathbf A(\mathbf r')\cdot d\mathbf r' + \mathbf B \cdot \mathbf a \times \mathbf r
\end{equation}
and that a $2\pi$ rotation is implemented by the identity matrix. This choice of constants ensures that the commutation relations between translation and rotations about the origin are the same as at zero field as we show in the Appendix~\ref{app:rotation}.
This choice of gauge is fixed throughout this paper; later when we refer to a gauge choice, we are referring to the gauge of the vector potential.
So far, we have only considered lattice degrees of freedom; orbital and spin degrees of freedom can be included by an extra unitary transformation in the action of $\hat{g}$, {which does not change $\lambda_g$}.
We will include these degrees of freedom in later sections.

\par

Eq.~(\ref{eqn:lambda}), which serves as the definition of $\lambda_g$, is the first key result of this manuscript.
Combining it with the spatial action of the symmetry yields the explicit form of the magnetic symmetry operator:
\begin{align}
\label{eqn:Gg1}
     g&=e^{i\sum_{\mathbf x'}\lambda_g (\mathbf x')c^\dagger_{\mathbf x'}c_{\mathbf x'}} \sum_\mathbf{x} c^\dagger_{\hat g \mathbf x}c_{\mathbf x} \\
     &=\sum_\mathbf{x} e^{i\lambda_g (\hat g \mathbf x)} c^\dagger_{\hat g\mathbf x}c_{\mathbf x} ,
     \label{eqn:Gg2}
\end{align}
where $\lambda_g$ is determined by Eq.~(\ref{eqn:lambda}). 
The second equality holds when $g$ acts on the single-particle Hilbert space.

We now explain why changing $\lambda_g$ up to a constant does not change the representation of the magnetic symmetry operators defined in Eq.~(\ref{eqn:Gg2}).
These operators furnish projective representations of the space group. 
A projective representation $\rho$ of a group satisfies the following multiplication rule
\begin{equation}
    \rho(h_1)\rho(h_2)=\omega(h_1,h_2) \rho(h_1h_2),
\end{equation}
where $h_1$, $h_2$ are group elements and $\omega(h_1,h_2)$ is called the 2-cocycle. 
If $\omega(h_1,h_2)\equiv 1$, then $\rho$ is an ordinary linear representation. In general, the magnetic symmetry operators in Eq.~(\ref{eqn:Gg2}) will have non-trivial 2-cocycles, as we show in the next sections. The $U(1)$ gauge freedom in Eq.~(\ref{eqn:lambda}) corresponding to the gauge transformation $\lambda_g \mapsto \lambda_g+C_g$, $g \mapsto e^{iC_g}g$ leaves the representation in the same group cohomology class, i.e. the transformed projective representation is equivalent to the previous one. Essential properties of projective representations are presented in Appendix.~\ref{app:projective}.
\par
In the next two subsections, we apply this formalism to two examples, first rederiving Zak's magnetic translation group and then reviewing the symmetries of the square lattice in a magnetic field.

\subsection{\label{sec:mag_sym_1_zak} Zak's magnetic translation group}
In Refs.~\onlinecite{zak1964magnetic} and~\onlinecite{zak1964magnetic2} Zak introduced the continuous magnetic translation symmetries. We reproduce Zak's result by taking the continuum limit {of Eq.~(\ref{eqn:Gg2})}. 

\par
Consider a two-dimensional infinite plane without a lattice and denote operators that translate by ${\mathbf \Delta}=\Delta_x \hat{\mathbf{x}}+\Delta_y \hat{\mathbf{y}}$ by $\hat T(\mathbf \Delta)\equiv \hat T(\Delta_x,\Delta_y)$,
where $\hat{\mathbf{x}}$ and $\hat{\mathbf{y}}$ denote the unit vectors.

We first work in the symmetric gauge: $\mathbf{A}(\mathbf r)=\frac B2(-r_y,r_x)$. 
Then from Eq.~(\ref{eqn:lambda}):
\begin{align}
    \label{eq:lambdaT_symmetric}
    \lambda_{T({\mathbf \Delta})}(\mathbf r)= \frac B2(\Delta_x r_y-\Delta_y r_x)
\end{align}
For continuous translations, we replace the sum $\sum_\mathbf{x} c^\dagger_{\hat T(\mathbf \Delta) \mathbf x}c_{\mathbf x}$ in Eq.~(\ref{eqn:Gg1}) with $e^{-ip_x\Delta_x-ip_y\Delta_y}$. Then the magnetic translation by vector $\mathbf \Delta$ is
\begin{align}
    T(\mathbf \Delta)&=e^{i(\frac12 B\Delta_x(r_y+\Delta_y)-\frac12 B\Delta_y(r_x+\Delta_x))}e^{-i(p_x\Delta_x+p_y\Delta_y)} \nonumber \\
    &=e^{-i((p_x-\frac12 Br_y)\Delta_x+(p_y+\frac12 Br_x)\Delta_y)}, 
\end{align}
where the Baker–Campbell–Hausdorff formula is considered.
Therefore, the generators of the magnetic translations in $\hat{\mathbf x}$ and $\hat{\mathbf y}$ directions are
\begin{align}
    K_x&=p_x-\frac12 Br_y \nonumber \\
    K_y&=p_y+\frac12 Br_x,
\end{align}
which is exactly Zak's definition from his 1964 paper~\cite{zak1964magnetic}.\par

In the remainder of the manuscript it will be easier to use the Landau gauge $\mathbf A(\mathbf r)=(-B r_y,0)$.
Repeating the calculation of $\lambda_g$ in the Landau gauge yields

\begin{align}
    \lambda_{T(\mathbf \Delta)}(\mathbf r)=-B\Delta_y r_x
    \label{eqn:lambdaT_Landau}
\end{align}

One important property of the magnetic translation operators is the gauge-invariant noncommutativity:
\begin{equation}
    T(\Delta_x \hat{\mathbf{x}}) T(\Delta_y \hat{\mathbf{y}})= T(\Delta_y \hat{\mathbf{y}}) T(\Delta_x \hat{\mathbf{x}}) e^{iB\Delta_x\Delta_y} \label{eqn:TxTyDelta},
\end{equation}
which reproduces the Aharonov-Bohm phase. 
More generally, for two translations $\mathbf a_1$ and $\mathbf a_2$, the gauge invariant multiplication equation is~\cite{zak1964magnetic}
\begin{equation}
\label{eqn:Ta1Ta2}
    T(\mathbf a_1) T(\mathbf a_2) =  T(\mathbf a_1+\mathbf a_2) e^{\frac i2 \mathbf B \cdot (\mathbf a_1 \times \mathbf a_2)}
\end{equation}
The gauge invariant phase term $e^{\frac i2 \mathbf B \cdot (\mathbf a_1 \times \mathbf a_2)}$ is the 2-cocycle of magnetic translations, which shows the magnetic translation operators form a non-trivial projective representation of the translation group.

\subsection{\label{sec:mag_sym_2_landau} Magnetic symmetries of the square lattice} 
\begin{table*}[t]
\begin{tabular}{c|c|c|c|c|c|c|c}
\hline
$g$ &$T(\Delta_x\hat{\mathbf x})$ &$T(\Delta_y\hat{\mathbf y})$& $C_2(\bar x, \bar y)$ &$C_4(\bar x, \bar y)$ &$I(\bar x, \bar y)$ &$Um_{x}(\bar x)$ &$Um_{y}(\bar y)$\\
\hline
&&&&&&&\\[-0.5em]
$\hat{g}=\{R_g|\tau_g\}$ &$\{0|(\Delta_x,0)\}$ &$\{0|(0,\Delta_y)\}$ &$\{\hat{C}_2|(2\bar x,2\bar y)\}$ &$\{\hat{C}_4|(\bar x+\bar y,\bar y-\bar x)\}$ &$\{\hat{I}|(2\bar x,2\bar y)\}$ &$U\{\hat{m}_{x}|(2\bar x,0)\} $ &$U\{\hat{m}_{y}|(0,2\bar y)\} $\\
\hline
&&&&&&&\\[-0.5em]
$\lambda_g(x,y)$&0&$-B\Delta_y x$ &$-2B\bar y (x-\bar x)$ &$-B(x-\bar x)(y-\bar y)$ &$-2B\bar y (x-\bar x)$ &$0$ &$-2B\bar y x$\\
&&&&$+B\bar y((y-{\bar y})-(x-\bar x))$&&\\
\hline
\end{tabular}
\caption{The gauge transformation $\lambda_g(x,y)$ for symmetries of the square lattice  in Landau gauge. For each symmetry $g$ in the first row, the second row lists the symmetry in the notation $\lbrace R_g | \tau_g \rbrace$, where $\hat{g}: \mathbf{r} \mapsto R_g \mathbf{r} + \tau_g$. The third row provides $\lambda_g$ from Eq.~(\ref{eqn:lambda}).}
\label{tab:lambda}
\end{table*}

As a second example, we consider discrete symmetries of the two-dimensional square lattice using the Landau gauge $\mathbf A(\mathbf r)=(-B r_y,0)$. 
When $B=0$, the square lattice is invariant under the layer group $p4/mmm$, which is generated by a four-fold rotation symmetry and the mirrors $m_x$ and $m_z$. Without a magnetic field, the system is also invariant under time-reversal symmetry, $\cal T$.

When $B\neq 0$, only the symmetries that leave the magnetic field invariant (four-fold rotations and $m_z$) remain; the resulting layer group is  $p4/m$.
To determine how these symmetries act on the electron creation/annihilation operators, one must compute the gauge transformation $\lambda_g$ from Eq.~(\ref{eqn:lambda}).
We summarize the results in Table~\ref{tab:lambda}. 
Notice that $\lambda_g$ depends on the rotation or inversion center; 
thus, it is necessary to introduce the notation 
\begin{equation}
\label{eqn:defCnxy}
    C_n(\bar{x},\bar{y}) \equiv T(\bar{x},\bar{y})C_n T(-\bar{x},-\bar{y})
\end{equation}
to denote an $n$-fold rotation about the point $(\bar x,\bar y)$; we use $C_n \equiv C_n(\bar{x}=0,\bar{y}=0)$ to denote a rotation about the origin.
We adopt analogous notation for inversions and reflections about different points and planes.

The symmetries $m_x$, $m_{(110)}$ and $\mathcal{T}$ flip the magnetic field and thus are not symmetries at finite $B$. 
However, the product of these symmetries with a magnetic flux shifting operator can leave the system invariant at special values of flux, as we now describe.

A lattice Hamiltonian coupled to a magnetic field is periodic in $B$:
the period corresponds to the minimal magnetic field such that every possible closed hopping path encloses an integer multiple of $2\pi$ flux.
Let $\phi$ denote the magnetic flux per unit cell and $\Phi = 2\pi n$ its periodicity, where $n$ is an integer. 
Following Ref.~\cite{herzog2020hofstadter}, we define the unitary matrix $U$ that shifts $\phi \mapsto \phi + \Phi$ by
\begin{align}
    {U}&=e^{i\sum_{\mathbf x'}\lambda_{U} (\mathbf x')c^\dagger_{\mathbf x'}c_{\mathbf x'}} \\
    \lambda_{U} (\mathbf r) &= \int ^{\mathbf r}_{\mathbf r_0} \widetilde{ \mathbf A}(\mathbf r) \cdot d\mathbf r,
\end{align}
where $\mathbf r_0$ is a reference lattice point and $\tilde{\mathbf{A}}$ is the magnetic vector potential corresponding to $\Phi$ flux, i.e., $\nabla \times \widetilde{\mathbf A}=\Phi$.

Notice that for any symmetry $g$ that flips $\phi \mapsto -\phi$, the product $Ug$ is a symmetry in the special case where $\phi = \Phi/2$.
In the case of the square lattice, the products $Um_x$, $Um_y$ and $U{\cal T} $ are recovered as symmetries of the system at the special value of $\phi = \Phi/2$. We list the gauge transformations for $Um_x$ and $Um_y$ at $\phi = \Phi/2$ in Table~\ref{tab:lambda}.

In the special case of a square lattice and Landau gauge, $\lambda_U= -\Phi yx,$ where $x = (\mathbf{r} - \mathbf{r}_0) \cdot \hat{\mathbf{x}}$.
Since $\Phi$ is a multiple of $2\pi$ and $x,y$ are integers, this phase is also a multiple of $2\pi$. The flux translation matrix is given by $U = \mathbb{I}$, where $\mathbb{I}$ is the identity matrix.

In summary, we have explicitly extended Zak's translation operators in a magnetic field to the discrete symmetries of the square lattice.
In Appendix~\ref{app:triangular} we generalize the results to the symmetries of the triangular lattice.

\subsection{BBH model}
\begin{figure}
    \centering
    \includegraphics[width=0.7\linewidth]{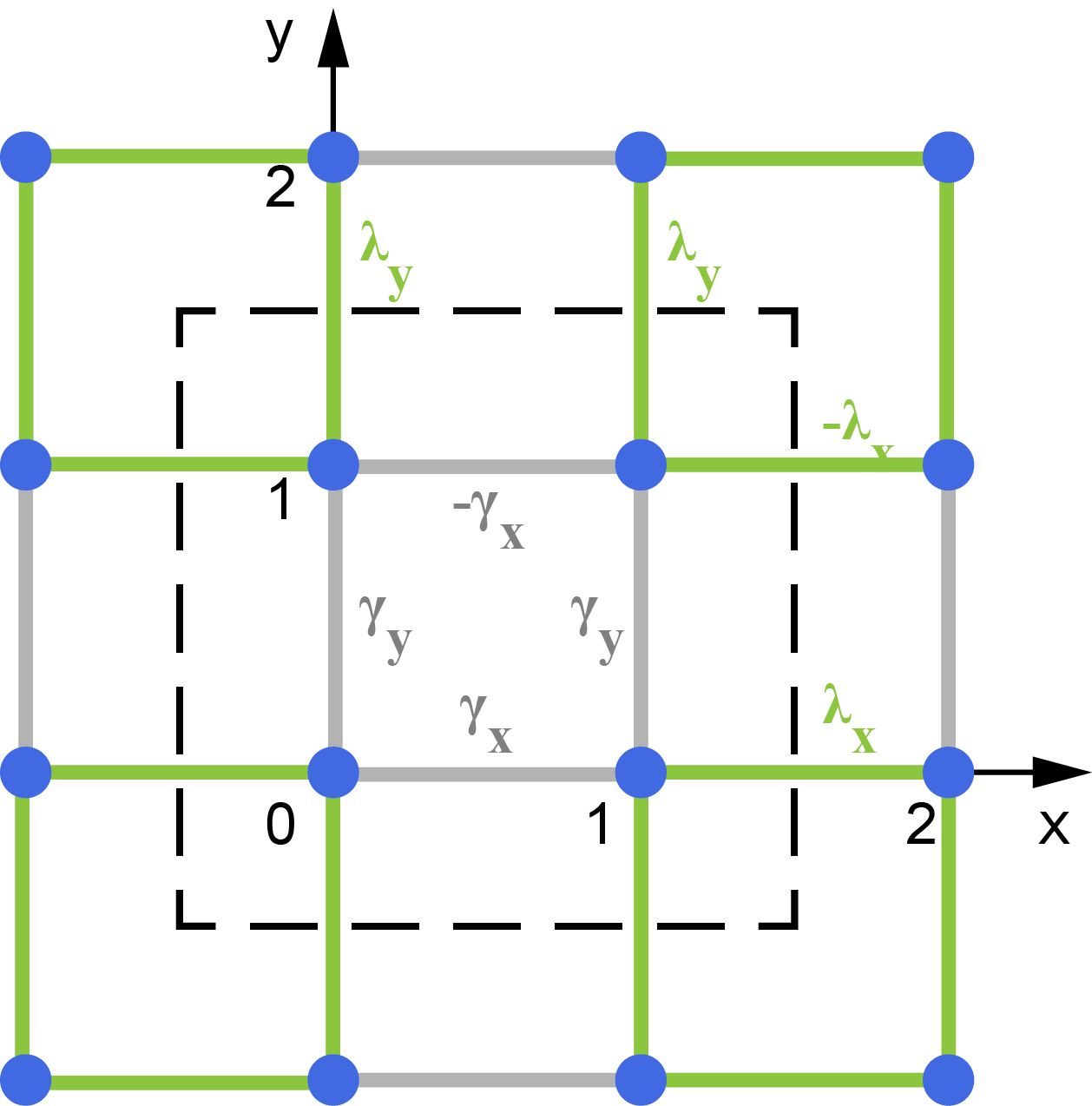}
    \caption{Lattice and hopping terms of the BBH model. The black dashed square indicates the unit cell. Blue dots indicate atoms, each with one orbital. The origin is at the left-bottom atom in the unit cell {indicated by $0$}. The hopping amplitudes $\gamma_{x/y}$ and $\lambda_{x/y}$ are real; the minus signs result from the magnetic flux $\phi=4\pi$, i.e., applying Eq.~(\ref{eq:peierls}) in Landau gauge.}
    \label{fig:BBHmodell}
\end{figure}

We apply the results of the previous section to derive the symmetry operators in the Benalcazar-Bernevig-Hughes (BBH) model ~\cite{benalcazar2017electric,benalcazar2017quantized}.
The model describes spinless electrons on a square lattice. 
The Hamiltonian consists of nearest-neighbor hopping terms, whose amplitudes $\lambda_{x/y}$ and  $\gamma_{x/y}$ are depicted in Fig.~\ref{fig:BBHmodell}. 
Since $\lambda_{x/y} \neq \gamma_{x/y}$, each unit cell contains four atoms.
{Further, each square plaquette has $\pi$ flux, for a total flux $\phi = 4\pi$ per unit cell.}
The flux periodicity is $\Phi = 8\pi$, corresponding to $2\pi$ flux per square plaquette.

We now derive the symmetry operations in the presence of the magnetic field; these commutation relations were stated in Refs.~\onlinecite{benalcazar2017electric,benalcazar2017quantized}, but here we derive them as an application of our formalism.

We start with the mirror symmetries: 
in zero field, the Hamiltonian is invariant under $m_x(\bar{x})$ and $m_y(\bar{y})$ where $\bar{x}, \bar{y}$ are half-integers. (The Hamiltonian is not invariant under reflections about lines containing the origin because $\gamma_{x/y} \neq \lambda_{x/y}$.)
These mirror reflections flip the sign of the magnetic field and thus generically are not symmetries of the Hamiltonian at finite flux.
However, since $\phi = 4\pi = \Phi/2$, the combined operations $Um_x(\bar{x})$ and $Um_y(\bar{y})$ are symmetries.
We showed in the previous section that in the Landau gauge, the flux shifting operator $U = \mathbb{I}$ for this model.
Therefore, at $\phi = 4\pi$, $m_x(\bar{x})$ and $m_y(\bar{y})$ are in fact symmetries of the Hamiltonian.
The effect of the magnetic field is to change their commutation relations: using Table~\ref{tab:lambda} with $U = \mathbb{I}$ yields
$ m_{x}(\bar x)m_{y}(\bar y) =m_{y}(\bar y)m_{x}(\bar x) e^{4iB\bar x\bar y} $.
Since $B=\pi$ and $\bar x,\bar y$ are half-integers
\begin{equation}
    \{m_{x}(\bar x), m_{y}(\bar y)\}=0,
\end{equation}
i.e., mirror symmetries in the BBH model anti-commute.

We now consider a four-fold rotation.
When $\gamma_x=\gamma_y$, $\lambda_x=\lambda_y$,
the BBH model has a four-fold rotation symmetry $C_4(\frac{1}{2}, \frac{1}{2})$, as well as other four-fold rotation axes related by translation.
Since $\phi=\Phi/2$, the system also has an effective time-reversal symmetry, $U\cal T$; since we established $U=\mathbb{I}$ for this model, $\mathcal{T}$ is a symmetry even at this finite field and acts by complex conjugation.
In the absence of a magnetic field, time-reversal pairs eigenstates with $\pm i$ rotation eigenvalues. We now show that in the presence of a magnetic field, time-reversal pairs eigenstates of $C_4(\bar{x},\bar{y})$ in a more complicated way.
\par

Since our origin is chosen such that all lattice sites have integer coordinates, $x,y\in \mathbb Z$, the phase $e^{i\lambda_{C_4}}$ in Table~\ref{tab:lambda} takes values of $\pm e^{-i\pi/4}$, so that $e^{-2i\lambda_{C_4}}=i$.
Therefore, given an eigenstate $|\xi\rangle$ of $C_4(\bar x,\bar y)$ with eigenvalue $\xi$, $\cal T|\xi\rangle$ is also an eigenstate of $C_4(\bar x,\bar y)$:
\begin{align}
    C_4(\bar x,\bar y){\cal T}|\xi\rangle &= e^{i\lambda_{C_4}}\hat{C_4}(\bar x,\bar y){\cal T}|\xi\rangle \nonumber\\
    &= {\cal T}e^{-i\lambda_{C_4}}\hat{C_4}(\bar x,\bar y)|\xi\rangle \nonumber\\
    &= {\cal T}e^{-2i\lambda_{C_4}}C_4(\bar x,\bar y)|\xi\rangle \nonumber\\
    &= {\cal T} i \xi |\xi\rangle =-i\xi^*{\cal T}|\xi\rangle.
    \label{eqn:BBHC4T}
\end{align}
Thus, $\cal T$ pairs $C_4(\bar x,\bar y)$ eigenstates with eigenvalues $\xi$ and $-i\xi^*$. This is an example of symmetry operators acting in unusual ways at finite field.

\section{Momentum space representations}

We now define how the magnetic symmetry operators act in momentum space.
This requires first defining how the symmetries act on Bloch wave functions and then labelling the Bloch wave functions by irreducible representations (irreps) of the symmetry group at each momentum point.
However, in the presence of magnetic flux, we cannot immediately define the Bloch wave functions because Bloch's theorem does not apply when the translation operators do not commute.


To apply Bloch's theorem, we define an enlarged ``magnetic unit cell,'' chosen to contain an integer multiple of $2\pi$ flux. The translation vectors that span the magnetic unit cell are referred to as magnetic translation vectors. From Eq.~(\ref{eqn:TxTyDelta}), the magnetic translation operators commute and thus can be simultaneously diagonalized, forming an abelian subgroup $\mathbb T_M$ of the full translation group $\mathbb T$. 
Consequently, Bloch's theorem applies to the magnetic unit cell and eigenstates of the Hamiltonian can be labelled by wave vectors in the magnetic Brillouin zone.

In Sec.~\ref{sec:mag_sym_3_momentum}, we define the Fourier transformed electron creation and annihilation operators in the magnetic unit cell. The operators necessarily have a ``sublattice'' index because the magnetic unit cell contains more than one non-magnetic unit cell.

In Sec.~\ref{sec:kirrep}, we address how to label the Bloch wave functions by irreps of the little co-group at each momentum.
Here we encounter another subtle point: 
since the little co-group is defined as a quotient group obtained from the space group mod magnetic translations, the little co-group only has a group structure if the magnetic translation group is a normal subgroup of the space group.
Thus, not all magnetic unit cells are equal: to label Bloch wave functions by irreps of the little co-group, we must choose a magnetic unit cell such that $\mathbb{T}_M$ is a normal subgroup.
After addressing this issue, we explain how to find the irreducible projective representations of the little co-group.

\subsection{\label{sec:mag_sym_3_momentum} Symmetries in the magnetic Brillouin zone}
 \begin{figure}
     \centering
     \includegraphics[width=\linewidth]{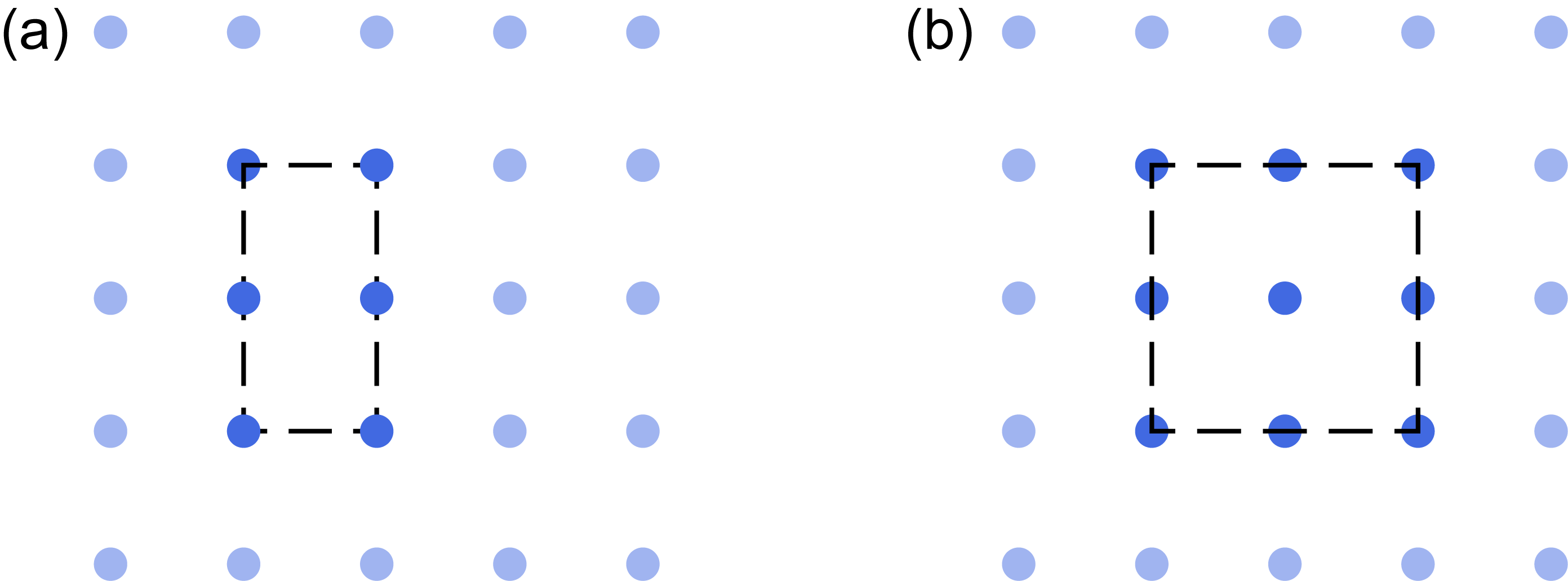}
     \caption{ Examples of (a) a $q$-by-$1$ unit cell and (b) a $q$-by-$q$ unit cell, taking $q=2$.}
     \label{fig:unitcell}
 \end{figure}

We first consider the minimal magnetic unit cell in Landau gauge, which is a $q$-by-$1$ unit cell (see Fig.~\ref{fig:unitcell}(a)). 
For this choice of unit cell, the magnetic translation group $\mathbb{T}_M$ is generated by $T(\hat{\mathbf{x}})$ and $T(q\hat{\mathbf{y}})$.

Now consider the layer group $p2$, generated by $C_2$ and lattice translations, for which 
$\mathbb{T}_M$ is a normal subgroup.
$C_2$ acts identically to the non-magnetic case, {mapping a Bloch wave function at $\mathbf{k}$ to one at $-\mathbf{k}$.}
However, $T(\hat{\mathbf y})$ acts in an unusual way, by mapping $k_x$ to $k_x+\phi$. This can be understood as follows:
let $|\mathbf k\rangle$ be an eigenstate of $T(\hat{\mathbf x})$ such that $T(\hat{\mathbf x}) |\mathbf k\rangle = e^{ik_x}|\mathbf k\rangle$. Then $T(\hat{\mathbf y})|\mathbf k\rangle$ is also an eigenstate of $T(\hat{\mathbf x})$, with eigenvalue $k_x + \phi$, i.e.,
\begin{equation}
    T(\hat{\mathbf x}) \left[ T(\hat{\mathbf y}) |\mathbf k\rangle \right] = e^{i(k_x+\phi)}T(\hat{\mathbf y})|\mathbf k\rangle
    \label{eq:Ty}
\end{equation}
Thus, $T(\hat{\mathbf{y}})$ shifts the eigenvalue of $T(\hat{\mathbf{x}})$ by $e^{i\phi}$.
{Nonetheless, both $C_2$ and $T(\hat{\mathbf{y}})$ have the usual property that a Bloch state at $\mathbf{k}$ is mapped to another Bloch state at $\mathbf{k}'$.}

This is not the case for the layer group $p4$, {with respect to which $\mathbb{T}_M$ is not a normal subgroup}.
As we will show below, the symmetry operator $C_4$ mixes a Bloch state at $\mathbf{k}$ into a linear combination of Bloch states at other momenta, forming a $q^2$-dimensional representation.
Thus, we are motivated to consider a $q$-by-$q$ unit cell (see Fig.~\ref{fig:unitcell}(b)), where, although the magnetic unit cell is larger, the symmetry matrices are the same size as in the $q$-by-$1$ case.
In Appendix~\ref{app:sym_q1_qq}, we show that the representations obtained from these two choices of magnetic unit cell are the same up to a unitary transformation. 
However, the $q$-by-$q$ unit cell is a more suitable to apply topological quantum chemistry because the corresponding magnetic translation group is a normal Abelian subgroup of the layer group $p4$.
We now consider the $q$-by-$1$ and $q$-by-$q$ unit cells in detail for the group $p4$ to illustrate these points.

\subsubsection{\label{sec:mag_sym_3_momentum_qby1}$q$-by-$1$ unit cell for $p4$}

We first consider the $q$-by-$1$ unit cell shown in Fig.~\ref{fig:unitcell}(a). The coordinates of lattice sites are labeled by $(x,y)=(R_x,qR_y+j)$ where $R_x,R_y\in \mathbb Z$, $j=0,\dots,q-1$. The Fourier transformed electron creation and annihilation operators are defined by
\begin{align}
    c^\dagger_{\mathbf R, j,\alpha}&=\frac{q}{(2\pi)^2}\int d\mathbf k e^{i (k_xR_x+k_yqR_y)}c^\dagger_{\mathbf k, j,\alpha} \label{eqn:FT_qby1_cd} \\
    c_{\mathbf R, j,\alpha}&=\frac{q}{(2\pi)^2}\int d\mathbf k e^{-i (k_xR_x+k_yqR_y)}c_{\mathbf k, j,\alpha},
\label{eqn:FT_qby1_c}
\end{align}
where $\alpha$ labels orbital degrees of freedom on each site. For now, we ignore the $\alpha$ degree of freedom, but will add it later when necessary. The magnetic Brillouin zone is a torus with $k_x\in [0,2\pi)$, $k_y\in [0,2\pi/q)$.
\par

Using the Fourier transforms in Eqs.~(\ref{eqn:FT_qby1_cd}) and (\ref{eqn:FT_qby1_c}), we find the action of the symmetry operators in momentum space.
A translation by one (non-magnetic) lattice vector in the $\hat{\mathbf y}$ direction is implemented by 
\begin{align}
    T(\hat{\mathbf y})&=\frac{q}{(2\pi)^2}\sum_j\int d\mathbf k~ e^{ik_y} c^\dagger_{\mathbf k+(\phi,0), j}c_{\mathbf k, j-1} 
    \label{eq:Tyq}
\end{align}
Unlike the non-magnetic case, $T(\hat{\mathbf y})$ does not leave each $\mathbf{k}$ point invariant: it maps $(k_x,k_y)$ to $(k_x+\phi,k_y)$. 
Translations by the magnetic lattice vectors do leave $\mathbf{k}$ invariant.

We now consider the four-fold rotation operator.
Using the function $\lambda_{C_4}$ in Table~\ref{tab:lambda}, 
\begin{align}
    C_4&=\frac{q}{(2\pi)^2}\int d\mathbf k \sum_{j,j'}\sum_{n=0}^{q-1} \frac1q e^{i(\phi jj'-(k_y+2\pi n /q)j-k_xj')} \nonumber \\
    &c^\dagger_{(k_x,k_y),j}c_{(k_y+2\pi n/q,~-k_x~\text{mod}~ 2\pi/q),j'}
    \label{eq:C4q}
\end{align}
Thus, the situation for $C_4$ is much worse than for $T(\hat{\mathbf y})$: $C_4$ does not rotate one $\mathbf{k}$ point to another, but instead mixes a state at $(k_x,k_y)$ into a linear combination of states at the different points $(k_y+2\pi n/q,-k_x)$ $n=0,1,\dots,q-1$.

\subsubsection{\label{sec:mag_sym_3_momentum_qbyq}$q$-by-$q$ unit cell for $p4$}
We now consider the $q$-by-$q$ unit cell shown in Fig.~\ref{fig:unitcell}(b). The coordinates of lattice sites are labeled by $(x,y)=(qR_x+j_x,qR_y+j_y)$ where $R_x,R_y\in \mathbb Z$ label a magnetic unit cell and $j_x,j_y=0,\dots,q-1$ label the coordinates of atoms within.
The Fourier transformed electron creation and annihilation operators are defined by
\begin{align}
\label{eqn:FT_qbyq_cd}
    c^\dagger_{\mathbf R, \mathbf{j},\alpha}&=(\frac{q}{2\pi})^2\int d\mathbf k e^{i (k_xqR_x+k_yqR_y)}c^\dagger_{\mathbf k, \mathbf{j},\alpha} \\
    c_{\mathbf R, \mathbf{j},\alpha}&=(\frac{q}{2\pi})^2\int d\mathbf k e^{-i (k_xqR_x+k_yqR_y)}c_{\mathbf k, \mathbf{j},\alpha}
    \label{eqn:FT_qbyq_c}
\end{align}
Again we omit the orbital degrees of freedom $\alpha$ in this section. The magnetic Brillouin zone is a torus with $k_x, k_y\in [0,2\pi/q)$.\par
Using the Fourier transforms in Eqs.~(\ref{eqn:FT_qbyq_cd}) and (\ref{eqn:FT_qbyq_c}) and plugging $\lambda_g$ from Table~\ref{tab:lambda} into Eq.~(\ref{eqn:Gg2}), the magnetic $T(\hat{\mathbf y})$ and $C_4$ symmetries are~\cite{herzog2020hofstadter} 
\begin{align} 
    T(\hat{\mathbf y})&=(\frac{q}{2\pi})^2\int d\mathbf k~e^{ik_y}\sum_{j_x,j_y}e^{-i\phi j_x} c^\dagger_{\mathbf k, j_x,j_y}c_{\mathbf k, j_x,j_y-1}
    \label{eq:Tyqq}
\end{align}
and
\begin{align}
    C_4&=(\frac{q}{2\pi})^2\int d\mathbf k~\sum_{j_x,j_y}e^{-i\phi j_x j_y} e^{-i(C_4\mathbf k \cdot \mathbf j-\mathbf k \cdot \mathbf j')} \nonumber\\
    &\qquad c^\dagger_{(-k_y,k_x), j_x,j_y}c_{(k_x,k_y), j_x',j_y'}
    \label{eq:C4qq}
\end{align}
where $\mathbf j'=(j_x',j_y')$ is a function of $j_x,j_y$ that satisfies $j_x'=j_y \mod q$ and  $j_y'=-j_x \mod q$.
In Eqs.~(\ref{eq:Tyqq}) and (\ref{eq:C4qq}), the action of the symmetry operator on $\mathbf{k}$ is identical to its action in the absence of a magnetic field, i.e., translation leaves $\mathbf{k}$ invariant and a rotation in space rotates $\mathbf{k}$.
This is an improvement over the $q$-by-$1$ magnetic unit cell (Eqs.~(\ref{eq:Tyq}) and (\ref{eq:C4q})), {for which a rotation mixed a Bloch state into a linear combination of several Bloch states.}

\subsection{\label{sec:kirrep}Irreps at high symmetry points}

We now address how to determine irreps of the symmetry group at each momentum.
A Bloch wave function at a particular momentum $\mathbf{k}$ transforms as a representation of the little group at $\mathbf{k}$, denoted $G_\mathbf{k}$, which consists of all the space group operations that leave $\mathbf{k}$ invariant up to a reciprocal lattice vector:
\begin{equation}
    \label{eq:defGk}
    G_\mathbf{k} = \lbrace g \in G | g\mathbf{k} \equiv \mathbf{k} \rbrace,
\end{equation} 
where $\equiv$ is defined by equality up to a reciprocal lattice vector.
Since the lattice translations are always represented by Bloch phases in the representations, it is useful to label the wave functions by irreps of the little co-group, defined as
\begin{equation}
    \label{eq:deftildeGK}
    \widetilde{G}_\mathbf{k}=G_\mathbf{k}/{\mathbb T}_M
\end{equation}
As mentioned above, for the little co-group to satisfy the definition of a group, $\mathbb{T}_M$ must be a normal subgroup of $G_\mathbf{k}$, i.e., for all $g\in G_\mathbf{k}$, $t\in \mathbb{T}_M$, $g^{-1} t g \in \mathbb{T}_M$.
One can check that for the $q$-by-$1$ unit cell, the magnetic translation group is a normal subgroup of the layer group $p2$, but it is not normal for the layer groups containing three- or four-fold rotations (because, for example, $C_4^{-1} T(\hat{\mathbf{x}}) C_4 = T(\hat{\mathbf{y}})^{-1}$, which is not in the magnetic translation group for the $q$-by-$1$ unit cell.)
Thus, we use the $q$-by-$q$ unit cell for layer groups with three- or four-fold rotations.

Thus, under magnetic flux, the little co-groups and their irreps differ from their zero-flux analogues in two important ways:
first, in the presence of magnetic flux, the little co-groups include sublattice translation symmetries; and second, the irreps of little co-groups in the presence of magnetic flux are projective representations corresponding to the 2-cocyle defined by the flux.



We now study some examples: in Tables~\ref{tab:p2}, \ref{tab:p4} and \ref{tab:p4TI} we summarize the projective irreps at high symmetry points for the layer groups $p2$, $p4$, $p4/m'$ at flux $\phi=\pi$. 
For later convenience we have assumed there is spin-orbit coupling, i.e., $C_n^n=-1$.
Notice that the character tables are not square, which is a general feature of projective representations.
The projective irreps corresponding to a particular 2-cocycle can be considered as a subset of non-projective representations of a larger group; 
the character table of that larger group will be square. 

To ensure that we have found all the projective irreps, we use the theorem by Schur~\cite{bradley2010mathematical} stating that for irreducible projective representations with a particular $2$-cocyle, 
\begin{equation}
\label{eqn:schur}
    \sum_{\rho} \left(\text{dim}(\rho)\right)^2=|\widetilde{G}_{\mathbf k}|,
\end{equation}
where the sum runs over all projective irreps $\rho$ of $\widetilde{G}_{\mathbf k}$ with the specified $2$-cocyle and $\widetilde{G}_{\mathbf k}$ is the little co-group defined above. (Notice this formula does not apply to anti-unitary groups.)

The calculation of the irreps of little co-groups are shown in Appendix~\ref{app:irreps} with the (anti)-commutation relations for the magnetic symmetries shown in Appendix~\ref{app:rotation}. 
In the remainder of this section, we sketch the calculation for the simplest non-trivial case, layer group $p2$ at $\pi$ flux.

For the $2$-by-$1$ unit cell, the group of magnetic lattice translations is ${\mathbb T}_M=\{T(n_1{\hat{\mathbf x}}+2n_2{\hat{\mathbf y}})|n_1,n_2\in \mathbb Z\}$ and the Brillouin zone is $[-\pi,\pi)\times[-\pi/2,\pi/2)$. 
We now determine the high-symmetry points.
Since $C_2$ symmetry maps $(k_x,k_y)$ to $(-k_x,-k_y)$,  there are four momenta that are symmetric under $C_2$ up to a magnetic reciprocal lattice vector: $(0,0),(0,\pi/2),(\pi,0),(\pi,\pi/2)$.
Since $T(\hat{\mathbf y})$ maps $(k_x,k_y)$ to $ (k_x+\pi,k_y)$ (Eq.~(\ref{eq:Ty})), $T(\hat{\mathbf y})C_2$ maps $(k_x,k_y)$ to $ (-k_x+\pi,-k_y)$.
Therefore, there are four $T(\hat{\mathbf y})C_2$ symmetric momenta, $(\pm \pi/2,0)$ and $(\pm \pi/2, \pi/2)$.

We derive in Appendix~\ref{app:irreps} that the $C_2$ eigenvalues at $(\pi,0)$ are the same as $(0,0)$, while the $C_2$ eigenvalues at $(0,\pi/2)$ are opposite of $(\pi,\pi/2)$. The same relations hold for the $T(\hat{\mathbf y})C_2$ symmetric points.
In conclusion, there are two independent $C_2$ symmetric points, $\Gamma=(0,0)$ and $Y=(0,\pi/2)$, and we find that each has two one-dimensional irreps labeled by $C_2$ eigenvalue $+i$, $-i$.
There are also two independent $T(\hat{\mathbf y})C_2$ symmetric points, $X=(\pi/2,0)$ and $M=(\pi/2,\pi/2)$, and each has two one-dimensional irreps labeled by $T(\hat{\mathbf y})C_2$ eigenvalue $+i$, $-i$. 
Since each little co-group contains the identity element and either $C_2$ or $T(\hat{\mathbf y})C_2$, $|\widetilde{G}_{\mathbf k}|=2$ for these points. Thus, Eq.~(\ref{eqn:schur}) is satisfied, which means we have found all the projective irreps.

\begin{table}[]
    \centering
    \begin{tabular}{c|c|c|c|c|c|c|c|c}
    &\multicolumn{2}{c|}{$X(\pi/2,0)$} &\multicolumn{2}{c|}{$Y(0,\pi/2)$} &\multicolumn{2}{c|}{$\Gamma(0,0)$} &\multicolumn{2}{c}{$M(\pi/2,\pi/2)$} \\
    \hline
     Irrep & $X_1^{(p2)}$ & $X_2^{(p2)}$ & $Y_1^{(p2)}$ & $Y_2^{(p2)}$ & $\Gamma_1^{(p2)}$ & $\Gamma_2^{(p2)}$ & $M_1^{(p2)}$ & $M_2^{(p2)}$\\
         \hline
      $C_2$  &  &   & $+i$ & $-i$  & $+i$ & $-i$  & &  \\
      $T(\hat{\mathbf y})C_2$  & $+i$ & $-i$ &&&& & $+i$ & $-i$ 
    \end{tabular}
    \caption{High symmetry momenta (first row) and the irreps (second row) of their little co-group for the group $p2$. The third and fourth rows list the eigenvalue of the indicated symmetry; the row is blank if the symmetry is not in the little co-group.}
    \label{tab:p2}
\end{table}
    
\begin{table}[]
    \begin{tabular}{c|c|c|c|c}
    &\multicolumn{2}{c|}{$X(\pi/2,0)$} &\multicolumn{2}{c}{$Y(0,\pi/2)$}\\ 
    \hline
     Irrep & $X_1$ & $X_2$ & $Y_1$ & $Y_2$\\
         \hline
      $C_2$  & $i\sigma_z$ & $-i\sigma_z$ & $i\sigma_z$ & $-i\sigma_z$\\
      $T(\hat{\mathbf x})$  & $\sigma_x$ & $\sigma_x$  & $\sigma_z$ & $\sigma_z$\\
      $T(\hat{\mathbf y})$  & $\sigma_z$ & $\sigma_z$  & $\sigma_y$ & $\sigma_y$\\
      $T(\hat{\mathbf x})T(\hat{\mathbf y})$  & $-i\sigma_y$ & $-i\sigma_y$  & $-i\sigma_x$ & $-i\sigma_x$  \\ \hline 
      \end{tabular}
      \vspace{10pt}
      
      \begin{tabular}{c|c|c|c|c}
      \hline
      &\multicolumn{4}{c}{$\Gamma(0,0)$} \\ \hline
      Irrep & $\Gamma_1$ & $\Gamma_2$ & $\Gamma_3$ & $\Gamma_4$ \\ \hline
      $C_4T(\hat{\mathbf x})$ & $\begin{pmatrix}
      1&\\&i \end{pmatrix}$
      & $\begin{pmatrix}
      i&\\&-1 \end{pmatrix}$
      & $\begin{pmatrix}
      -1&\\&-i \end{pmatrix}$
      & $\begin{pmatrix}
      -i&\\&1 \end{pmatrix}$\\
      $T(\hat{\mathbf x})T(\hat{\mathbf y})$ & $i\sigma_z$ & $i\sigma_z$ & $i\sigma_z$ & $i\sigma_z$ 
    \end{tabular}
      \vspace{10pt}
      
      \begin{tabular}{c|c|c|c|c}
      \hline
      &\multicolumn{4}{c}{$M(\pi/2,\pi/2)$} \\ \hline
      Irrep &  $M_1$ & $M_2$ &$M_3$ &$M_4$\\ \hline
      $C_4$  &$\begin{pmatrix} \epsilon &\\&\epsilon^*\end{pmatrix}$ &$\begin{pmatrix} -\epsilon^* &\\&\epsilon\end{pmatrix}$ &$\begin{pmatrix} -\epsilon &\\&-\epsilon^*\end{pmatrix}$ &$\begin{pmatrix} \epsilon^* &\\&-\epsilon\end{pmatrix}$ \\
      $T(\hat{\mathbf x})T(\hat{\mathbf y})$  & $i\sigma_z$ & $i\sigma_z$ & $i\sigma_z$ &$i\sigma_z$ 
    \end{tabular}
    \caption{High symmetry momenta (first row) and the irreps (second row) of their little co-group for the group $p4$. Subsequent rows list the eigenvalue of the indicated symmetry with $\epsilon=e^{i\pi/4}$; the row is blank if the symmetry is not in the little co-group.}
    \label{tab:p4}
\end{table}
    
\begin{table}[]
    \begin{tabular}{c|c|c|c|c|c|c|c}
    &{$X(\pi/2,0)$} &{$Y(0,\pi/2)$} &\multicolumn{2}{c|}{$\Gamma(0,0)$} &\multicolumn{3}{c}{$M(\pi/2,\pi/2)$} \\ \hline
        Irrep &$X_1X_2$ &$Y_1Y_2$ &$\Gamma_1\Gamma_4$&$\Gamma_2\Gamma_3$ &$M_1M_1$ &$M_3M_3$ &$M_2M_4$  
    \end{tabular}
    \caption{High symmetry momenta (first row) and the irreps (second row) of their little co-group for the group $p4/m'$. The notation $\Pi_i\Pi_j$ indicates that $\Pi_i$ and $\Pi_j$ are paired by $\cal TI$ symmetry, where $\Pi_i$ is an irrep of the corresponding little co-group with respect to layer group $p4$, shown in Table~\ref{tab:p4}.}
    \label{tab:p4TI}
\end{table}

\section{\label{sec:TQC}Topological quantum chemistry in a magnetic field}


Finally we turn to the theory of TQC.
TQC classifies topological crystalline insulators (TCIs) by enumerating all trivial phases in each space group, where
a trivial phase is defined as one where exponentially localized Wannier functions exist and transform locally under all symmetries.
A group of bands can be identified as a TCI if it is not in the space of trivial phases.


Together, the Wannier functions corresponding to a single band (or group of bands) transform as a representation of the full space group, called a band representation \cite{zak1980symmetry,zak1981band,michel1999connectivity,michel2000elementary,michel2001elementary,bradlyn2017topological,cano2018building}. 
TQC labels each band representation by how its Bloch wave functions transform under symmetry at high symmetry momenta, i.e., by a set of irreps of the little co-group at each high symmetry momentum; this label is known as a symmetry indicator~\cite{po2017symmetry}.
Symmetry indicators provide a practical way to identify many TCIs: specifically, a group of bands whose irreps at high symmetry momenta are not consistent with any of the trivial phases must be topological.

In Sec.~\ref{sec:magneticWannier} we describe how to construct a basis of symmetric magnetic Wannier functions.
We use this basis in Sec.~\ref{sec:inducedrep} to derive how the space group symmetries act on the Wannier functions; the symmetry matrices comprise the band representation.
Fourier transforming the band representation yields its symmetry indicator.

\subsection{Magnetic Wannier functions}
\label{sec:magneticWannier}



We now describe how to construct a basis of symmetric Wannier functions for a magnetic unit cell.
Given a site $\mathbf{q}$, which will serve as a Wannier center, the site-symmetry group $G_\mathbf{q}$ is defined as the set of symmetries that leave $\mathbf{q}$ invariant, i.e., $G_\mathbf{q} = \lbrace g\in G | g\mathbf{q} = \mathbf{q} \rbrace$.
The site-symmetry group defines a coset decomposition of the space group,
\begin{equation}
\label{eqn:cosetdecomp}
    G=\bigcup_{\alpha} g_\alpha G_{\mathbf q} \ltimes {\mathbb T}_M,
\end{equation}
where $G$ is the space group, ${\mathbb T}_M$ is the {magnetic} lattice translation group, and $\alpha=1, \dots, n$, where $n=|G/{\mathbb T}_M|/|G_{\mathbf q}|$ is the multiplicity of the Wyckoff position containing $\mathbf{q}$.
The symmetries $g_\alpha$ are coset representatives.
The choice of coset representatives is not unique; a different choice will yield a band representation related to the original by a unitary transformation, while the symmetry indicator is unchanged.

\begin{figure}[b]
    \centering
    \includegraphics[width=\linewidth]{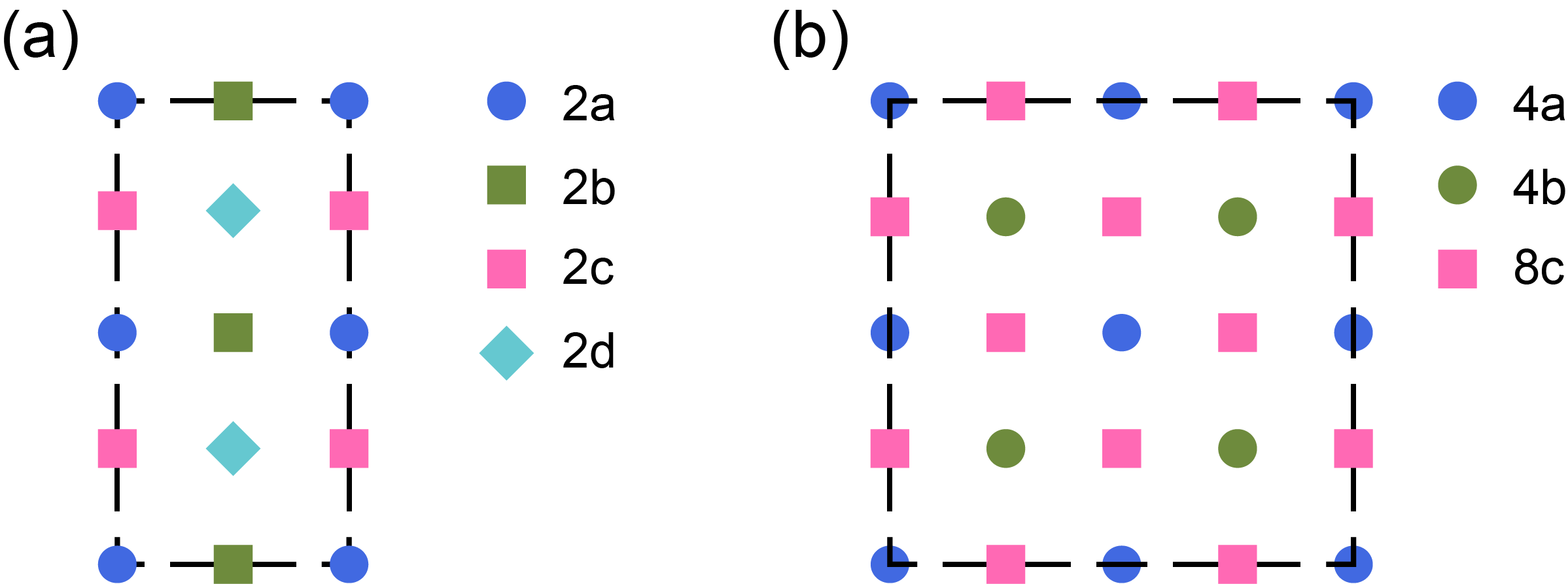}
    \caption{(Color online) Wyckoff positions in a magnetic flux $\pi$ for (a) the $2$-by-$1$ unit cell for group $p2$ and (b) the $2$-by-$2$ unit cell for the groups $p4$ and $p4/m'$. Each Wyckoff position is labelled by its multiplicity and a lowercase letter. The general Wyckoff position, whose site-symmetry group consists of only the identity, is not shown. }
    \label{fig:WC}
\end{figure}

The coset representatives define positions ${\mathbf q}_\alpha = g_\alpha {\mathbf q}$ that form the orbit of $\mathbf q$ within the magnetic unit cell.
Together, these points are part of the same Wyckoff position, whose multiplicity $n$ is equal to the number of points in the orbit of $\mathbf{q}$ in the magnetic unit cell.
Unlike the case of zero magnetic field, the set of coset representatives $g_\alpha$ includes pure translations within the magnetic unit cell.
Fig.~\ref{fig:WC} shows the Wyckoff positions for the groups $p2$, and $p4$, $p4/m'$.

Suppose there are $n_{\mathbf q}$ orbitals centered at $\mathbf q$. These orbitals are described by $n_{\mathbf q}$  Wannier functions $|W_{i1}\rangle$, where $i=1\dots n_{\mathbf q}$. The Wannier functions transform under symmetries $g\in G_{\mathbf q}$ as a projective representation $\rho$ of $G_{\mathbf q}$, 
\begin{equation}
\label{eqn:rep_sitegroup}
    g|W_{i1}\rangle =\sum_{j=1}^{n_{\mathbf q}}[\rho(g)]_{ji}|W_{j1}\rangle.
\end{equation}
Applying the representatives $g_\alpha$ in the coset decomposition of the space group $G$ in Eq.~(\ref{eqn:cosetdecomp}) to $|W_{i1}\rangle$ gives another Wannier function 
\begin{equation}
\label{eqn:Walpha}
    |W_{i\alpha}\rangle = g_\alpha |W_{i1}\rangle,
\end{equation}
localized at ${\mathbf q}_\alpha$. All these Wannier functions $|W_{i\alpha}\rangle$, where $i=1\dots n_{\mathbf q}$ and $\alpha=1\dots n$, form an induced representation of $G$, as we now explain.

\par

\subsection{Induced representation}
\label{sec:inducedrep}

In this section, we derive how the space group symmetries act on the Wannier functions.
This provides an explicit construction of a band representation with Wannier functions as a basis.
Fourier transforming the band representation gives the irreps of the little co-group at each high-symmetry point, i.e., the symmetry indicator.

Consider a group element $h g_{\alpha} \in G$.
The coset decomposition in Eq.~(\ref{eqn:cosetdecomp}) implies that $hg_\alpha$ can be written in the form
\begin{equation}
\label{eqn:hg_coset}
    hg_\alpha = e^{if_{\alpha\beta}(h)}\{E|{\mathbf t}_{\alpha\beta}(h)\}g_\beta g
\end{equation}
where $\mathbf t_{\alpha\beta}(h) = h{\mathbf q}_\alpha -{\mathbf q}_\beta$ and $ \{E|\mathbf{t}_{\alpha\beta}\}\in \mathbb{T}_M$, $g\in G_{\mathbf q}$, and the coset representative $g_\beta$ are uniquely determined by the coset decomposition.
The remaining phase factor $f_{\alpha\beta}(h)$ is due to the non-trivial 2-cocycles. For the two-dimensional systems without magnetic field, $f_{\alpha\beta}(h)\equiv0$. For the case with magnetic field, in general $f_{\alpha\beta}(h)$ is nonzero. 

The phase factor $f_{\alpha\beta}(h)$ is the new ingredient that appears in a magnetic field and is a key result of the present work; it does not appear in the non-magnetic theory (for example, it does not appear in Eq.~(6) in Ref.~\onlinecite{cano2018building}).
This phase factor is gauge invariant because it results from the commutations between rotations and translations (see Appendix~\ref{app:rotation}). We briefly give two examples to show how this phase factor appears.


As a first example, consider the layer group $p1$ with a $2$-by-$2$ unit cell, corresponding to $\pi$ flux.
{Starting from a Wannier function centered at a general position $\mathbf{q} = (x,y)$, the coset representatives in Eq.~(\ref{eqn:cosetdecomp}) can be chosen as $g_1=\{E|0\}$, $g_2 = T(\hat{\mathbf x}) $, $g_3 = T(\hat{\mathbf y})$, $g_4=T(\hat{\mathbf x})T(\hat{\mathbf y})$. }
Now consider the left-hand-side of Eq.~(\ref{eqn:hg_coset}) with $h=T(\hat{\mathbf y})$, $g_\alpha=T(\hat{\mathbf x})$.
Then on the right-hand-side of Eq.~(\ref{eqn:hg_coset}), $g_\beta=T(\hat{\mathbf x})T(\hat{\mathbf y})$, $g=E$, and ${\mathbf t}_{\alpha\beta}=0$. Since $T(\hat{\mathbf y})T(\hat{\mathbf x}) = e^{i\pi} T(\hat{\mathbf x})T(\hat{\mathbf y})$, $f_{\alpha\beta}(h)=\pi$.

As a second example, consider layer group $p4$ with a $2$-by-$2$ unit cell, corresponding again to $\pi$ flux. 
{Starting from a Wannier function centered at $\mathbf{q} = (\frac{1}{2}, \frac{1}{2})$, the coset representatives in Eq.~(\ref{eqn:cosetdecomp}) can be chosen as $g_1=\{E|0\}$, $g_2 = T(\hat{\mathbf x}) $, $g_3 = T(\hat{\mathbf y})$, $g_4=T(\hat{\mathbf x})T(\hat{\mathbf y})$.}
Now consider the left-hand-side of Eq.~(\ref{eqn:hg_coset}) with $h=C_4$, $g_\alpha=T(\hat{\mathbf x})$. 
The coset decomposition uniquely determines $g_\beta=T(\hat{\mathbf x})T(\hat{\mathbf y})$, $g=C_4(\frac{1}{2},\frac{1}{2})$, {and $\mathbf{t}_{\alpha\beta} = (-2,0)$ on the right-hand-side of Eq.~(\ref{eqn:hg_coset}).}
Since $ C_4 T(\hat{\mathbf x}) = e^{i3\pi/4}\{E|(-2,0)\} T(\hat{\mathbf x})T(\hat{\mathbf y})C_4(1/2,1/2) $,
the extra phase term $f_{\alpha\beta}(h)=3\pi/4$.

As discussed at the start of Sec.~\ref{sec:TQC}, the set of Wannier functions {centered at all $\mathbf q_\alpha$} form a basis for a band representation, which we denote $\rho_G$.
Given a space group symmetry $h\in G$, Eq.~(\ref{eqn:hg_coset}) determines the matrix elements of $\rho_G(h)$ in the basis of Wannier functions defiend in Eq.~(\ref{eqn:Walpha}) by:
\begin{align}
\label{eqn:inducedR}
     \rho_{G}(h) |W_{i\alpha}(\mathbf r-\mathbf t)\rangle &= e^{if_{\alpha\beta}(h)}[\rho(g)]_{ji}|W_{j\beta}(\mathbf r-R\mathbf t-{\mathbf t}_{\alpha\beta})\rangle 
\end{align}
where $R$ is the rotational part of $h$, $\rho(g)$ is the given representation defined in Eq.~(\ref{eqn:rep_sitegroup}), $\mathbf t_{\alpha\beta}(h) = h{\mathbf q}_\alpha -{\mathbf q}_\beta$ and sum over $j=1\dots n_{\mathbf q}$ is implied.


Substituting the Fourier transformed Wannier functions, 
\begin{align}
     |W_{j\beta}(\mathbf r-\mathbf t) \rangle = \int d\mathbf k e^{i{\mathbf k}\cdot{\mathbf t}} |a_{j\beta}(\mathbf k,\mathbf r)\rangle, \\
     |a_{j\beta}(\mathbf k,\mathbf r)\rangle = \sum_{\mathbf t} e^{-i{\mathbf k}\cdot{\mathbf t}} |W_{j\beta}(\mathbf r-\mathbf t) \rangle, 
\end{align}
into Eq.~(\ref{eqn:inducedR}) yields
\begin{align}
\label{eqn:inducedk}
     \rho_{G}(h) |a_{i\alpha}(\mathbf k,\mathbf r)\rangle &= e^{if_{\alpha\beta}(h)-iR{\mathbf k}\cdot {\mathbf t}_{\alpha\beta}}[\rho(g)]_{ji}|a_{j\beta}(R\mathbf k,\mathbf r)\rangle 
\end{align}



From Eq.~(\ref{eqn:inducedk}), a representation of the little co-group (defined in Sec.~\ref{sec:kirrep}) is determined from $\rho_G$ by restricting each matrix $\rho_G(h\in \widetilde{G}_\mathbf{k})$ to only the rows and column corresponding to Fourier-transformed Wannier functions at $\mathbf{k}$. The set of irreps obtained at all $\mathbf{k}$ determines the symmetry indicator following the procedure we introduced in Ref.~\onlinecite{fang2021filling}, which is summarized in Appendix~\ref{app:TQC}.

We now derive the symmetry indicator classification for a few examples.

\subsection{Examples}
\label{sec:applications}

We apply our formalism of TQC in a magnetic flux to three magnetic layer groups with $\pi$ flux: $p2$, $p4$, and $p4/m'$. 
In each case, we discuss the stable symmetry indicator classification; the derivations are in Appendix~\ref{app:TQC}.
We further apply our formalism to derive the Chern number indicators in $\pi$ flux for groups $p3$, $p4$ and $p6$ without spin-orbit coupling. They are shown in Appendix~\ref{app:Chern_Cn}.
\par
\subsubsection{p2}
For layer group $p2$ with $\pi$ flux, we choose a $2$-by-$1$ magnetic unit cell, following the discussion in Sec.~\ref{sec:mag_sym_3_momentum}. 
The symmetry indicator has a $\mathbb{Z}_4$ classification.
The indicator for a particular group of bands is
\begin{multline}
\label{eqn:indexp2}
    \text{index} = \#\Gamma_1-\#\Gamma_2+\#X_1-\#X_2 
    +\#Y_1-\#Y_2 
    \\
    +\#M_1-\#M_2-N \mod 4
\end{multline}
where $\#\Pi_i$ indicates the number of times the irrep $\Pi_i$ at the high symmetry point $\Pi$ appears in the bands and $N=\#\Gamma_1+\#\Gamma_2=\#X_1+\#X_2=\#Y_1+\#Y_2=\#M_1+\#M_2$ is the filling per magnetic unit cell.
\par
{This indicator Eq.~(\ref{eqn:indexp2}) is exactly the same as the Chern number indicator Eq.~(3) in Ref.~\onlinecite{matsugatani2018universal} in the special case of $\pi$-flux and spinful electrons, i.e.,}
\begin{equation}
\label{eqn:Chern}
    e^{i\pi (C/q-\bar{\rho})}=(-)^{2SN}w_{C_2}^\Gamma w_{C_2}^Y w_{T(\hat{\mathbf y})C_2}^X w_{T(\hat{\mathbf y})C_2}^M
\end{equation}
where {at flux $\pi$}, $q=2$; $\bar{\rho}=N/2$ is the filling per non-magnetic primitive unit cell; $S=1/2$ is the spin (angular momentum) quantum number; and $w_{g}^{\Pi}$ is the product of eigenvalues of the symmetry $g$ for all filled bands at momentum $\Pi$.

\subsubsection{p4 }
For layer group $p4$ with $\pi$ flux, we choose $2$-by-$2$ unit cell, following the discussion in Sec.~\ref{sec:mag_sym_3_momentum}. 
The symmetry indicator has a  $\mathbb{Z}_8$ classification.
The indicators for a particular group of bands are determined by:
\begin{multline}
\label{eqn:indexp4}
\text{index}=2\# \Gamma_1+4\# \Gamma_2-2\# \Gamma_3+\# M_1+3\# M_2\\
-3\# M_3-\# M_4+4\# X_1 \mod 8
\end{multline}
To understand this indicator, we compare the new index to the symmetry indicator formula for Chern number in Ref.~\onlinecite{fang2012bulk}:
\begin{equation}
\label{eqn:Chernnumber}
    e^{i\frac{\pi}{2} C}=(-)^{2SN}w_{C_4}^\Gamma  w_{C_2}^X w_{C_4}^M,
\end{equation}
which, in terms of irreps, is given by (derivation in Appendix~\ref{app:Chern})
\begin{align}
    \label{eqn:ChernumberIrrep}
    C = 2N &+\#\Gamma_1+\#\Gamma_3-\#\Gamma_2-\#\Gamma_4 \nonumber \\
    &+2(\#M_2+\#M_4) \mod 4
\end{align}
$N$ is always an even integer due to the two-dimensional irreps (shown in Table~\ref{tab:p4}). We conclude 
\begin{equation}
    C = \text{index} \mod 4
\end{equation}
In fact this $\mathbb Z_8$ index is exactly the Chern number mod $8$. This can be seen by considering a $\sqrt{2}$-by-$\sqrt{2}$ unit cell which has lattice vectors $\mathbf a_1=(1,1),~ \mathbf a_2=(-1,1)$; in this basis, Eq.~(\ref{eqn:indexp4}) is identified as the $\mathbb Z_8$ Chern number indicator in Ref.~\onlinecite{matsugatani2018universal}.

\subsubsection{p4/m'}
For layer group $p4/m'$ with $\pi$ flux, we choose $2$-by-$2$ unit cell, following the discussion in Sec.~\ref{sec:mag_sym_3_momentum}. 
The symmetry indicator has a $\mathbb{Z}_2$ classification.
The indicator for a particular group of bands is determined by:
\begin{align}
\label{eqn:index1p4TI}
\text{index}&= N/4 \mod 2,
\end{align}
where $N/4\equiv \bar \rho$ is the filling per original unit cell.
Notice each band in this group is four-fold degenerate (see Table~\ref{tab:p4TI} in Sec.~\ref{sec:kirrep}),
and hence $\bar \rho \in \mathbb Z$.

The group $p4/m'$ is generated by a four-fold rotation and the product of time-reversal and inversion symmetry $\cal TI$. As is well known, $\cal TI$ prevents a non-vanishing Chern number~\cite{bernevig2013topological} and the absence of $\cal T$ prevents the existence of strong topological insulator~\cite{fu2006time}. 
Since $\cal T$ and $\cal I$ are not separately symmetries, there is no mirror symmetry and hence no mirror Chern number. Thus, our stable index is a new phase that only exists in systems with magnetic flux.

This phase is realized in the model we present in Sec.~\ref{sec:Model}.
However, it does not realize an anomalous gapless boundary state because when the boundary is opened, the sublattice translation symmetries that protect the phase are broken. 
\par

\section{\label{sec:Model} Application to a quadrupole insulator}


{In this section, we apply our results to a model on the square lattice. At zero flux, this model is a quadrupole insulator that exhibits corner states. Since the symmetries that protect the corner states are preserved in the presence of a perpendicular magnetic field, the corner states must survive when magnetic flux is introduced. We use the formalism developed in the previous sections to verify the presence of corner states using symmetry indicators.
Finally, we show that at a critical magnetic flux, the bulk gap closes and the corner states disappear, as shown in the Hofstadter butterfly spectrum in Fig.~\ref{fig:Hof}. We use the symmetry indicators to verify that when the corner states disappear, the symmetry indicator vanishes.}

{Our results provide a new probe of the higher order topology in the model, i.e., the presence of a gap closing phase transition in the presence of a magnetic field, which may be easier to observe than probing the corner states directly.
}

\subsection{\label{sec:Model_model} Model}
\begin{figure}
    \centering
    \includegraphics[width=\linewidth]{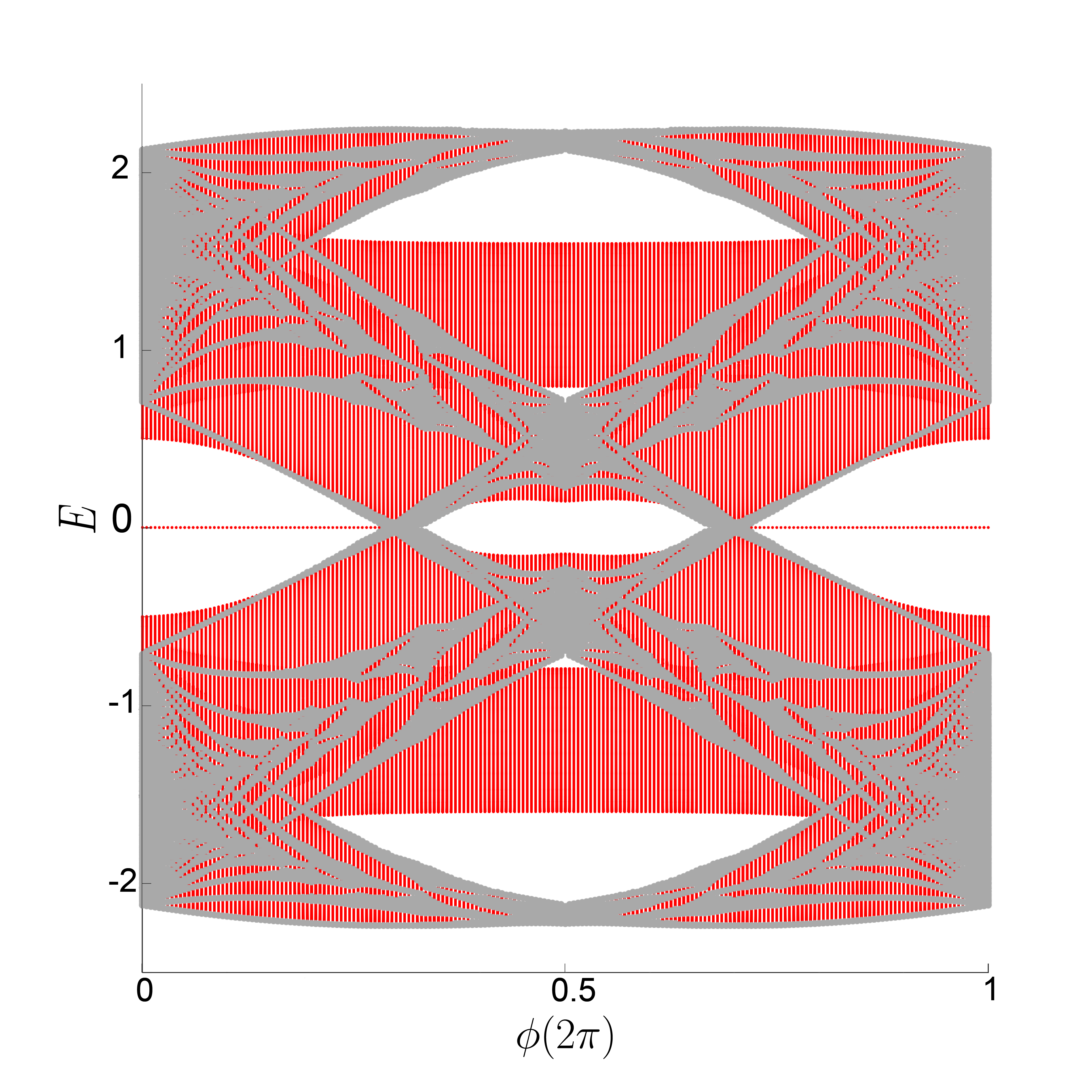}
    \caption{(Color online) Hofstadter spectrum of the OAL model. The grey states are calculated with periodic boundary conditions and show the bulk gap closing at a critical flux. The red states are calculated with open boundary conditions and show the disappearance of the corner states upon bulk gap closure. 
    The spectrum is computed for dimensions $L_x=200$, $L_y=10$ and parameters $\lambda=1$, $\gamma=0.5$.}
    \label{fig:Hof}
\end{figure}

{We study a model proposed by Wieder et al in Ref.~\onlinecite{wieder2020strong}}
at zero flux that has the same momentum space Hamiltonian as the {$C_4$ symmetric} Bernevig-Benalcazar-Hughes (BBH) model which is at $4\pi$ flux per unit cell.  \cite{benalcazar2017quantized,benalcazar2017electric}
{This model was given as an example in Fig.~3 and Appendix~A of Ref.~\onlinecite{wieder2020strong}}.
Yet the two models have some fundamental differences: while the BBH model has four atoms per unit cell and one orbital per atom, Wieder's model has one atom sitting at the origin of the unit cell and four orbitals per atom. 
Since the position of atoms in the unit cell will be important when we include magnetic flux, the two models have different Hofstadter spectra. 
{Further, the BBH model describes spinless fermions, while Wieder's model describes fermions with spin-orbit coupling. As a result, the symmetry representations for the two models are different.}

In momentum space the Hamiltonian is
\begin{align}
    H(\mathbf k)&=(v_m+t_1(\cos(k_x)+\cos(k_y)))\Gamma_3 \nonumber \\
    &+t_2(\cos(k_x)-\cos(k_y))\Gamma_4 \nonumber \\
    &+u \sin(k_x)\Gamma_1+u \sin(k_y)\Gamma_2,
\end{align}
where $\Gamma_1=\tau_y\sigma_y$, $\Gamma_2=\tau_y\sigma_x$, $\Gamma_3=\tau_z$, and $\Gamma_4=\tau_x$.
{The Pauli matrices $\tau$ and $\sigma$ together span the orbital space of each atom.}
In the limit $t_1=t_2=u/\sqrt 2=\lambda/\sqrt 2$ and $v_m=\sqrt 2 \gamma$, this Hamiltonian is equivalent to the BBH Hamiltonian after a basis change. In this section, we adopt these parameters and set $\lambda=1$ and $\gamma=0.5$ so that the system is a quadrupole insulator at zero flux.

The generators of the symmetries of this Hamiltonian take the matrix forms:
\begin{align}
\label{eqn:C4z}
    C_{4z}&=\tau_z\left(\frac{{\mathbb I}_\sigma-\sigma_z}{2}\right),\\
    M_x&=\sigma_x,\\
    {\cal TI}&=\sigma_y {\cal K},
\end{align}
where $\cal K$ is the complex conjugation. There is also a chiral symmetry that anti-commutes with the Hamiltonian
\begin{equation}
    \Gamma_5=\tau_y\sigma_z
    \label{eq:chiral}
\end{equation}

{To incorporate the effect of a magnetic field, we need the real space Hamiltonian, given by:}
\begin{align}
    H&=\sum_{i,j\in \mathbb Z}\mathbf t_x c^\dagger_{(i+1,j)} c_{(i,j)}+\mathbf t_y c^\dagger_{(i,j+1)} c_{(i,j)}+h.c. \nonumber \\
    &\quad +v_m\Gamma_3 c^\dagger_{(i,j)} c_{(i,j)}
    \label{eqn:Ham}
\end{align}
where $\mathbf{t}_{x,y}$ are hopping matrices given by
\begin{align}
    \mathbf t_x=(t_1\Gamma_3+t_2\Gamma_4)/2-iu\Gamma_1/2\\
    \mathbf t_y=(t_1\Gamma_3-t_2\Gamma_4)/2-iu\Gamma_2/2
\end{align}

When a magnetic field in the $z$-direction is turned on, the Hamiltonian in Eq.~(\ref{eqn:Ham}) requires the Peierls substitution \cite{hofstadter1976energy}. Working in Landau gauge, $\mathbf A(x,y)=(-\phi y,0)$ where $\phi=B$ is the flux per unit cell and the substitution is given by
\begin{align}
    c^\dagger_{(i+1,j)} c_{(i,j)} &\mapsto e^{-i\phi j}c^\dagger_{(i+1,j)} c_{(i,j)}\\
    c^\dagger_{(i,j+1)} c_{(i,j)} &\mapsto c^\dagger_{(i,j+1)} c_{(i,j)}
\end{align}
The momentum space Hamiltonian at finite flux can be obtained by Fourier transforming Eq.~(\ref{eqn:Ham}) using the convention in Eqs.~(\ref{eqn:FT_qby1_cd}) and (\ref{eqn:FT_qby1_c}) 
when the flux is rational $\phi=2\pi p/q$.

In Fig.~\ref{fig:Hof}, we numerically compute the Hofstadter spectrum for this model.


\subsection{\label{sec:Model_analysis} Symmetry analysis}
The model has a $2\pi$ periodicity in $\phi$, the flux per unit cell. At zero flux and $\pi$-flux the system is invariant under the symmetry group $p4/m'mm$, while at other fluxes the symmetry group is $p4$. 
Using the formalism developed in this manuscript, we apply TQC in a magnetic field to compute the symmetry indicators at $\pi$ flux.
Indicators at other fluxes are discussed in Appendix~\ref{app:symmetry}.
Ultimately, we will show that the symmetry indicator at $\pi$ flux corresponds to an absence of corner states, from which we deduce there must be a gap closing phase transition at a critical flux between zero and $\pi$.

At $\pi$-flux, the magnetic unit cell is $2$-by-$2$ and the Brillouin zone is $[-\pi/2,\pi/2]\times[-\pi/2,\pi/2]$.  According to Sec.~\ref{sec:mag_sym_3_momentum}, the four-fold rotation symmetry operators at $\Gamma=(0,0)$ and $M=(\pi/2,\pi/2)$ are 
\begin{align}
    D(C_{4z},\Gamma)&=
    \begin{pmatrix}
    1&&&\\
    &&1&\\
    &1&&\\
    &&&-1
    \end{pmatrix}\otimes C_{4z}, \\
    D(C_4,M))&=
    \begin{pmatrix}
    1&&&\\
    &&-1&\\
    &1&&\\
    &&&1
    \end{pmatrix}\otimes C_{4z},
\end{align}
{where the first matrix acts on the sublattice basis,} and the $C_{4z}$ matrix acts on the orbital basis as defined in Eq.~(\ref{eqn:C4z}).
The magnetic translation symmetries at $\mathbf k$ are implemented by
\begin{align}
D(T(\hat{\mathbf x}),\mathbf k)&=e^{ik_x}
    \begin{pmatrix}
    &1&&\\
    1&&&\\
    &&&1\\
    &&1&
    \end{pmatrix} \otimes \tau_0\sigma_0 \\
    D(T(\hat{\mathbf y}),\mathbf k)&=e^{ik_y}
    \begin{pmatrix}
    &&1&\\
    &&&-1\\
    1&&&\\
    &-1&&
    \end{pmatrix} \otimes \tau_0\sigma_0,
\end{align}
where $\tau_0$ and $\sigma_0$ are identity matrices.

The irreps of the occupied bands are listed in Table~\ref{tab:BRatpi}. Each band is four-fold degenerate 
because $({\cal TI})^2=-1$ and $\{T(\hat{\mathbf x}),T(\hat{\mathbf y})\}=0$, as explained in Appendix~\ref{app:irreps}. 
Using the computed irreps in Table~\ref{tab:BRatpi}, the symmetry indicators are listed in Table~\ref{tab:indatpi}.


\begin{table}
    \centering
    \begin{tabular}{c|c|c|c|c}
         band index& $1$ & $2$ &$3$ &$4$  \\
         \hline
        irrep at $\Gamma$ &$\Gamma_2\Gamma_3$ &$\Gamma_1\Gamma_4$ &$\Gamma_2\Gamma_3$ &$\Gamma_1\Gamma_4$ \\
        irrep at $X$ &$X_1X_2$ &$X_1X_2$ &$X_1X_2$ &$X_1X_2$ \\
        irrep at $M$ &$M_2M_4$ &$M_3M_3$ &$M_1M_1$ &$M_2M_4$
    \end{tabular}
    \caption{Band representation of the four four-fold degenerate bands at $\pi$-flux. The ordering of band index is from lowest energy to highest energy, i.e., half-filling corresponds to filling bands 1 and 2.
    Each irrep $\Pi_i\Pi_j$ is four-dimensional and defined in Table~\ref{tab:p4TI}. }
    \label{tab:BRatpi}
\end{table}
\begin{table}
    \begin{tabular}{c|c|c|c|c}
         band index& $n=1,4$ & $n=2,3$ & $1\oplus2$ &$3\oplus4$  \\
         \hline
        $\mathbb Z_2$ phase (Eq.~(\ref{eqn:index1p4TI}))  &$1$ &$1$ &$0$ &$0$\\
        \hline
        $e_{4a} \mod 8$ &&&$2$&$2$\\
        $e_{4b} \mod 8$ &&&$0$&$0$\\
        $e_{8c} \mod 4$ &&&$0$&$0$ \\
        \hline
        $e_{1a'} \mod 4$ &$0$&$2$&$2$&$2$\\
        $e_{1b'} \mod 4$ &$0$&$2$&$2$&$2$\\
        $e_{2c'} \mod 2$ &$2$&$0$&$2$&$2$ \\
    \end{tabular}
    \caption{Symmetry indicators at $\pi$-flux.
    {The second column corresponds to each four-fold degenerate band individually, while the last two columns correspond to sums of bands.}
    {The second row shows} the strong topological index in Eq.~(\ref{eqn:index1p4TI}) is $1\mod 2$ for each band, while for two occupied/empty bands the index is $0\mod 2$.
    Since symmetric and exponentially localized Wannier functions exist for the two occupied or two empty bands,
    {in the next three rows, $e_\mathbf{q}$ indicates the number of Wannier functions centered at the Wyckoff position $\mathbf{q}$}, computed using Eqs.~(\ref{eqn:4a}) -- (\ref{eqn:8c}) in Appendix~\ref{app:TQC}.
    If the sublattice translation symmetry within each magnetic unit cell is broken, 
    the number of Wannier functions centered at the indicated Wyckoff positions in the lower symmetry group are shown in Fig.~\ref{fig:WCWC}.
    }
    \label{tab:indatpi}
\end{table}

\begin{figure}[h]
    \centering
    \includegraphics[width=0.8\linewidth]{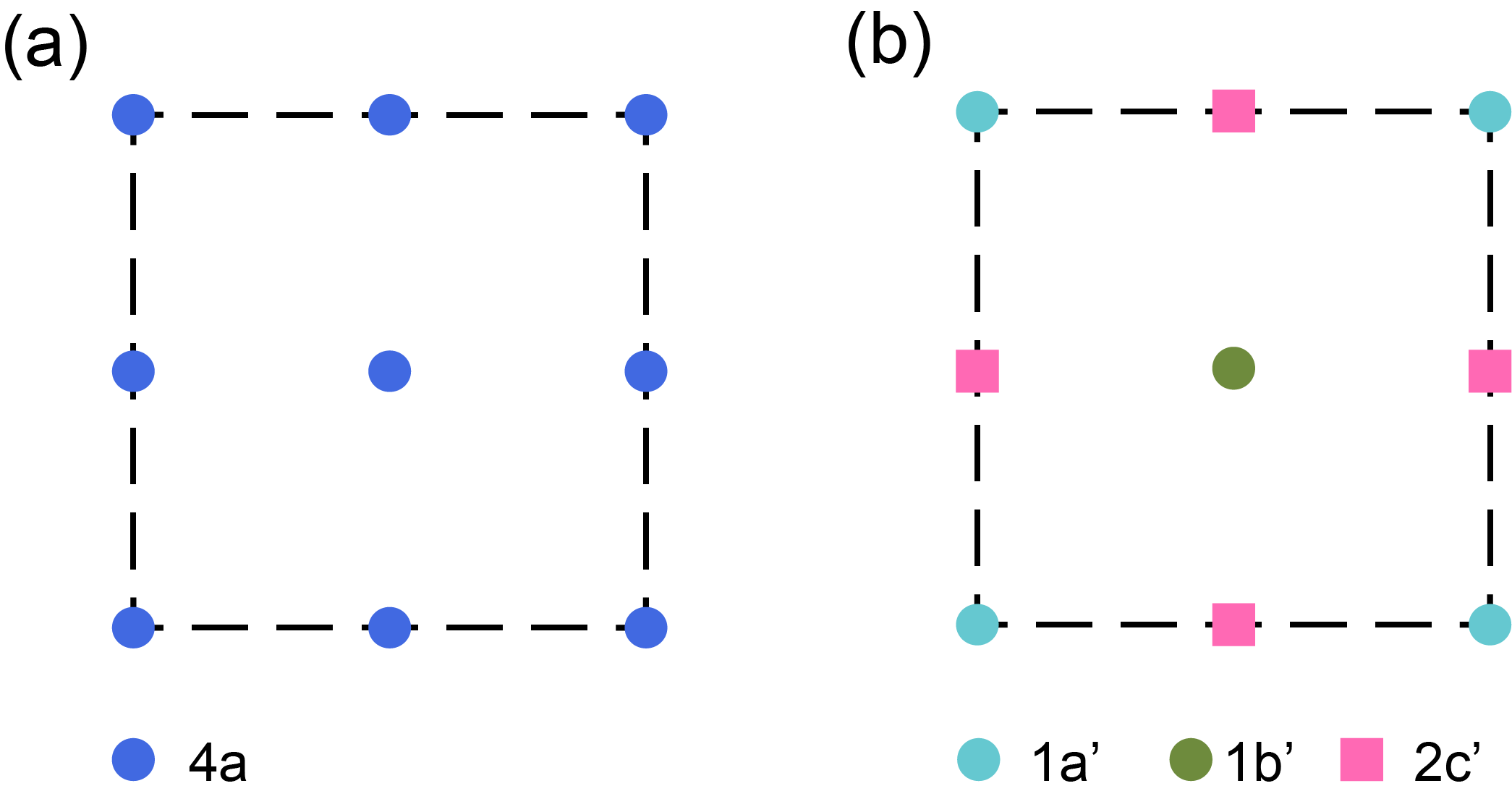}
    \caption{(a) Wyckoff position $4a$ in $2$-by-$2$ unit cell with space group $G$ splits into (b) Wyckoff positions $1a'$, $1b'$, $2c'$ in the same unit cell with symmetry group $G/(\mathbb T/ \mathbb T_M)$, i.e. no sublattice translations. }
    \label{fig:WCWC}
\end{figure}

Below the gap at half filling, the two occupied bands together ($1\oplus2$ in Table~\ref{tab:indatpi}) are topologically trivial. They admit symmetric exponentially localized Wannier functions located at the $4a$ Wyckoff position. Since the atoms are also at $4a$ Wyckoff position, there is no corner charge. This analysis agrees with the Hofstadter spectrum shown in Fig.~\ref{fig:Hof}.

Open boundary conditions break the lattice translation symmetries and, in particular, break the sublattice translation symmetries within the magnetic unit cell.
Once the sublattice translation symmetries are broken, the little co-group (Eq.~(\ref{eq:deftildeGK})) is identical to the non-magnetic case. 
Thus, the crystalline symmetry protected phases with open boundary condition should be labelled by the usual symmetry indicators in zero flux but with respect to the enlarged magnetic unit cell; these indicators are computed in Ref.~\onlinecite{fang2021filling}. 
The results are shown in the lower half of Table~\ref{tab:indatpi}. 
In this reduced symmetry group, the magnetic $4a$ Wyckoff position splits into positions: $1a'=(0,0)$, $1b'=(1,1)$, $2c'=(1,0),(0,1)$, 
as shown in Fig.~\ref{fig:WCWC}.
\par

\subsection{Corner states}
The spectrum with periodic boundary condition has a gap at half filling at $\phi=0$ and $\phi=\pi$. This gap closes at some $\phi^*$ between $0$ and $\pi$ as shown in Fig.~\ref{fig:Hof}. For the spectrum with open boundary condition, there are higher-order topological states
when $-\phi^*<\phi<\phi^*$ that are corner localized.
Due to the chiral symmetry (\ref{eq:chiral}), they are at zero energy in this model. The corner states can be understood by the non-zero quadrupole moment~\cite{benalcazar2017electric}
or the non-zero filling anomaly~\cite{benalcazar2019quantization,schindler2019fractional,wieder2020strong,fang2021filling}.
\par
The corner states have four-fold degeneracy, consistent with the non-magnetic symmetry analysis in Refs.~\onlinecite{fang2021filling, fang2021classification}.
The corner states with open boundary condition always come in a group of $d$ states. This degeneracy $d$ is determined by the point group of the finite system. Let $w$ be the general Wyckoff position of the point group, with multiplicity $n_w$. Denote the site symmetry group $G_{w}$.
It has only one irrep, $\rho(G_{w})$. The degeneracy of corner states is~\cite{fang2021classification}
\begin{equation}
\label{eqn:deg}
    d=\text{dim}(\rho(G_{w}))\times n_w
\end{equation}
where dim$(\rho(G_{w}))=2$ for spinful systems with time-reversal symmetry that squares to $-1$, otherwise dim$(\rho(G_{w}))=1$.

In the present case, at zero flux the system is invariant under the symmetry group $p4/m'mm$, while at any small flux the symmetry group reduces to $p4$.
For $p4/m'mm$ and $p4$, the point groups of the finite size system are $4/m'mm$ and $4$ respectively. Each has a general Wyckoff position $w$ with $n_w=4$ and $\text{dim}(\rho(G_w))=1$; thus, Eq.~(\ref{eqn:deg}) yields a degeneracy of $4$~\cite{fang2021classification}. 
Since the degeneracy of corner states is the same for zero flux and finite flux, the corner states do not split when the magnetic flux is introduced
(The chiral symmetry in this model pins the corner states to zero energy, but even in the absence of chiral symmetry, the nonzero filling anomaly will remain the same for $0\leq\phi<\phi^*$.)

We have also shown from the symmetry indicators that at half filling and $\pi$ flux, the system is in the trivial phase, without corner states. 
Thus, the corner states must terminate at $\phi=\phi^*$ by either a bulk or edge gap closing.
There is indeed a bulk gap closing at flux $\phi^*$ as Fig.~\ref{fig:Hof} shows.
This is consistent with the Wannier centers of the occupied bands, which can be deduced from the symmetry indicators: 
the Wannier centers are at the $4b$ Wyckoff position at zero flux and the $4a$ Wyckoff position at $\pi$ flux (see Table~\ref{tab:indatpi} and Appendix~\ref{app:symmetry}). Symmetries prevent four Wannier functions from moving continuously between the $4a$ and $4b$ positions~\cite{fang2021filling,song2017d}. A discontinuous jump of the Wannier centers implies the bulk gap closes between zero and $\pi$ flux.
\par

In Appendix~\ref{app:symmetry} we compute the symmetry indicator at intermediate flux $\phi=2\pi/5<\phi^*$ and $\phi=4\pi/5>\phi^*$ to verify that symmetry indicators are consistent with the presence and absence of corner states between zero and $\pi$. 
In Appendix \ref{app:Wilson} we show that the presence and absence of corner states also agrees with the nested Wilson loops
~\cite{benalcazar2017electric}.

\par

\section{\label{sec:conclusion} Conclusion}

In conclusion, we derived a general framework to apply TQC and the theory of symmetry indicators to crystalline systems at rational flux per unit cell. 
Applying our results to some simple examples at $\pi$ flux revealed new symmetry indicators that did not appear at zero flux. Finally, the symmetry indicators enable us to study a quadrupole insulator at finite field, which reveals a gap closing topological-to-trivial phase transition as a function of magnetic field.
Observing this phase transition could be particularly promising in moir\'e systems where higher flux is attainable for reasonable magnetic fields~\cite{das2022observation,xie2021fractional,pierce2021unconventional,yu2022correlated}.


While preparing our work, we became aware of a related study  \cite{herzog2022hofstadter}, 
which gives criteria for when such a bulk gap closing at finite flux can be predicted from the band structure at zero flux.
The two bulk gap closings between zero and $\Phi = 2\pi$ flux of our model in Sec.~\ref{sec:Model} are indicated by the real space invariant in Ref.~\onlinecite{herzog2022hofstadter}.
\par

We note that the Zeeman term is neglected in this manuscript. When the Zeeman term is present, the periodicity in the flux direction is broken. Thus there is no magnetic time-reversal symmetry or other mirror symmetries that flip magnetic flux. At large magnetic field where Zeeman term dominates, the two dimensional system must be in the trivial atomic limit where Wannier centers locate at the atom positions.

{Our work is also restricted to a spatially constant magnetic field. It would be interesting to extend our results to a spatially varying periodic magnetic field that maintains a commensurate flux per unit cell. This more general theory might be relevant to magnetically ordered crystals.}
\par

As a final note, we draw a connection between our results and the theory of phase space quantization, where one seeks a symmetric and exponentially localized Wannier basis that can continuously reduce to points in the classical phase space by setting the Planck constant $h\rightarrow 0$~\cite{von2018mathematical}. However, 
such a basis can never be found
due to the Balian-Low theorem, which forbids the existence of exponentially localized translational symmetric basis for any single particle~\cite{benedetto1994differentiation}.
The magnetic Wannier functions in two dimensions share a similar translation group structure to the one-dimensional quantum phase space and the non-vanishing Chern number for any single magnetic band also forbids Wannierization~\cite{zak1997balian}, as we explain in Appendix~\ref{app:Wannier}.
Thus, there is an interesting open question: 
since in two dimensional magnetic systems, it is possible to find a Wannier basis for a group of bands, we conjecture that the continuous quantization of phase space may be realized by constructing Wannier functions for groups of particles.

\section{Acknowledgements}
We acknowledge useful conversations with Andrei Bernevig, Aaron Dunbrack, Sayed Ali Akbar Ghorashi, Jonah Herzog-Arbeitman, and Oskar Vafek.
We thank Jonah, Andrei, and their collaborators for sharing their unpublished manuscript. 

Our manuscript is based upon work supported by the National Science Foundation under Grant No. DMR-1942447.
The work was performed in part at the Aspen Center for Physics, which is supported by National Science Foundation grant PHY-1607611.
J.C. also acknowledges the support of the Flatiron Institute, a division of the Simons Foundation.

\bibliography{reference.bib}
\begin{appendix}

\section{\label{app:Eqlambda}Derivation of $\lambda_g$ in Eq.~(\ref{eqn:lambda})}
{In this appendix, we derive Eq.~(\ref{eqn:lambda}) starting from Eq.~(\ref{eqn:covariance2}), which we rewrite here for convenience:}
\begin{equation}
\label{eqn:A1}
    \int_{\mathbf r_1}^{\mathbf r_2} \mathbf A(\mathbf r) \cdot d\mathbf r +\int_{\hat g \mathbf r_1}^{\hat g \mathbf r_2} \nabla \lambda (\mathbf r') \cdot d\mathbf r' = \int_{\hat g \mathbf r_1}^{\hat g \mathbf r_2} \mathbf A(\mathbf r') \cdot d\mathbf r'
\end{equation}
Substituting $\mathbf r = \hat g^{-1} \mathbf r'$ in the first term on the left side of Eq.~(\ref{eqn:A1}) yields
\begin{align}
     \int_{\mathbf r_1}^{\mathbf r_2} \mathbf A(\mathbf r) \cdot d\mathbf r
    &= \int_{\hat g \mathbf r_1}^{\hat g \mathbf r_2} \mathbf A( \hat g^{-1} \mathbf r') \cdot d(\hat g^{-1}\mathbf r') \nonumber \\
    &= \int_{\hat g \mathbf r_1}^{\hat g \mathbf r_2} \mathbf A( \hat g^{-1} \mathbf r')\cdot R_g^{-1}d\mathbf r'  \nonumber \\
    &= \int_{\hat g \mathbf r_1}^{\hat g \mathbf r_2} R_g\mathbf A( \hat g^{-1} \mathbf r') \cdot d\mathbf r'
    \label{eqn:A2}
\end{align}
In the second equality we have defined the rotation $R_g$ by the action $\hat{g}: \mathbf{r} \mapsto R_g \mathbf{r} + \tau_g$, from which it follows that $\hat{g}^{-1}: \mathbf{r} \mapsto R_g^{-1} \mathbf{r} - R_g^{-1}\tau_g$, and, consequently, $d(\hat{g}^{-1}\mathbf{r}') = R_g^{-1} d\mathbf{r}'$, since $d (R_g^{-1}\tau_g)=0$.
Eqs.~(\ref{eqn:A1}) and (\ref{eqn:A2}) together are exactly the integral form of Eq.~(\ref{eqn:lambda}).

\section{\label{app:rotation}Gauge invariant commutation relations between translation and rotation operators}

In this Appendix, we derive the commutation relations for magnetic rotation and translation symmetry operators in real space. We show these relations are gauge invariant. The results are used in Secs.~\ref{sec:mag_sym} and \ref{sec:TQC} and Appendix~\ref{app:irreps}.

We denote an $n$-fold magnetic rotation operator centered about $(\bar x_1, \bar x_2)$ as $C_n(\bar x_1,\bar x_2)$.
For rotations about the origin, we simplify the notation by dropping the argument $(\bar{x}_1,\bar{x}_2)$, i.e., $C_n \equiv C_n(\bar{x}_1=0,\bar{x}_2=0)$. We define $C_n(\bar x,\bar y)$ by conjugating magnetic translation operators with $C_n$
\begin{equation}
\label{eqn:defCnxy}
    C_n(\bar{x}_1,\bar{x}_2) \equiv T(\bar{x}_1\hat{\mathbf x}_1+\bar{x}_2\hat{\mathbf x}_2)C_n T(-\bar{x}_1\hat{\mathbf x}_1-\bar{x}_2\hat{\mathbf x}_2)
\end{equation}
where $\hat{\mathbf x}_1$ and $\hat{\mathbf x}_2$ are unit vectors in the $x_1$ and $x_2$ directions.
This definition is a general form for any gauge choice of vector potential.
In this definition $(C_n(\bar x_1,\bar x_2))^n=C_n^n$ is the identity. 
The definition is independent of the ordering of translations. For example, 
\begin{align*}
    C_n(\bar x_1,\bar x_2)
    &=T(\bar x_2 \hat{\mathbf x}_2)T(\bar x_1 \hat{\mathbf x}_1)C_n T(-\bar x_1 \hat{\mathbf x}_1)T(-\bar x_2 \hat{\mathbf x}_2)\\
    &=T(\bar x_1 \hat{\mathbf x}_1)T(\bar x_2 \hat{\mathbf x}_2)C_n T(-\bar x_2 \hat{\mathbf x}_2)T(-\bar x_1 \hat{\mathbf x}_1)
\end{align*}

\par

We now derive commutation relations between the magnetic rotations and translations.
First, we derive a gauge invariant commutation relation between a translation $T(\mathbf a)$ and $C_n$, 
\begin{equation}
\label{eqn:CnT_TCn}
    C_n T(\mathbf a) = T(R_n \mathbf a) C_n,
\end{equation}
which is the same as the commutation relation in zero field.
We now prove Eq.~(\ref{eqn:CnT_TCn}); Ref.~\onlinecite{herzog2022magnetic} gives a different proof of Eq.~(\ref{eqn:CnT_TCn}).

Eq.~(\ref{eqn:Gg2}) provides explicit operator forms for $C_n$ and $T(\mathbf a)$:
\begin{align}
    C_n&=\sum_{\mathbf x} e^{i\lambda_{C_n}(R_n \mathbf x)}c^\dagger_{R_n \mathbf x}c_{\mathbf x} \\
    T(\mathbf a) &=  \sum_{\mathbf x} e^{i\lambda_{T(\mathbf a)} (\mathbf x+\mathbf a)}c^\dagger_{ \mathbf x+\mathbf a}c_{\mathbf x} 
\end{align}
The gauge dependence of the operators are encoded in the definition of $\lambda_g$. 
{The explicit expression of $\lambda_{C_n}$ and $\lambda_{T(\mathbf a)}$ from Eqs.~(\ref{eqn:lambda}) and (\ref{eqn:lambdaT}) in our definition are
\begin{align}
\label{eqn:FCn}
     \lambda_{C_n}(\mathbf x) &= \int^{\mathbf x}_{\mathbf r_0} 
     \mathbf A(\mathbf x') - R_n\mathbf A(R_n^{-1}\mathbf x')
     \cdot d\mathbf x' \nonumber\\
     &= \left[ \int^{\mathbf x}_{\mathbf r_0}- \int^{R_n^{-1}\mathbf x}_{R_n^{-1}\mathbf r_0} \right]
     \mathbf A(\mathbf x') 
     \cdot d\mathbf x' \\
     \lambda_{T(\mathbf a)}(\mathbf x) &= \int_{\mathbf x-\mathbf a}^{\mathbf x} \mathbf A(\mathbf x')\cdot d\mathbf x' + \mathbf B \cdot \mathbf a \times \mathbf x
     \label{eqn:FTa}
\end{align}
where $\mathbf r_0$ is a reference point that will cancel below. }
\par
Plugging into Eq.~(\ref{eqn:CnT_TCn}) and considering the action on single particle states, 
the equation holds when the phase terms on both sides are equal, i.e., 
\begin{multline}
    \lambda_{C_n}(R_n(\mathbf x+\mathbf a))+\lambda_{T(\mathbf a)}({\mathbf x+\mathbf a}) \\
    =\lambda_{T(R_n\mathbf a)}({R_n(\mathbf x+\mathbf a)})+\lambda_{C_n}(R_n\mathbf x).
\end{multline}
\par
Applying Eq.~(\ref{eqn:FCn}) yields
\begin{align}
    &\lambda_{C_n}(R_n(\mathbf x+\mathbf a))-\lambda_{C_n}(R_n\mathbf x) \nonumber \\
    &= \left[ \int^{R_n(\mathbf x+\mathbf a)}_{\mathbf r_0}- \int^{\mathbf x+\mathbf a}_{R_n^{-1}\mathbf r_0} - \int^{R_n\mathbf x}_{\mathbf r_0} + \int^{\mathbf x}_{R_n^{-1}\mathbf r_0} \right]
     \mathbf A(\mathbf x') 
     \cdot d\mathbf x' \nonumber \\
    &= \left[ \int^{R_n(\mathbf x+\mathbf a)}_{R_n \mathbf x}+\oint_{R_n\cal C} -\int^{\mathbf x+\mathbf a}_{\mathbf x}-\oint_{\cal C} \right] \mathbf A(\mathbf x) \cdot d\mathbf r \nonumber \\
    &= \left[ \int^{R_n(\mathbf x+\mathbf a)}_{R_n \mathbf x} -\int^{\mathbf x+\mathbf a}_{\mathbf x}\right] \mathbf A(\mathbf x) \cdot d\mathbf r,
\end{align}
{where the path $\cal C$ is a triangle whose vertices are $R_n^{-1}r_0$, $\mathbf x$, $\mathbf x+\mathbf a$. The last equality follows because the two closed loops enclose the same amount of flux.}
Then applying Eq.~(\ref{eqn:FTa}) yields
\begin{align}
    \lambda_{T(R_n\mathbf a)}&({R_n(\mathbf x+\mathbf a)})-\lambda_{T(\mathbf a)}({\mathbf x+\mathbf a}) \nonumber \\
   &=\left[ \int^{R_n(\mathbf x+\mathbf a)}_{R_n \mathbf x} -\int^{\mathbf x+\mathbf a}_{\mathbf x}\right] \mathbf A(\mathbf x) \cdot d\mathbf r.
\end{align}
The two expressions are equal, proving Eq.~(\ref{eqn:CnT_TCn}).

Combined with the gauge invariant relation for magnetic translations (Eq.~(\ref{eqn:Ta1Ta2}) in the main text),
\begin{equation}
\label{eqn:Ta1Ta2E}
    T(\mathbf a_1) T(\mathbf a_2) =  T(\mathbf a_1+\mathbf a_2) e^{\frac i2 \mathbf B \cdot (\mathbf a_1 \times \mathbf a_2)},
\end{equation} 
Eq.~(\ref{eqn:CnT_TCn}) implies that the phases that appear in the commutation relations between $C_n(\bar x_1,\bar x_2)$ and $T(\mathbf a)$ are all gauge invariant. For example, for two-fold and four-fold magnetic rotations on a square lattice
\begin{align}
     C_2(\bar x,\bar y) T(\Delta\hat{\mathbf x})&= T(-\Delta\hat{\mathbf x}) C_2(\bar x,\bar y) e^{-2iB\Delta \bar y} \nonumber \\
     C_2(\bar x,\bar y) T(\Delta\hat{\mathbf y})&= T(-\Delta\hat{\mathbf y}) C_2(\bar x,\bar y) e^{2iB\Delta \bar x} \nonumber \\
     C_4(\bar x,\bar y) T(\Delta\hat{\mathbf x})&= T(\Delta\hat{\mathbf y}) C_4(\bar x,\bar y) e^{-iB\Delta (\bar x+\bar y)} \nonumber \\
     C_4(\bar x,\bar y) T(\Delta\hat{\mathbf y})&= T(-\Delta\hat{\mathbf x}) C_4(\bar x,\bar y) e^{iB\Delta (\bar x-\bar y)} 
\end{align}
and 
\begin{align}
    C_2(\bar{x}_1,\bar{y}_1)C_2(\bar{x}_2,\bar{y}_2)&=C_2(-\bar{x}_2,-\bar{y}_2)C_2(-\bar{x}_1,-\bar{y}_1) \nonumber \\
    & \qquad e^{4iB(\bar{x}_1\bar{y}_1-\bar{x}_2\bar{y}_2+\bar{x}_2\bar{y}_1-\bar{x}_1\bar{y}_2)} \nonumber \\
    C_4(\bar{x}_1,\bar{y}_1)C_4(\bar{x}_2,\bar{y}_2)&=C_4(-\bar{y}_2,\bar{x}_2)C_4(\bar{y}_1,-\bar{x}_1) \nonumber \\
    &\qquad e^{iB((\bar{x}_1+\bar{x}_2)^2-(\bar{y}_1-\bar{y}_2)^2)}
\end{align}
These relations are used when we study irreducible representations in Sec.~\ref{sec:kirrep} and topological quantum chemistry in Sec.~\ref{sec:TQC}.

\section{\label{app:projective} Projective representation}
In general, a projective representation $\rho$ of a group $G$ satisfies
\begin{equation}
    \rho(g_1)\rho(g_2)=\omega(g_1,g_2) \rho(g_1g_2)
\end{equation}
for all $g_1,g_2\in G$,
where $\omega(g_1,g_2)$ is called the 2-cocycle (or Schur multiplier, or factor system). When $\omega(g_1,g_2)\equiv 1$, the representation is an ordinary linear representation.

Under a gauge transformation of $\rho$ by $\rho'(g)=f(g)\rho(g)$, 2-cocyles satisfy the cocycle condition $\omega'(g_1,g_2)= f(g_1g_2)f(g_1)^{-1}f(g_2)^{-1}\omega(g_1,g_2) $, which defines an equivalence class $[\omega(g_1,g_2)]$ of cocycles. The classification of the equivalence classes is determined by the group cohomology.

In the remaining part of this section, we explain the standard way to derive irreducible projective representations through the group extension.
The magnetic symmetries can be viewed as projective representations of the non-magnetic symmetries. These projective representations are projected from the ordinary representations of a larger group
that can be described as a group extension of the non-magnetic group. For example, the continuous magnetic translations form the Heisenberg group $Heis$. It is extended from the non-magnetic translations $\mathbb R^2$ by the abelian group $U(1)$. There are many irreducible representations of $Heis$, but only some of them are compatible with the 2-cocycles, which are the phases due to gauge transformations of the magnetic symmetries. For example, the trivial irreducible representation is not compatible with the particular 2-cocycles. 

We now describe the group extension for the translation symmetries on a square lattice as a concrete example.
Consider a model with only nearest neighbor hopping.
Then the phase introduced by the magnetic translations in Eq.~(\ref{eqn:TxTyDelta}) with commensurate magnetic flux $\phi=2\pi p/q$ belongs to $\mathbb Z_q$.
We first consider the extension of full translation group $\mathbb T=\mathbb Z\times \mathbb Z$ by $\mathbb Z_q$.
This real space Heisenberg group is defined by the central group extension, which is a short exact sequence
\begin{equation}
    \begin{tikzcd}
  1\arrow[r] &\mathbb{Z}_q \arrow[r,"i"] &Heis \arrow[r,"\pi"] &\mathbb{Z}\times \mathbb Z \arrow[r] &1
\end{tikzcd}
\end{equation}
where each arrow is a group homomorphism and $\text{im}(i)=\text{ker}(\pi)$.
Next we consider the momentum space Heisenberg group $Heis_k$ (for the $q$-by-$q$ unit cell), which is the sublattice translation group $\mathbb T/ \mathbb T_M$ extended by $\mathbb Z_q$. It's defining central group extension is 
\begin{equation}
    \begin{tikzcd}
  1\arrow[r] &\mathbb{Z}_q \arrow[r,"i"] &Heis_k \arrow[r,"\pi"] &\mathbb{Z}_q\times \mathbb Z_q \arrow[r] &1
\end{tikzcd}
\end{equation}
It can be shown that this $Heis_k$ is an extra special group of order $q^3$. It has $q^2$ $1$-dimensional irreducible representations and $q-1$ $q$-dimensional irreducible representations~\cite{gorenstein2007finite}. The abelian group $\mathbb Z_q$ acts trivially on the $1$-dimensional irreps. Thus, the projective representations are projected from the $q$-dimensional representations. 
Therefore, the momentum space states satisfying magnetic translation symmetries are $q$-fold degenerate. For example, at $\pi$ flux, $Heis_k$ is isomorphic to the group $D_4$ and the irreducible projective representation of translations $\mathbb{Z}_2\times \mathbb Z_2$ is projected from the only two-dimensional irreducible representation $E$ of $D_4$.
\par
For more complicated cases with magnetic rotation symmetries, in principle one can use the same formalism to get the corresponding extended group and the irreducible projective representations of the unextended symmetry group. However, practically it is easier to derive the irreducible projective representations by studying the (anti-)commutation relations as shown in Appendix~\ref{app:irreps}.

\section{\label{app:triangular} Triangular lattice symmetries in a magnetic field}

\begin{table*}[ht]
\begin{tabular}{c|c|c|c|c|c}
\hline
$g$ &$T(\Delta_1\hat{\mathbf x}_1)$ &$T(\Delta_2\hat{\mathbf x}_2)$& $C_2(\bar r_1, \bar r_2)$ &$C_6(\bar r_1, \bar r_2)$ &$C_3(\bar r_1, \bar r_2)$ \\
\hline
&&&&&\\[-0.5em]
$\hat{g}=\{R_g|\tau_g\}$ &$\{0|(\Delta_1,0)\}$ &$\{0|(0,\Delta_2)\}$ &$\{\hat{C}_2|(2\bar r_1,2\bar r_2)\}$ &$\{\hat{C}_6|(\bar r_1+\bar r_2,-\bar r_1)\}$ &$\{\hat{C}_3|(2\bar r_1+\bar r_2,\bar r_2-\bar r_1)\}$ \\
\hline
&&&&&\\[-0.5em]
$\lambda(r_1,r_2)$&0&$-\phi \Delta_2 r_1$ &$-2\phi\bar r_2 (r_1-\bar r_1)$ &$-\phi ((r_1-\bar r_1) (r_2-\bar r_2)+\frac{(r_1-\bar r_1)^2}{2})$ &$-\phi ((r_1-\bar r_1) (r_2-\bar r_2)+\frac{(r_2-\bar r_2)^2}{2})$ \\
&&&&$+\phi\bar r_2(r_2-\bar r_2)$&$-\phi \bar r_2((r_1-\bar r_1)- (r_2-\bar r_2))$\\
\hline
\end{tabular}
\caption{The gauge transformation $\lambda_g(x,y)$ for symmetries of the six-fold symmetric triangular lattice in Landau gauge. For each symmetry $g$ in the first row, the second row lists the symmetry in the notation $\lbrace R_g | \tau_g \rbrace$ and the third row provides $\lambda_g$ from Eq.~(\ref{eqn:lambda}).}
\label{tab:lambda_triangle}
\end{table*}


The Landau gauge in triangular and hexagonal lattices is defined differently than for the square lattice. Therefore, we discuss the magnetic symmetries of these lattices separately here.\par

We consider the lattice vectors $\mathbf a_1=(1,0)$, $\mathbf a_2=(\frac12,\frac{\sqrt{3}}{2})$ in Cartesian coordinates. The reciprocal lattice vectors are $\mathbf b_1=(1,-1/\sqrt{3})$, $\mathbf b_2=(0,2/\sqrt{3})$ in Cartesian coordinates, where $\mathbf a_i \cdot \mathbf b_j=\delta_{ij}$.
The vector potential for the magnetic field in Landau gauge is $\mathbf A(\mathbf r)=-\phi \mathbf b_1(\mathbf b_2\cdot \mathbf r)$, where $\phi$ is the flux per primitive unit cell. It is helpful to note the gradient in these coordinates is $\nabla=\mathbf b_1 \partial_{r_1}+\mathbf b_2 \partial_{r_2}$.

From Eq.~(\ref{eqn:lambda}) we can compute the gauge transformation terms $\lambda(r_1,r_2)$, where $\mathbf r=r_1\mathbf a_1+r_2\mathbf a_2$.
The results for several spatial symmetries $\hat g$ compatible with the triangular lattice are summarized in Table~\ref{tab:lambda_triangle}. 

\section{\label{app:sym_q1_qq}Equivalent representations for the $q$-by-$1$ and $q$-by-$q$ unit cells}
In this appendix, we argue that the symmetry operators for the $q$-by-$1$ and $q$-by-$q$ unit cells form equivalent  $q^2 \times q^2$-dimensional representations. We study $C_4$ and $T(\hat {\mathbf y})$ as examples.
\par


\subsection{$q$-by-$1$ unit cell}

As shown in Sec.~\ref{sec:mag_sym_3_momentum_qby1}, in the $q$-by-$1$ unit cell the $C_4$ symmetry mixes a Bloch eigenstate at momentum $(k_x,k_y)$ into a linear combination of eigenstates at momenta $(k_y+2\pi l/q,-k_x)$, $l=0,1,\dots q-1$.
Each momentum can be classified by the number of other momenta it mixes with under $C_4$: a generic $\mathbf{k}$ mixes into $4q$ distinct momenta; $(\pi,0)$ mixes into $2q$ momenta (including $(0,\pi/q)$); and $(0,0)$ or $(\pi/q,\pi/q)$ mix into $q$ distinct momenta.

We now compute the symmetry operators for the highest symmetry momenta, $(0,0)$ and $(\pi/q,\pi/q)$.
The $T(\hat{\mathbf y})$ and $C_4$ symmetries will be implemented by a $q^2 \times q^2$ matrix: the first factor of $q$ comes from the $q$ original unit cells in the magnetic unit cell, while the second factor comes from $q$ different momenta that mix into each other under $C_4$.
Specifically, following Eqs.~(\ref{eq:Tyq}) and (\ref{eq:C4q}), at the high-symmetry points $\mathbf{k} = \Gamma$ or $M$, we choose the basis
$\{c^\dagger_{(l\phi,0),j}|0\rangle\}$
or $\{c^\dagger_{(\pi/q + l\phi,\pi/q),j}|0\rangle\}$, respectively, where $l,j=0,1,\dots,q-1$.

As a concrete example, we write the matrix form of the symmetry operators at $\phi = \pi$ flux, i.e., $q=2$.
A symmetry operator $g$ defined in the above basis takes the general form
\begin{equation}
    g=\frac{q}{(2\pi)^2} \int_0^{\frac {2\pi}{q}}\int_0^{\frac {2\pi}{q}} d\mathbf k \sum_{l,j,l',j'=0,1} c^{\dagger}_{\mathbf k,l,j} D(g,\mathbf k)_{l,j;l',j'} c_{\mathbf k,l',j'}
\end{equation}
where $c_{\mathbf k,l,j}=c_{\mathbf k+l(2\pi/q,0),j}$ and $D(g,\mathbf k)_{l,j;l',j'}$ is a $4\times 4$ matrix. (This form does not apply at momenta other than $(0,0)$ or $(\pi/2,\pi/2)$; at other momenta the $C_4$ symmetry requires a larger matrix.)

Taking the specific symmetries $g=T(\hat{\mathbf y})$ or $C_4$,
\begin{equation}
    D(T(\hat{\mathbf y}),\mathbf k)=e^{ik_y}
    \begin{pmatrix}
    &&&1\\
    &&1&\\
    &1&&\\
    1&&&
    \end{pmatrix},
    \label{eq:Tymatq}
\end{equation}
whose eigenvalues are $e^{ik_y}\{+1,+1,-1,-1\}$;
\begin{equation}
    D(C_4,\Gamma)=\frac12
    \begin{pmatrix}
    1&1&1&1\\
    1&-1&-1&1\\
    1&-1&1&-1\\
    1&1&-1&-1
    \end{pmatrix},
    \label{eq:C4matqG}
\end{equation}
whose eigenvalues are $\{+1,+1,-1,-1\}$; and
\begin{equation}
    D(C_4,(\pi/2,\pi/2))=\frac12
    \begin{pmatrix}
    1&-i&1&-i\\
    -i&1&i&-1\\
    1&i&1&i\\
    -i&-1&i&1
    \end{pmatrix},
    \label{eq:C4matqM}
\end{equation}
whose eigenvalues are $\{+1,+1,+i,-i\}$.\par

\subsection{$q$-by-$q$ unit cell}
For the $q$-by-$q$ unit cell, the $C_4$ operator behaves similar to the non-magnetic case, in that it mixes each Bloch wave function into a single other Bloch wave function. The highest symmetry points are $\Gamma=(0,0)$ and $M=(\pi/q,\pi/q)$.
\par
As in the previous section, we consider the flux $\phi = \pi$ as an example. An operator $g$ takes the general form
\begin{equation}
    g=(\frac{q}{2\pi})^2 \int d\mathbf k \sum_{j_x,j_y,j_x',j_y'} c^{\dagger}_{\mathbf k,j_x,j_y} D(g,\mathbf k)_{j_x,j_y;j_x',j_y'} c_{\mathbf k,j_x',j_y'}
\end{equation}
This form of the matrix representation $D(g,\mathbf k)$ is valid for $T(\hat{\mathbf y})$ at any $\mathbf k$ and for $C_4$ at $\Gamma=(0,0)$ and $M=(\pi/q,\pi/q)$. From Eqs.~(\ref{eq:Tyqq}) and (\ref{eq:C4qq}),
\begin{equation}
    D(T(\hat{\mathbf y}),\mathbf k)=e^{ik_y}
    \begin{pmatrix}
    &&1&\\
    &&&-1\\
    1&&&\\
    &-1&&
    \end{pmatrix}
\end{equation}
whose eigenvalues are $e^{ik_y}\{+1,+1,-1,-1\}$;
\begin{equation}
    D(C_4,(0,0))=
    \begin{pmatrix}
    1&&&\\
    &&1&\\
    &1&&\\
    &&&-1
    \end{pmatrix},
\end{equation}
whose eigenvalues are $\{+1,+1,-1,-1\}$; and
\begin{equation}
    D(C_4,(\pi/2,\pi/2))=
    \begin{pmatrix}
    1&&&\\
    &&-1&\\
    &1&&\\
    &&&1
    \end{pmatrix},
\end{equation}
whose eigenvalues are $\{+1,+1,+i,-i\}$.

These eigenvalues are identical to the eigenvalues of the symmetry operators derived for the $q$-by-1 unit cell in Eqs.~(\ref{eq:Tymatq}), (\ref{eq:C4matqG}) and (\ref{eq:C4matqM}),
which implies that the representations of the symmetry operators are the same for a $q$-by-$1$ unit cell and a $q$-by-$q$ unit cell. 

In conclusion: the two natural choices of magnetic unit cell yield unitarily equivalent symmetry representations, but the $q$-by-$q$ unit cell allows the symmetry operators to act in a more familiar way because their action on $\mathbf{k}$ is identical to that in zero field.

\section{\label{app:irreps}Irreps at high symmetry momenta for $p2, p4, p4/m'$}
The irreps of little co-groups at high symmetry momenta for $p2$, $p4$ and $p4/m'$ with $\pi$ flux are calculated in detail in this section. The results are summarized in Tables~\ref{tab:p2}, \ref{tab:p4}, and \ref{tab:p4TI} in Sec.~\ref{sec:kirrep}.
\par
For simplicity, we denote $T_x\equiv T(\hat{\mathbf x})$, $T_y\equiv T(\hat{\mathbf y})$ in this section.

In this appendix, we will repeatedly use the gauge invariant commutation relation
\begin{equation}
\label{eqn:CnT_TCnappE}
     C_n T(\mathbf a) = T(R_n \mathbf a) C_n
\end{equation}
which is derived in Appendix~\ref{app:rotation} and
\begin{equation}
\label{eqn:TxTyappE}
    T_xT_y=T_yT_x e^{i\phi}=-T_yT_x
\end{equation}
which is a consequence of the Aharonov-Bohm phase.
We will also use $T_x^2=e^{2ik_x}$, $T_y^2=e^{2ik_y}$ to derive the (anti-)commutation relations of rotation and translation symmetries.

\subsection{$p2$}
We first study the $p2$ with $\phi=\pi$ and a $2$-by-$1$ unit cell. The magnetic lattice translation group is $\mathbb T_M=\{(n_1,2n_2)|n_1,n_2\in \mathbb Z\}$. The little co-group is a subgroup of $p2/{\mathbb T}_M$, which is generated by $C_2$ and $T_y$ symmetries. 
\par
For the $2$-by-$1$ unit cell, the Brillouin zone is $[-\pi,\pi]\times[-\pi/2,\pi/2]$. In momentum space, $T_y$ maps $(k_x,k_y) \mapsto  (k_x+\pi,k_y)$ because Eq.~(\ref{eqn:TxTyappE}) implies
\begin{align}
    T_xT_y|k_x,k_y\rangle&=-T_yT_x|k_x,k_y\rangle \nonumber \\
    &=e^{i(k_x+\pi)}T_y|k_x,k_y\rangle 
    \label{eq:Tyonkx}
\end{align}
Therefore, given a Bloch wave function $|0,0;\xi\rangle$ at $(0,0)$ with $C_2$ eigenvalue $\xi$, we can construct a state $T_y|0,0;\xi\rangle$ at $(\pi,0)$ with the same $C_2$ eigenvalue:
\begin{align}
    C_2T_y |0,0;\xi\rangle &= T_yT_y^{-2}C_2|0,0;\xi\rangle \nonumber \\
    &=\xi T_y^{-1}|0,0;\xi\rangle, 
\end{align}
{where the first equality is due to the commutation relation Eq.~(\ref{eqn:CnT_TCnappE}) and the second equality follows because the $T_y^2$ eigenvalue of a Bloch wave function at $k_y$ is $e^{2ik_y}$.}
Similarly, for each Bloch state at $(0,\pi/2)$ with $C_2$ eigenvalue $\xi$, there is a state at $(\pi,\pi/2)$ with eigenvalue $-\xi$:
\begin{align}
    C_2 T_y|0,\pi/2; \xi\rangle &= T_yT_y^{-2} C_2 |0,\pi/2; \xi\rangle \nonumber\\
    &= \xi T_yT_y^{-2} |0,\pi/2; \xi\rangle\nonumber\\
    &= -\xi T_y|0,\pi/2; \xi\rangle
\end{align}
It follows that although there are four $C_2$ symmetric momenta, only two of them have independent eigenvalues, $\Gamma=(0,0)$ and $Y=(0,\pi/2)$. 
Each Bloch wave function at those points can have $C_2$ eigenvalue $+i$ or $-i$.
\par

Due to the unusual behavior of $T_y$ in Eq.~(\ref{eq:Tyonkx}), the four momenta invariant under $T_yC_2$ are $(\pm \pi/2,0)$ and $(\pm \pi/2,\pi/2)$. Similarly only two of them are independent. We choose the two points to be $X=(\pi/2,0)$ and $M=(\pi/2,\pi/2)$. Since $(T_yC_2)^2=-1$, there are two irreps for each point and each irrep has $T_yC_2$ eigenvalue $+i$, $-i$.
\par
We summarize the irreps at all independent high symmetry points for the group $p2$ in Table~\ref{tab:p2}.
Notice that we have used $C_2^2=-1$, corresponding to spinful electrons, in the above derivations. If $C_2^2=+1$, the $C_2$ eigenvalues will be $+1$, $-1$.
\par

\subsection{$p4$}
We now study the group $p4$ group with $\phi=\pi$ and a $2$-by-$2$ unit cell. The lattice translation group is $\mathbb T_M=\{(2n_1,2n_2)|n_1,n_2\in \mathbb Z\}$. The little co-group is a subgroup of $p4/{\mathbb T}_M$ generated by $C_4$ and $T_x$, $T_y$ symmetries.
\par

For the $2$-by-$2$ unit cell, the Brillouin zone is $[-\pi/2,\pi/2]\times[-\pi/2,\pi/2]$. $T_x$ and $T_y$ both map $(k_x,k_y)$ to itself. Since $\{T_x,T_y\}=0$, each irrep at generic $\mathbf k$ is at least two-dimensional.
\par
The two high symmetry points invariant under $C_4$ are $(0,0)$ and $(\pi/2,\pi/2)$, while the two high symmetry points invariant under $C_2$ {but not $C_4$} are $(\pi/2,0)$ and $(0,\pi/2)$. We study each point separately.
\par
At $X=(\pi/2,0)$, the (anti-)commutation relations derived from Eqs.~(\ref{eqn:CnT_TCnappE}) and (\ref{eqn:TxTyappE}), $[C_2,T_y]=0$, $\{C_2,T_x\}=0$, $\{T_x,T_y\}=0$ lead to two two-dimensional irreps $X_1$, $X_2$ as shown in Table~{\ref{tab:p4}}.
\par
At $Y=(0,\pi/2)$, the (anti-)commutation relations {derived from Eqs.~(\ref{eqn:CnT_TCnappE}) and (\ref{eqn:TxTyappE}), }  $[C_2,T_x]=0$, $\{C_2,T_y\}=0$, $\{T_x,T_y\}=0$ lead to two two-dimensional irreps $Y_1$, $Y_2$ as shown in Table~{\ref{tab:p4}}.
\par
At $\Gamma=(0,0)$, $[C_4T_x,T_xT_y]=0$ and $\{T_xT_y,T_x\}=0$ due to Eqs.~(\ref{eqn:CnT_TCnappE}) and (\ref{eqn:TxTyappE}).
For a Bloch eigenstate $|\xi,\eta\rangle$ with eigenvalue $\xi=\pm 1$ or $\pm i$ of $C_4T_x$ and eigenvalues $\eta=\pm i$ of $T_xT_y$, $T_x$ maps it to an eigenstate of $C_4T_x$ with eigenvalue $\eta\xi$ and an eigenstate of $T_xT_y$ with eigenvalue $-\eta$:
\begin{align}
    C_4T_x T_x|\xi,\eta\rangle &= T_y C_4T_x |\xi,\eta\rangle \nonumber \\
    &= T_x T_x^{-2} T_xT_y C_4T_x |\xi,\eta\rangle \nonumber \\
    &=T_x e^{-2ik_x} \eta \xi |\xi,\eta\rangle \nonumber \\
    &=\eta \xi T_x |\xi,\eta\rangle 
\end{align}
where the first equality follows Eq.~(\ref{eqn:CnT_TCnappE}) and
\begin{align}
    T_xT_y T_x|\xi,\eta\rangle &= -T_xT_xT_y |\xi,\eta\rangle \nonumber \\
    &=-\eta T_x|\xi,\eta\rangle 
\end{align}
{due to Eq.~(\ref{eqn:TxTyappE}).}
The two equations pair up eigenvalues and lead to four two-dimensional irreps $\Gamma_1$, $\Gamma_2$, $\Gamma_3$, $\Gamma_4$ as shown in  Table~{\ref{tab:p4}}. 
\par
At $M=(\pi/2,\pi/2)$, there are (anti-)commutation relations $[C_4,T_xT_y]=0$ and $\{T_x,T_xT_y\}=0$.
For an eigenstate $|\xi,\eta\rangle$ with eigenvalue $\xi=e^{i\pi/2}, e^{3i\pi/2}, e^{-3i\pi/2},$ or $e^{-i\pi/2}$ of $C_4$ and eigenvalue $\eta=+i,-i$ of $T_xT_y$, $T_x$ maps it to an eigenstate of $C_4$ and $T_xT_y$ with eigenvalues $-\eta\xi$ and $-\eta$, respectively:
\begin{align}
    C_4 T_x|\xi,\eta\rangle &= T_y C_4 |\xi,\eta\rangle \nonumber \\
    &= T_x T_x^{-2} T_xT_y C_4 |\xi,\eta\rangle \nonumber \\
    &=T_x e^{-2ik_x} \eta \xi |\xi,\eta\rangle \nonumber \\
    &=-\eta \xi T_x |\xi,\eta\rangle 
\end{align}
where the first line comes from Eq.~(\ref{eqn:CnT_TCnappE}) and
\begin{align}
    T_xT_y T_x|\xi,\eta\rangle &= -T_xT_xT_y |\xi,\eta\rangle \nonumber \\
    &=-\eta T_x|\xi,\eta\rangle 
\end{align}
due to Eq.~(\ref{eqn:TxTyappE}).
The commutation relations imply four two-dimensional irreps $M_1$, $M_2$, $M_3$, $M_4$ as shown in Table.~{\ref{tab:p4}}.
\par

\subsection{$p4/m'$}
We study the group $p4/m'$ with $\phi=\pi$ and a  $2$-by-$2$ unit cell. The lattice translation group and Brillouin zone are the same as for the $p4$ case.
We first prove that each band is four-fold degenerate:
since $T_x^2=e^{2ik_x}$, $T_x$ has eigenvalues $\eta e^{ik_x}$, $\eta=\pm 1$. A Bloch eigenstate at $k$, $|k;\eta\rangle$, is mapped by $\cal TI$ to another state at $k$ with the same eigenvalue:
\begin{align}
\label{eqn:TxTIeig}
    T_x {\cal TI} |k;\eta\rangle &= {\cal TI} T_x^{-2}T_x |k;\eta\rangle \nonumber \\ 
    &= {\cal TI} \eta e^{-ik_x}|k;\eta\rangle \nonumber\\
    &=\eta e^{ik_x} {\cal TI} |k;\eta\rangle
\end{align}
where the first line can be derived from Table~\ref{tab:lambda}.
In the spinful case where $({\cal TI})^2=-1$, there is a Kramers degeneracy, i.e. ${\cal TI} |k;\eta\rangle$ and $|k;\eta\rangle$ are two different states. The anticommutation $\{T_x,T_y\}=0$ implies that $T_y$ flips the eigenstate of $T_x$:
\begin{align}
\label{eqn:TxTyeig}
    T_x T_y |k;\eta\rangle &= -T_y T_x |k;\eta\rangle \nonumber \\
    &=-\eta e^{ik_x }T_y|k;\eta\rangle
\end{align}
Combining Eqs.~(\ref{eqn:TxTIeig}) and (\ref{eqn:TxTyeig}), we conclude that each eigenstate is at least four-fold degenerate.
\par
We now study the irreps at the high symmetry points $\Gamma$, $X$, $Y$, $M$. The role of $\cal TI$ is to pair the two-dimensional irreps of the $p4$ group. Following the irrep lebelling scheme used for $p4$ in the previous section, the irreps are summarized in Table~\ref{tab:p4TI} and justified as follows:
\par
At $X=(\pi/2,0)$, the commutation relations $[C_2,{\cal TI}]=0$ and $[T_y,{\cal TI}]=0$ lead to one irrpe $X_1X_2$, and similarly at $Y$ for $Y_1Y_2$.
At $\Gamma=(0,0)$, the commutation relations $[C_4,{\cal TI}]=0$ leads to one irrep $\Gamma_1\Gamma_2$. Finally, at $M=(\pi/2,\pi/2)$, the relations $[C_4,{\cal TI}]=0$ and $\{T_xT_y,{\cal TI}\}=0$ lead to three four-dimensional irreps $M_1M_1$, $M_3M_3$, and  $M_2M_4$.
\par

\section{\label{app:TQC} Deriving the symmetry indicator}

In this appendix, we describe how to find the symmetry indicator classification of a space group and how to apply it to a group of bands to determine in which topological class the bands belong.

At the crux of the theory is the ``EBR matrix'' for the space group. Each row of the matrix corresponds to a particular choice of $\mathbf{q}$ and an irrep $\rho$ of $G_\mathbf{q}$ (Sec.~\ref{sec:inducedrep}). Each column corresponds to a particular irrep of the little co-group at a particular high symmetry momentum (Sec.~\ref{sec:kirrep}). The entry in the matrix indicates the number of times the irrep appears in the band representation $\rho_G$ induced from $\rho$, as we will define in Eq.~(\ref{eqn_vAn}) \cite{cano2021band,song2020fragile,song2020twisted}.

Let $A$ be an integer EBR matrix of the symmetry group under consideration.
Since a group of topologically trivial bands transforms identically to a sum of Wannier functions, its irreps at high symmetry points satisfy
\begin{equation}
    v=A{n},
    \label{eqn_vAn}
\end{equation}
where $v_j$ is the number of times the $j^\text{th}$ irrep appears in the band structure~\cite{bradlyn2017topological}.

Let the Smith normal form of $A$ be given by
\begin{equation}
    A = U^{-1}DV^{-1},
    \label{eqn_smithA}
\end{equation}
where $D$ is a diagonal positive integer matrix with diagonal entries $(d_1, \dots, d_M, 0, \dots 0)$, i.e., the first $M$ entries are positive and the remaining entries are zero, and $U,V$ are integer matrices invertible over the integers.
The stable topological classification for the space group is given by
\begin{equation}
\label{eqn_Zdm}
    \mathbb{Z}_{d_1} \times \cdots \times \mathbb{Z}_{d_M}.
\end{equation}

We seek a formula that expresses the topological invariant (i.e., the element of $\mathbb{Z}_{d_m}$ of a particular group of bands) in terms of the little co-group irreps at high symmetry points. This index is given by \cite{po2017symmetry,song2018quantitative,song2020fragile,song2020twisted,cano2021band}
\begin{equation}
    \text{index} = \left( Uv \right)_m \mod d_m
    \label{eqn_index}
\end{equation}
where $1\leq m\leq M$, and $d_m>1$.

The number of Wannier functions that are centered at a  particular maximal Wyckoff position $w$ can be determined by the following formula~\cite{fang2021filling,fang2021classification}:
\begin{multline}
    e_w =  \sum_{i\in w} \text{dim}(\rho_i) \left[ VD^pUv\right]_i  \\
    \mod  \text{gcd}\lbrace \left( \sum_{i\in w} \text{dim}(\rho_i) V_{ij} \right) | _{j > M} \rbrace.
    \label{eqn_smithew}
\end{multline}
The sum over ${i\in w}$ indicates the sum over EBRs induced from a representation $\rho_i$ of the site symmetry group of the Wyckoff position $w$; $D^p$ is the pseudo-inverse of $D$, a diagonal matrix with diagonal entries $(d_1^{-1},...,d_M^{-1},0,...,0)$;
and gcd indicates the greatest common divisor.

In the following, we compute the EBR matrix and Smith decomposition for relevant groups discussed in the main text.

\subsection{$p2$}
The basis for band representations (columns) and the basis for coefficients of EBRs (rows) are
\begin{equation}
    \left( \Gamma_1^{(p2)},\Gamma_2^{(p2)},Y_1^{(p2)},Y_2^{(p2)},X_1^{(p2)},X_2^{(p2)},M_1^{(p2)},M_2^{(p2)}\right),
\end{equation}
where $\Pi^{(p2)}_i$ is defined in Table~\ref{tab:p2}, and
\begin{equation}
    \left( ^1\bar E^{ 4a},^2\bar E^{4a},^1\bar E^{4b},^1\bar E^{4b},^1\bar E^{4c},^2\bar E^{4c},^1\bar E^{4d},^2\bar E^{4d}\right).
\end{equation}
where $^j \bar E^{nw}$ is an irrep of the site symmetry group $C_2$ of the Wyckoff position $nw$. $^1\bar E$ and $^2\bar E$ have $C_2$ eigenvalue $+i$ and $-i$ respectively.
In this basis, the EBR matrix is
\begin{equation}
    \label{eqn:p2_A}
    A = \begin{pmatrix}
  2 & 0 & 1 & 1 & 1 & 1 & 1 & 1 \\
 0 & 2 & 1 & 1 & 1 & 1 & 1 & 1 \\
 1 & 1 & 2 & 0 & 1 & 1 & 1 & 1 \\
 1 & 1 & 0 & 2 & 1 & 1 & 1 & 1 \\
 1 & 1 & 1 & 1 & 2 & 0 & 1 & 1 \\
 1 & 1 & 1 & 1 & 0 & 2 & 1 & 1 \\
 1 & 1 & 1 & 1 & 1 & 1 & 2 & 0 \\
 1 & 1 & 1 & 1 & 1 & 1 & 0 & 2 
    \end{pmatrix}
\end{equation}
The Smith normal form matrices are
\begin{equation}
    \label{eqn:p2_D}
    D = 
\begin{pmatrix}
 1 & 0 & 0 & 0 & 0 & 0 & 0 & 0 \\
 0 & 1 & 0 & 0 & 0 & 0 & 0 & 0 \\
 0 & 0 & 1 & 0 & 0 & 0 & 0 & 0 \\
 0 & 0 & 0 & 1 & 0 & 0 & 0 & 0 \\
 0 & 0 & 0 & 0 & 4 & 0 & 0 & 0 \\
 0 & 0 & 0 & 0 & 0 & 0 & 0 & 0 \\
 0 & 0 & 0 & 0 & 0 & 0 & 0 & 0 \\
 0 & 0 & 0 & 0 & 0 & 0 & 0 & 0 \\
\end{pmatrix}
\end{equation}

\begin{equation}
    \label{eqn:p2_U}
    U = \begin{pmatrix}
  2 & 2 & -1 & 0 & -1 & 0 & -1 & 0 \\
 1 & 2 & -1 & 0 & -1 & 0 & 0 & 0 \\
 -1 & -1 & 1 & 0 & 0 & 0 & 1 & 0 \\
 -1 & -1 & 0 & 0 & 1 & 0 & 1 & 0 \\
 -3 & -5 & 2 & 0 & 2 & 0 & 2 & 0 \\
 -1 & -1 & 0 & 0 & 1 & 1 & 0 & 0 \\
 -1 & -1 & 1 & 1 & 0 & 0 & 0 & 0 \\
 -1 & -1 & 0 & 0 & 0 & 0 & 1 & 1 \\
    \end{pmatrix}
\end{equation}

\begin{equation}
    \label{eqn:p2_V}
    V = \begin{pmatrix}
  1 & 0 & 0 & 0 & 1 & -1 & -1 & -1 \\
 0 & 0 & 0 & 0 & -1 & -1 & -1 & -1 \\
 0 & -1 & 1 & 0 & -2 & 0 & 1 & 0 \\
 0 & 0 & 0 & 0 & 0 & 0 & 1 & 0 \\
 0 & -1 & 0 & 1 & -2 & 1 & 0 & 0 \\
 0 & 0 & 0 & 0 & 0 & 1 & 0 & 0 \\
 0 & 1 & 0 & 0 & 2 & 0 & 0 & 1 \\
 0 & 0 & 0 & 0 & 0 & 0 & 0 & 1 \\
    \end{pmatrix}
\end{equation}
\par
The stable indicator is given by Eq.~(\ref{eqn:indexp2}) in the main text:
\begin{multline}
    \text{index} = \#\Gamma_1-\#\Gamma_2+\#X_1-\#X_2 
    +\#Y_1-\#Y_2 
    \\
    +\#M_1-\#M_2-N \mod 4.
\end{multline}
The symmetry indicators for Wannier centers are
\begin{align}
    e_{2a} &= 2N-\# X_1-\# Y_1-\# M_1 \mod 2\\
    e_{2b} &=-N/2+\# Y_1 \mod 2\\
    e_{2c} &=-N/2+\# X_1 \mod 2\\
    e_{2d} &=-N/2+\# M_1 \mod 2
\end{align}

\subsection{$p4$}
The basis for band representations (columns) and the basis for coefficients of EBRs (rows) are
\begin{equation}
    \left( \Gamma_1,\Gamma_2,\Gamma_3,\Gamma_4,M_1,M_2,M_3,M_4,X_1,X_2\right),
\end{equation}
where $\Pi_i$ is defined in Table~\ref{tab:p4}, and
\begin{equation}
    \left( ^1\bar E^{ 4a}_1,^1\bar E^{4a}_2,^2\bar E^{4a}_2,^2\bar E^{4a}_1,
    E^{4b}_1,^1\bar E^{4b}_2,^2\bar E^{4b}_2,^2\bar E^{4b}_1, 
    ^1\bar E^{8c},^2\bar E^{8c}\right).
\end{equation}
where $^j \bar E^{nw}$ is an irrep of the site symmetry group of the Wyckoff position labelled by $nw$. $^1\bar E$ and $^2\bar E$ have $C_2$ eigenvalues $+i$ and $-i$ respectively, while $^1\bar E_1$, $^1\bar E_2$, $^2\bar E_2$, and $^2\bar E_1$ have $C_4$ eigenvalues $e^{i\pi/4}$, $e^{i3\pi/4}$, $e^{-3i\pi/4}$, $e^{-i\pi/4}$ respectively.
In this basis, the EBR matrix is
\begin{equation}
    \label{eqn:p2_A}
    A = \begin{pmatrix}
 1 & 0 & 1 & 0 & 1 & 0 & 0 & 1 & 1 & 1 \\
 0 & 1 & 0 & 1 & 1 & 1 & 0 & 0 & 1 & 1 \\
 1 & 0 & 1 & 0 & 0 & 1 & 1 & 0 & 1 & 1 \\
 0 & 1 & 0 & 1 & 0 & 0 & 1 & 1 & 1 & 1 \\
 1 & 0 & 0 & 1 & 1 & 0 & 1 & 0 & 1 & 1 \\
 1 & 1 & 0 & 0 & 0 & 1 & 0 & 1 & 1 & 1 \\
 0 & 1 & 1 & 0 & 1 & 0 & 1 & 0 & 1 & 1 \\
 0 & 0 & 1 & 1 & 0 & 1 & 0 & 1 & 1 & 1 \\
 1 & 1 & 1 & 1 & 1 & 1 & 1 & 1 & 3 & 1 \\
 1 & 1 & 1 & 1 & 1 & 1 & 1 & 1 & 1 & 3 \\
    \end{pmatrix}
\end{equation}
The Smith normal form matrices are
\begin{equation}
    \label{eqn:p2_D}
    D = 
\begin{pmatrix}
 1 & 0 & 0 & 0 & 0 & 0 & 0 & 0 & 0 & 0 \\
 0 & 1 & 0 & 0 & 0 & 0 & 0 & 0 & 0 & 0 \\
 0 & 0 & 1 & 0 & 0 & 0 & 0 & 0 & 0 & 0 \\
 0 & 0 & 0 & 1 & 0 & 0 & 0 & 0 & 0 & 0 \\
 0 & 0 & 0 & 0 & 1 & 0 & 0 & 0 & 0 & 0 \\
 0 & 0 & 0 & 0 & 0 & 1 & 0 & 0 & 0 & 0 \\
 0 & 0 & 0 & 0 & 0 & 0 & 1 & 0 & 0 & 0 \\
 0 & 0 & 0 & 0 & 0 & 0 & 0 & 8 & 0 & 0 \\
 0 & 0 & 0 & 0 & 0 & 0 & 0 & 0 & 0 & 0 \\
 0 & 0 & 0 & 0 & 0 & 0 & 0 & 0 & 0 & 0 \\
\end{pmatrix}
\end{equation}

\begin{equation}
    \label{eqn:p2_U}
    U = \begin{pmatrix}
 0 & -1 & -1 & 0 & 1 & 1 & 0 & 1 & 0 & 0 \\
 -1 & -1 & -1 & 0 & 1 & 1 & 1 & 1 & 0 & 0 \\
 0 & -1 & -1 & 0 & 1 & 1 & 1 & 2 & -1 & 0 \\
 -1 & -1 & -1 & 0 & 2 & 1 & 1 & 2 & -1 & 0 \\
 0 & 1 & 1 & 0 & 0 & -1 & 0 & -1 & 0 & 0 \\
 0 & 1 & 1 & 0 & -1 & -1 & -1 & -1 & 1 & 0 \\
 -1 & 0 & 1 & 0 & 0 & -1 & 0 & -1 & 1 & 0 \\
 6 & 4 & 2 & 0 & -9 & -3 & -5 & -7 & 4 & 0 \\
 1 & 1 & 1 & 1 & -1 & -1 & -1 & -1 & 0 & 0 \\
 0 & 0 & 0 & 0 & -1 & -1 & -1 & -1 & 1 & 1 \\
    \end{pmatrix}
\end{equation}

\begin{equation}
    \label{eqn:p2_V}
    V = \begin{pmatrix}
 1 & 0 & 0 & -3 & 0 & 0 & -1 & -5 & -1 & -1 \\
 0 & 1 & 0 & -2 & 0 & 0 & -1 & -3 & -1 & -1 \\
 0 & 0 & 1 & -1 & 0 & 0 & 0 & -1 & -1 & -1 \\
 0 & 0 & 0 & 1 & 0 & 0 & 0 & 1 & -1 & -1 \\
 0 & 0 & 0 & 1 & 1 & 0 & 0 & 2 & 1 & 0 \\
 0 & 0 & 0 & -2 & 0 & 1 & -1 & -4 & 1 & 0 \\
 0 & 0 & 0 & -1 & 0 & 0 & 0 & -2 & 1 & 0 \\
 0 & 0 & 0 & 0 & 0 & 0 & 0 & 0 & 1 & 0 \\
 0 & 0 & 0 & 2 & 0 & 0 & 1 & 4 & 0 & 1 \\
 0 & 0 & 0 & 0 & 0 & 0 & 0 & 0 & 0 & 1 \\
    \end{pmatrix}
\end{equation}
\par
The stable indicators is given by Eq.~(\ref{eqn:indexp4}) in the main text,
\begin{multline}
\text{index}=2\# \Gamma_1+4\# \Gamma_2-2\# \Gamma_3+\# M_1+3\# M_2\\
-3\# M_3-\# M_4+4\# X_1 \mod 8,
\end{multline}
which corresponds to the $8$ in the diagonal of $D$.
\par
The symmetry indicators for Wannier centers are
\begin{align}
    e_{4a} &=2\left(2\#\Gamma_2+2\# \Gamma_3+\# M_2+\# M_4+2\# X_2 \right) \mod 8\\
    e_{4b} &=2\left(2\#\Gamma_2+2\# \Gamma_3-\# M_2-\# M_4-N/2 \right) \mod 8\\
    e_{8c} &= \#X_1 - \#X_2 \mod 4
\end{align}

\subsection{$p4/m'$}
The basis for band representations (columns) and the basis for coefficients of EBRs (rows) are
\begin{equation}
    \left( \Gamma_1\Gamma_4,\Gamma_2\Gamma_3,M_1M_1,M_3M_3,M_2M_4\right),
\end{equation}
where $\Pi_i$ is defined in Table~\ref{tab:p4TI}, and
\begin{equation}
    \left( \bar E^{4a}_{1/2}, \bar E^{4a}_{3/2}, {}^1\bar E^{4b}_{1/2} {}^2\bar E^{4b}_{3/2}, {}^1\bar E^{4b}_{3/2} {}^1\bar E^{4b}_{3/2}, {}^2\bar E^{4b}_{1/2} {}^2\bar E^{4b}_{1/2},  \bar E^{8c}  \right).
\end{equation}
where $\bar E^{nw}$ is an irrep of the site symmetry group of the Wyckoff position labelled by $nw$. $\bar E^{4a}_{j_z}$ has $C_4$ eigenvalues $e^{\pm i\frac{\pi}{2}j_z}$, while ${^1}\bar E^{4b}_{j_z}$ and ${^2}\bar E^{4b}_{j_z}$ correspond to eigenvalues $e^{ i\frac{\pi}{2}j_z}$ and $e^{- i\frac{\pi}{2}j_z}$ of $C_4(\frac12,\frac12)$ separately.
For the site-symmetry group of the $4a$ position, there are two irreps $\bar E^{4a}_{1/2}$ and $\bar E^{4a}_{3/2}$, while for the site-symmetry group of the $4b$ position, there are three irreps, ${}^1\bar E^{4b}_{1/2} {}^2\bar E^{4b}_{3/2}$, ${}^1\bar E^{4b}_{3/2} {}^1\bar E^{4b}_{3/2}$ and ${}^2\bar E^{4b}_{1/2} {}^2\bar E^{4b}_{1/2}$.
The difference between these comes from the unusual pairing of irreps with $C_4(\frac12,\frac12)$ eigenvalues $\xi$ and  $-i\xi^*$ as derived in Eq.~(\ref{eqn:BBHC4T}). 
For the site-symmetry group of the $8c$ position, there is only one irrep, $\bar E^{8c}$, with $C_2(\frac12,0)$ eigenvalues $\pm i$.

\par
In this basis, the EBR matrix is
\begin{equation}
    \label{eqn:p2_A}
    A = \begin{pmatrix}
 1 & 1 & 1 & 2 & 2 \\
 1 & 1 & 1 & 0 & 2 \\
 1 & 0 & 0 & 1 & 1 \\
 0 & 1 & 0 & 1 & 1 \\
 1 & 1 & 2 & 0 & 2 \\
    \end{pmatrix}
\end{equation}
The Smith normal form matrices are
\begin{equation}
    \label{eqn:p2_D}
    D = 
\begin{pmatrix}
 1 & 0 & 0 & 0 & 0 \\
 0 & 1 & 0 & 0 & 0 \\
 0 & 0 & 1 & 0 & 0 \\
 0 & 0 & 0 & 2 & 0 \\
 0 & 0 & 0 & 0 & 0 \\
\end{pmatrix}
\end{equation}

\begin{equation}
    \label{eqn:p2_U}
    U = \begin{pmatrix}
 0 & 0 & 1 & 0 & 0 \\
 0 & 0 & 0 & 1 & 0 \\
 1 & 0 & -1 & -1 & 0 \\
 1 & -1 & 0 & 0 & 0 \\
 -1 & -1 & 1 & 1 & 1 \\
    \end{pmatrix}
\end{equation}

\begin{equation}
    \label{eqn:p2_V}
    V = \begin{pmatrix}
 1 & 0 & 0 & -1 & -1 \\
 0 & 1 & 0 & -1 & -1 \\
 0 & 0 & 1 & 0 & 0 \\
 0 & 0 & 0 & 1 & 0 \\
 0 & 0 & 0 & 0 & 1 \\
    \end{pmatrix}
\end{equation}
\par
The stable indicator is given by Eq.~(\ref{eqn:index1p4TI}) in the main text,
\begin{align}
\text{index}_1&= N/4 \mod 2 .
\end{align}

The symmetry indicators for Wannier centers are
\begin{align}
    \label{eqn:4a}
    e_{4a} &= \frac{N}{2}+2(\# M_1M_1+\# M_3M_3) \mod 4\\
    \label{eqn:4b}
    e_{4b} &= -\frac{N}{4}+2(\# M_1M_1+\# M_3M_3) \mod 4\\
    \label{eqn:8c}
    e_{8c} &=0 \mod 2
\end{align}

\section{\label{app:Chern_Cn} Chern number indicators with $n$-fold rotation symmetry at $\pi$-flux and no spin-orbit coupling}
In this section we use our theory to derive the Chern number symmetry indicators for $C_4$, $C_6$ and $C_3$ rotational symmetric systems in $\phi=\pi$ flux. We consider the case without time-reversal symmetry or spin-orbit coupling.

\subsection{Irreps of the little co-groups}
\begin{table}[]
    \begin{tabular}{c|c|c|c|c|c|c}
    \hline
    &\multicolumn{6}{c}{$\Gamma(0,0)$} \\ 
    \hline
     Irrep & $\Gamma_1$ & $\Gamma_2$ & $\Gamma_3$ & $\Gamma_4$ & $\Gamma_5$ & $\Gamma_6$\\
         \hline
      $\xi$  &$1$ &$e^{i\pi/3}$ &$e^{i2\pi/3}$ &$-1$ &$-e^{i\pi/3}$ & $-e^{i2\pi/3}$ \\ \hline    
      \end{tabular}
      \caption{\label{tab_C6_Gamma_irrep} An irrep at $\Gamma$ is labeled with $\xi$ which has $C_6$ eigenvalues $\epsilon \xi$ and $\bar\epsilon \xi$, where $\epsilon = e^{i2\pi/3}$.}
      \vspace{10pt}

      \begin{tabular}{c|c|c|c}
      \hline
    &\multicolumn{3}{c}{$K(1/3,2/3)$} \\ 
    \hline
     Irrep & $K_1$ & $K_2$ & $K_3$ \\
         \hline
      $\xi$  &$1$ &$e^{i2\pi/3}$ &$e^{-i2\pi/3}$ \\ \hline 
      \end{tabular}
      \caption{\label{tab_C3_K_irrep} An irrep at $K$ is labeled with $\xi$ which corresponds to $C_3$ eigenvalues $\epsilon \xi$ and $\bar\epsilon \xi$, where $\epsilon = e^{i2\pi/3}$.}
      \vspace{10pt}

      \begin{tabular}{c|c|c}
      \hline
    &\multicolumn{2}{c}{$M(1/2,0)$} \\ 
    \hline
     Irrep & $M_1$ & $M_2$\\
         \hline
      $C_2$  & $\sigma_z$ & $-\sigma_z$ \\ \hline 
      $T(\hat{\mathbf a}_2)$  & $\sigma_z$ & $\sigma_z$ \\ \hline 
      \end{tabular}
      \caption{\label{tab_C2T2_M_irrep} The irreps of $M$ are given in the basis such that 
      $T(\hat{\mathbf a}_1)=i\sigma_y$ and $T(\hat{\mathbf a}_2)=\sigma_z$.}
\end{table}

\begin{table}[]  
      \begin{tabular}{c|c|c|c|c}
      \hline
      &\multicolumn{4}{c}{$\Gamma(0,0)$} \\ \hline
      Irrep & $\Gamma_1$ & $\Gamma_2$ & $\Gamma_3$ & $\Gamma_4$ \\ \hline
      $C_4T(\hat{\mathbf x})$ & $\begin{pmatrix}
      1&\\&i \end{pmatrix} \epsilon$
      & $\begin{pmatrix}
      i&\\&-1 \end{pmatrix} \epsilon$
      & $\begin{pmatrix}
      -1&\\&-i \end{pmatrix} \epsilon$
      & $\begin{pmatrix}
      -i&\\&1 \end{pmatrix} \epsilon$\\
      $T(\hat{\mathbf x})T(\hat{\mathbf y})$ & $i\sigma_z$ & $i\sigma_z$ & $i\sigma_z$ & $i\sigma_z$ 
    \end{tabular}
      \vspace{10pt}
      
      \begin{tabular}{c|c|c|c|c}
      \hline
      &\multicolumn{4}{c}{$M(\pi/2,\pi/2)$} \\ \hline
      Irrep &  $M_1$ & $M_2$ &$M_3$ &$M_4$\\ \hline
      $C_4$  &$\begin{pmatrix}
      1&\\&i \end{pmatrix} $
      & $\begin{pmatrix}
      i&\\&-1 \end{pmatrix} $
      & $\begin{pmatrix}
      -1&\\&-i \end{pmatrix} $
      & $\begin{pmatrix}
      -i&\\&1 \end{pmatrix} $\\
      $T(\hat{\mathbf x})T(\hat{\mathbf y})$  & $i\sigma_z$ & $i\sigma_z$ & $i\sigma_z$ &$i\sigma_z$ 
    \end{tabular}
     \vspace{10pt}
     
    \begin{tabular}{c|c|c}
    \hline
    &\multicolumn{2}{c}{$X(1/2,0)$} \\ 
    \hline
     Irrep & $X_1$ & $X_2$\\
         \hline
      $C_2$  & $\sigma_z$ & $-\sigma_z$ \\ \hline 
      $T(\hat{\mathbf y})$  & $\sigma_z$ & $\sigma_z$ \\ \hline 
      \end{tabular}
    \caption{High symmetry momenta (first row) and the irreps (second row) of their little co-group for the group $p4$. Subsequent rows list the eigenvalue of the indicated symmetry with $\epsilon=e^{i\pi/4}$.}
    \label{tab:p4_spinless}
\end{table}

Consider a two dimensional lattice system with $C_n$ rotation symmetry. We choose the $q$-by-$q$ unit cell for $2\pi p/q$ magnetic flux.
The little co-group at a high-symmetry momentum point $k$ is $\widetilde{G}_k = C_n \ltimes {\mathbb T}_k$, where $ {\mathbb T}_k = {\mathbb T}/{\mathbb T}_M$. ${\mathbb T}$ is generated by $T(\hat{\mathbf a}_1),T(\hat{\mathbf a}_2)$ and ${\mathbb T_M}$ is generated by $T(\hat{q\mathbf a}_1),T(q\hat{\mathbf a}_2)$; thus, $|{\mathbb T}_k|=q^2$.
The rotation group has $n$ elements, $|C_n|=n$.
Therefore, according to Eq.~(\ref{eqn:schur}), at $2\pi p/q$ flux there are $n$ distinct $q$-dimensional projective irreps. \par

At $\pi$ flux, we choose a basis such that the translations are represented by 
\begin{align}
    T(\hat{\mathbf a}_1)&=e^{i k_1}\sigma_x, \\
    T(\hat{\mathbf a}_2)&=e^{i k_2}\sigma_y, \\
    T(\hat{\mathbf a}_3)&=-e^{i (k_2-k_1)}\sigma_z,
\end{align}
where $\hat{\mathbf a}_3 =\hat{\mathbf a}_2- \hat{\mathbf a}_1 $. For the four-fold symmetric case ($n=4$), the lattice vectors are $q\mathbf a_1=q(1,0)$, $q\mathbf a_2=q(0,1)$, and the reciprocal lattice vectors are $\mathbf b_1/q=(1,0)/q$, $\mathbf b_2/q=(0,1)/q$ in Cartesian coordinates; for the three- and six-fold symmetric cases ($n=3,6$), the lattice vectors are $q\mathbf a_1=q(1,0)$, $q\mathbf a_2=q(\frac12,\frac{\sqrt{3}}{2})$, and the reciprocal lattice vectors are $\mathbf b_1/q=(1,-1/\sqrt{3})/q$, $\mathbf b_2/q=(0,2/\sqrt{3})/q$ in Cartesian coordinates.
\par
In this basis, the representation matrix for $C_n$ is obtained by solving the equation 
\begin{equation}
\label{eqn:H4}
    C_nT(\mathbf a) = T(R_n\mathbf a)C_n
\end{equation}
for any $C_n$-invariant momentum point. 
{We further require the $C_n$ matrices to satisfy $C_n^n=1$. Then there are $n$ solutions corresponding to the $n$ overall phases $\xi$, each of which is a distinct irreducible representation of $\widetilde{G}_k$. }
From this equation, we find the $C_6$, $C_3$ and $C_4$ matrices explicitly:

\begin{itemize}
    \item 
The $C_6$ symmetry operator at $\Gamma=(0,0)$ is 
\begin{equation}
\label{eqn_C6_irrep_pi}
    C_6 = -\frac12 \left( \sigma_0+i(\sigma_x+\sigma_y+\sigma_z) \right) \xi
\end{equation}
where $\xi = e^{i (j-1)\pi/3}, j = 1,...,6$ labels the irrep. 
\item The $C_3$ operator at $\Gamma$ and $K$ is the square of this $C_6$ matrix. 
\item The $C_4$ symmetry operator at $\Gamma=(0,0)$ is $(\sigma_x+\sigma_y)\xi/\sqrt{2}$, while the $C_4$ symmetry operator at $M=(1/2,1/2)$ is $(\sigma_0-i\sigma_z)\xi/\sqrt{2}$, where $\xi=1,i,-1,-i$.
\end{itemize}

The irreps obtained by this method for the point group $p6$ are listed in Tables~\ref{tab_C6_Gamma_irrep}, \ref{tab_C3_K_irrep} and \ref{tab_C2T2_M_irrep} for
$\Gamma$, $K$, and $M$, respectively.
The irreps for the point group $p3$ at $\Gamma=(0,0)$, $K=(1/3,2/3)$, $K'=(2/3,1/3)$ are all isomorphic to the irreps of $K$ in $p6$ shown in Table~\ref{tab_C3_K_irrep}.
The irreps for the point group $p4$ are listed in Table~\ref{tab:p4_spinless}. The irreps are isomorphic to those that appear for the spinful case in Table~\ref{tab:p4}
which could also be derived using Eq.~(\ref{eqn:H4}).

The discussion above pertains to irreps of the little co-group of a point in momentum space. 
We now derive the irreps of the site symmetry group of each Wyckoff position in real space. Whil the site symmetry groups are the same as for the non-magnetic cases, and therefore, the irreps remain the same, the elementary band representations induced from these irreps are fundamentally different from the non-magnetic cases.
We compute these induced representations using Eq.~(\ref{eqn:inducedk}) for $p4$, $p6$ and $p3$. The resulting EBR matrices are shown in the following.

\subsection{Chern number indicators}
\subsubsection{$p4$}
For wallpaper group $p4$, all the irreps in the spinless cases are isomorphic to the spinful cases. Therefore, the EBR matrix and the symmetry indicators are the same as the spinful case shown in Appendix~\ref{app:TQC}. We conclude that the indicator is
\begin{multline}
\text{index}=2\# \Gamma_1+4\# \Gamma_2-2\# \Gamma_3+\# M_1+3\# M_2 \\
-3\# M_3-\# M_4+4\# X_1 \mod 8,
\end{multline}
where the irreps $\Pi_j$ are defined in Table~\ref{tab:p4_spinless}.

\subsubsection{$p6$}
The basis for band representations (columns) and the basis for coefficients of EBRs (rows) are
\begin{equation}
    \left( \Gamma_1,\Gamma_2,\Gamma_3,\Gamma_4,\Gamma_5,\Gamma_6,K_1,K_2,K_3,M_1,M_2\right),
\end{equation}
where $\Pi_i$ is defined in Tables~\ref{tab_C6_Gamma_irrep}, \ref{tab_C3_K_irrep} and \ref{tab_C2T2_M_irrep}, and
\begin{equation}
    \left( \Gamma^{ 4a}_1,\Gamma^{ 4a}_2,\Gamma^{ 4a}_3,\Gamma^{ 4a}_4,\Gamma^{ 4a}_5,\Gamma^{ 4a}_6, \Gamma^{ 8b}_1,\Gamma^{ 8b}_2,\Gamma^{ 8b}_3,\Gamma^{ 12c}_1,\Gamma^{ 12c}_2\right),
\end{equation}
where $\Gamma^{nw}_j$ is an irrep of the site symmetry group of the Wyckoff position labelled by $nw$. $\Gamma^{4a}_j$ has $C_6$ eigenvalue $e^{i(j-1)\pi/3 }, j=1,...,6$, $\Gamma^{8b}_j$ has $C_3$ eigenvalue $e^{i(j-1)2\pi/3 }, j=1,2,3$, and $\Gamma^{12c}_j$ has $C_2$ eigenvalue $+1,-1$.
In this basis, the EBR matrix is
\begin{equation}
    \label{eqn:p6_A}
    A = \left(
\begin{array}{ccccccccccc}
 0 & 0 & 1 & 0 & 1 & 0 & 1 & 1 & 0 & 1 & 1 \\
 0 & 0 & 0 & 1 & 0 & 1 & 0 & 1 & 1 & 1 & 1 \\
 1 & 0 & 0 & 0 & 1 & 0 & 1 & 0 & 1 & 1 & 1 \\
 0 & 1 & 0 & 0 & 0 & 1 & 1 & 1 & 0 & 1 & 1 \\
 1 & 0 & 1 & 0 & 0 & 0 & 0 & 1 & 1 & 1 & 1 \\
 0 & 1 & 0 & 1 & 0 & 0 & 1 & 0 & 1 & 1 & 1 \\
 0 & 1 & 1 & 0 & 1 & 1 & 1 & 1 & 2 & 2 & 2 \\
 1 & 0 & 1 & 1 & 0 & 1 & 2 & 1 & 1 & 2 & 2 \\
 1 & 1 & 0 & 1 & 1 & 0 & 1 & 2 & 1 & 2 & 2 \\
 1 & 1 & 1 & 1 & 1 & 1 & 2 & 2 & 2 & 4 & 2 \\
 1 & 1 & 1 & 1 & 1 & 1 & 2 & 2 & 2 & 2 & 4 \\
\end{array}
\right)
\end{equation}
The Smith normal form matrices are
\begin{equation}
    \label{eqn:p6_D}
    D = 
\left(
\begin{array}{ccccccccccc}
 1 & 0 & 0 & 0 & 0 & 0 & 0 & 0 & 0 & 0 & 0 \\
 0 & 1 & 0 & 0 & 0 & 0 & 0 & 0 & 0 & 0 & 0 \\
 0 & 0 & 1 & 0 & 0 & 0 & 0 & 0 & 0 & 0 & 0 \\
 0 & 0 & 0 & 1 & 0 & 0 & 0 & 0 & 0 & 0 & 0 \\
 0 & 0 & 0 & 0 & 1 & 0 & 0 & 0 & 0 & 0 & 0 \\
 0 & 0 & 0 & 0 & 0 & 1 & 0 & 0 & 0 & 0 & 0 \\
 0 & 0 & 0 & 0 & 0 & 0 & 1 & 0 & 0 & 0 & 0 \\
 0 & 0 & 0 & 0 & 0 & 0 & 0 & 1 & 0 & 0 & 0 \\
 0 & 0 & 0 & 0 & 0 & 0 & 0 & 0 & 12 & 0 & 0 \\
 0 & 0 & 0 & 0 & 0 & 0 & 0 & 0 & 0 & 0 & 0 \\
 0 & 0 & 0 & 0 & 0 & 0 & 0 & 0 & 0 & 0 & 0 \\
\end{array}
\right)
\end{equation}

\begin{equation}
    \label{eqn:p6_U}
    U = \left(
\begin{array}{ccccccccccc}
 -1 & 0 & 0 & 0 & 0 & -1 & 1 & 1 & 1 & -1 & 0 \\
 -1 & 0 & 0 & 0 & 0 & 0 & 1 & 0 & 0 & 0 & 0 \\
 0 & 0 & -1 & 0 & 0 & -1 & 1 & 1 & 1 & -1 & 0 \\
 -1 & 0 & -1 & -1 & -1 & -1 & 1 & 1 & 1 & 0 & 0 \\
 -1 & 0 & 0 & -1 & -1 & -1 & 2 & 1 & 1 & -1 & 0 \\
 -1 & 0 & -1 & 0 & -1 & -2 & 1 & 1 & 1 & 0 & 0 \\
 1 & 0 & 0 & 1 & 0 & 0 & -1 & 0 & 0 & 0 & 0 \\
 1 & 0 & 3 & -1 & 2 & 5 & -2 & -3 & -4 & 2 & 0 \\
 2 & 0 & 10 & -4 & 6 & 16 & -5 & -9 & -13 & 6 & 0 \\
 1 & 1 & 1 & 1 & 1 & 1 & -1 & -1 & -1 & 0 & 0 \\
 0 & 0 & 0 & 0 & 0 & 0 & -1 & -1 & -1 & 1 & 1 \\
\end{array}
\right)
\end{equation}

\begin{equation}
    \label{eqn:p6_V}
    V = \left(
\begin{array}{ccccccccccc}
 1 & 0 & 0 & 0 & 0 & 0 & 0 & -1 & 5 & -1 & -1 \\
 0 & 1 & 0 & 0 & 0 & -1 & 0 & -2 & 7 & -1 & -1 \\
 0 & 0 & 1 & 0 & 0 & 0 & 0 & -2 & 9 & -1 & -1 \\
 0 & 0 & 0 & 1 & 0 & -1 & 0 & -3 & 11 & -1 & -1 \\
 0 & 0 & 0 & 0 & 1 & 0 & 0 & 0 & 1 & -1 & -1 \\
 0 & 0 & 0 & 0 & 0 & 0 & 0 & -4 & 15 & -1 & -1 \\
 0 & 0 & 0 & 0 & 0 & 0 & 1 & 0 & 0 & 1 & 0 \\
 0 & 0 & 0 & 0 & 0 & 0 & 0 & 2 & -8 & 1 & 0 \\
 0 & 0 & 0 & 0 & 0 & 0 & 0 & 1 & -4 & 1 & 0 \\
 0 & 0 & 0 & 0 & 0 & 1 & 0 & 3 & -10 & 0 & 1 \\
 0 & 0 & 0 & 0 & 0 & 0 & 0 & 1 & -4 & 0 & 1 \\
\end{array}
\right)
\end{equation}
\par
The stable indicator is
\begin{multline}
\text{index}=2\#\Gamma_1-2\#\Gamma_3-4\#\Gamma_4+6\#\Gamma_5+4\#\Gamma_6 \\
-5\#K_1+3\#K_2-\#K_3+6\#M_1 \mod 12,
\end{multline}
where the modulus corresponds to the $12$ in the diagonal of $D$.

\subsubsection{p3}
The basis for band representations (columns) and the basis for coefficients of EBRs (rows) are
\begin{equation}
    \left( \Gamma_1,\Gamma_2,\Gamma_3,K_1,K_2,K_3,K'_1,K'_2, K'_3\right),
\end{equation}
where $\Pi_i$ is defined in Table~ \ref{tab_C3_K_irrep} and
\begin{equation}
    \left( \Gamma^{ 4a}_1,\Gamma^{ 4a}_2,\Gamma^{ 4a}_3,
    \Gamma^{ 4b}_1,\Gamma^{ 4b}_2,\Gamma^{ 4b}_3,
    \Gamma^{ 4c}_1,\Gamma^{ 4c}_2,\Gamma^{ 4c}_3
    \right),
\end{equation}
where $\Gamma^{nw}_j$ is an irrep of the site symmetry group of the Wyckoff position labelled by $nw$. $\Gamma^{nw}_j$ has $C_3$ eigenvalue $e^{i(j-1)2\pi/3 }, j=1,2,3$.
In this basis, the EBR matrix is
\begin{equation}
    \label{eqn:p3_A}
    A = 
    \left(
\begin{array}{ccccccccc}
 0 & 1 & 1 & 1 & 1 & 0 & 1 & 1 & 0 \\
 1 & 0 & 1 & 0 & 1 & 1 & 0 & 1 & 1 \\
 1 & 1 & 0 & 1 & 0 & 1 & 1 & 0 & 1 \\
 0 & 1 & 1 & 1 & 0 & 1 & 0 & 1 & 1 \\
 1 & 0 & 1 & 1 & 1 & 0 & 1 & 0 & 1 \\
 1 & 1 & 0 & 0 & 1 & 1 & 1 & 1 & 0 \\
 0 & 1 & 1 & 0 & 1 & 1 & 1 & 0 & 1 \\
 1 & 0 & 1 & 1 & 0 & 1 & 1 & 1 & 0 \\
 1 & 1 & 0 & 1 & 1 & 0 & 0 & 1 & 1 \\
\end{array}
\right)
\end{equation}
The Smith normal form matrices are
\begin{equation}
    \label{eqn:p3_D}
    D = 
\left(
\begin{array}{ccccccccc}
 1 & 0 & 0 & 0 & 0 & 0 & 0 & 0 & 0 \\
 0 & 1 & 0 & 0 & 0 & 0 & 0 & 0 & 0 \\
 0 & 0 & 1 & 0 & 0 & 0 & 0 & 0 & 0 \\
 0 & 0 & 0 & 1 & 0 & 0 & 0 & 0 & 0 \\
 0 & 0 & 0 & 0 & 1 & 0 & 0 & 0 & 0 \\
 0 & 0 & 0 & 0 & 0 & 1 & 0 & 0 & 0 \\
 0 & 0 & 0 & 0 & 0 & 0 & 6 & 0 & 0 \\
 0 & 0 & 0 & 0 & 0 & 0 & 0 & 0 & 0 \\
 0 & 0 & 0 & 0 & 0 & 0 & 0 & 0 & 0 \\
\end{array}
\right)
\end{equation}

\begin{equation}
    \label{eqn:p6_U}
    U = 
\left(
\begin{array}{ccccccccc}
 0 & 1 & 0 & 0 & -1 & 0 & 0 & 1 & 0 \\
 -1 & 0 & 0 & 1 & 0 & 0 & 1 & 0 & 0 \\
 0 & 1 & 0 & 0 & -1 & -1 & 1 & 1 & 0 \\
 1 & 0 & 0 & 0 & 0 & 0 & -1 & 0 & 0 \\
 2 & 1 & 0 & -1 & -1 & 0 & -1 & 0 & 0 \\
 1 & 0 & 0 & -1 & -1 & 0 & 0 & 1 & 0 \\
 2 & -2 & 0 & -1 & 1 & 3 & -4 & -2 & 0 \\
 1 & 1 & 1 & -1 & -1 & -1 & 0 & 0 & 0 \\
 0 & 0 & 0 & -1 & -1 & -1 & 1 & 1 & 1 \\
\end{array}
\right)
\end{equation}

\begin{equation}
    \label{eqn:p3_V}
    V =
    \left(
\begin{array}{ccccccccc}
 1 & 0 & -2 & 0 & 0 & 0 & -3 & -1 & -1 \\
 0 & 1 & 0 & 0 & 0 & 0 & 1 & -1 & -1 \\
 0 & 0 & 0 & 0 & 0 & 0 & -1 & -1 & -1 \\
 0 & 0 & -1 & 1 & 0 & 0 & -2 & 1 & 0 \\
 0 & 0 & -2 & 0 & 1 & 0 & -4 & 1 & 0 \\
 0 & 0 & 0 & 0 & 0 & 0 & 0 & 1 & 0 \\
 0 & 0 & -1 & 0 & 0 & 1 & -2 & 0 & 1 \\
 0 & 0 & 1 & 0 & 0 & 0 & 2 & 0 & 1 \\
 0 & 0 & 0 & 0 & 0 & 0 & 0 & 0 & 1 \\
\end{array}
\right)
\end{equation}
\par
The stable indicator is
\begin{multline}
\text{index}=2\#\Gamma_1-2\#\Gamma_2-\#K_1+\#K_2+3\#K_3\\
+2\#K'_1-2\#K'_2 \mod 6,
\end{multline}
where the modulus corresponds to the $6$ in the diagonal of $D$.

\section{\label{app:Chern}
Symmetry indicator for the Chern number}
The Chern number in a four-fold symmetric system is given by Eq.~(\ref{eqn:Chernnumber}) ~\cite{fang2012bulk}:
\begin{equation}
\label{eqn:appChern}
    e^{i\frac{\pi}{2} C}=(-)^{2SN}w_{C_4}^\Gamma  w_{C_2}^X w_{C_4}^M,
\end{equation}
where $S=1/2$ for spinful systems and $w_{g}^{\Pi}$ is the product of eigenvalues of the symmetry $g$ for filled bands at momentum $\Pi$. 
We now rewrite this formula in terms of the irreps of the little co-groups for $p4$.
\par
From the irreps of $p4$ listed in Table~\ref{tab:p4}, it is evident that $w_{C_2}^X\equiv +1$ for both irreps $X_1$ and $X_2$. For the four irreps at $M$,
$w_{C_4}^{M_1}=w_{C_4}^{M_3}=+1$ and $w_{C_4}^{M_2}=w_{C_4}^{M_4}=-1$, where the superscript is now labeling the irrep.
\par
We now find the $C_4$ eigenvalues of irreps at $\Gamma$. We know from Appendix~\ref{app:irreps} that for an eigenstate $|\xi,\eta\rangle$ with an eigenvalue $\xi$ of $C_4T_x$, and $\eta$ of $T_xT_y$, there is a degenerate state $T_x|\xi,\eta\rangle$ with eigenvalues:
\begin{align}
    C_4T_x ~T_x|\xi,\eta\rangle &=\xi\eta  T_x|\xi,\eta\rangle \\
    T_xT_y ~T_x|\xi,\eta\rangle &=-\eta  T_x|\xi,\eta\rangle
\end{align}
Neither of these states is separately an eigenstate of $C_4$, but we can find a linear combination that is an eigenstate by solving the eigenvalue equation
\begin{align}
    C_4\left(\alpha |\xi,\eta\rangle+\beta T_x|\xi,\eta\rangle\right) = \lambda  \left(\alpha |\xi,\eta\rangle+\beta T_x|\xi,\eta\rangle\right)
\end{align}
Using the equations above and $T_x^2 = 1$, one finds $\lambda = \pm \sqrt{ \xi^2\eta}$.
Therefore, for the four irreps at $\Gamma$ listed in Table~\ref{tab:p4},  $w_{C_4}^{\Gamma_1}=w_{C_4}^{\Gamma_3}=+i$ and $w_{C_4}^{\Gamma_2}=w_{C_4}^{\Gamma_4}=-i$. 
\par
Plug these $w_{C_n}^{\Pi_i}$ of irreps into Eq.~(\ref{eqn:appChern}), we get Eq.~(\ref{eqn:ChernumberIrrep})
\begin{align}
    C = 2N &+\#\Gamma_1+\#\Gamma_3-\#\Gamma_2-\#\Gamma_4 \nonumber \\
    &+2(\#M_2+\#M_4) \mod 4.
\end{align}


\section{\label{app:symmetry} Symmetry analysis at other fluxes} %
\begin{table*}[ht]
    \centering
    \begin{tabular}{|c|c|c|c|c|c|c|c|c|}
    \hline
     $\phi/2\pi$   &$\#\Gamma_{\frac12}$&$\#M_{\frac12}$&$N$&$e_{1a'}\mod 4$&$e_{1b'}\mod 4$&$a_{1a'}$&$\eta \mod 4$ &Phase\\
    \hline
    $0$ &0&1&2&0&2&2&2 &OAL\\
    $1/2$ &2&3&8&2&2&2&0 &Trivial\\
    \hline
    \end{tabular}
    \caption{Evaluation of symmetry indicators in Eqs.~(\ref{eqn_a_1}) and (\ref{eqn_b_1}) at half filling of the OAL model at $\phi=0$ {(middle row)} and $\phi=\pi$ {(last row)}.}
    \label{tab:eigs}
\end{table*}
\begin{table*}[ht]
    \begin{tabular}{|c|c|c|c|c|c|c|c|c|c|c|}
    \hline
     ${\phi}/{2\pi}$   &$[X_2]$&$[M_1+M_3]$&$[M_2]$&$N$&$e_{1a'}\mod 4$&$e_{1b'}\mod 4$&$e_{2c'}\mod 2$&$a_{1a'}$&$\eta \mod 4$ &Phase\\
    \hline
    $1/5$ &0&0&1&50&0&2&0&2&2&OAL\\
    $2/5$ &0&0&0&50&2&0&0&2&0&Trivial\\
    \hline
    \end{tabular}
    \caption{Evaluation of symmetry indicators Eqs.~(\ref{eqn_a_2}), (\ref{eqn_b_2}), and (\ref{eqn_c_2}) at half filling of the OAL model at $\phi=2\pi/5$ {(middle row)} and $\phi=4\pi/5$ {(last row)}.}
    \label{tab:eigs2}
\end{table*}

In this section, we apply the non-magnetic symmetry indicators (i.e. ignoring the sublattice symmetries) to analyze fluxes $\phi/2\pi=0,1/5,2/5$. The Wyckoff positions are defined in Fig.~\ref{fig:WCWC}(b).
\par
For $\phi=0$, the layer group is $p4/m'mm$. The symmetry indicators of Wannier centers are~\cite{fang2021filling}
\begin{align}
\label{eqn_a_1}
    e_{1a'}&=N-2[M_{\frac12}] \mod 4\\
    e_{1b'}&=2[M_{\frac12}] \mod 4
\label{eqn_b_1}
\end{align}
where $N$ is the number of filled bands and $[M_{\frac12}] = \# M_\frac12-\# \Gamma_\frac12$, where $\# \Pi_\frac12$ indicates the number of times the two-dimensional irrep $E_\frac12$ ($C_4$ eigenvalues $e^{i\pi/4}$, $e^{-i\pi/4}$) appears in the valence bands at the high-symmetry point $\Pi = \Gamma, M$.\par

For $\phi=\frac{2\pi}{5},~\frac{4\pi}{5}$, the layer group is $p4$. The symmetry indicators are~\cite{fang2021filling}
 \begin{align}
     \label{eqn_a_2}
     e_{1a'} &= N-[X_2]+\frac32([M_1]+[M_3])+2[M_2] \mod 4\\
 \label{eqn_b_2}
     e_{1b'} &= [X_2]-\frac12 ([M_{1}]+[M_3])-2[M_2]~\mod 4\\
 \label{eqn_c_2}
     e_{2c'} &= -\frac12 ([M_1]+[M_3]) ~\mod 2
 \end{align}
Here we use the notation $[M_j]=\# M_j-\# \Gamma_j$, where $j=1,2,3,4$ corresponds to the irrep with $C_4$ eigenvalue $\exp(i\frac{\pi}{2}(j-\frac12))$, and $[X_1]=\#X_1-\#\Gamma_1-\#\Gamma_3$, $[X_2]=\#X_2-\#\Gamma_2-\#\Gamma_4$, where $X_{1,2}$ corresponds to the irrep with $C_2$ eigenvalues $+i$, $-i$.
$\# \Pi_j$ indicates the number of times the irrep $\Pi_j$ appears in the valence bands at the high-symmetry point $\Pi$.\par

The OAL phase and corner charges are indicated by the filling anomaly $\eta$. The filling anomaly is defined as the electron number difference between neutral and symmetric configurations for a symmetric finite system \cite{benalcazar2017quantized}. 
In the absence of polarization, nonzero $\eta$ implies that the ion charge at $1a'$ Wyckoff position is not equal to the electron charge corresponding to the Wannier functions centered at $1a'$ \cite{fang2021filling,watanabe2020corner}
\begin{align}
    \eta&=N_{e,\text{neutral}}-N_{e,\text{symmetric}} \mod 4 \nonumber \\
    &= a_{1a'}-e_{1a'} \mod 4
\end{align}
{where $a_{1a'}$ and $e_{1a'}$ are the ion charge and electron charge in units of $|e|$ at $1a'$ position. }
The modulus $4$ is specific for our model.
In the absence of polarization, the corner charge is given by $Q_c=\eta/4 \mod 1$.
\par

The non-magnetic symmetry indicators of the model described in Sec.~\ref{sec:Model} at $0$ and $\pi$-flux are shown in Table~\ref{tab:eigs}.
The indicators at $2\pi/5$ and $4\pi/5$-flux are in Table~\ref{tab:eigs2}. The symmetry indicators show that the system at $0$ and $2\pi/5$-flux is an OAL and at $4\pi/5$ and $\pi$ flux is trivial.
These results are consistent agree with the open-boundary Hofstadter spectrum in Fig.~\ref{fig:Hof}.\par

The filling anomaly cannot jump when $C_4$ symmetry is preserved, unless the bulk gap or the surface gap closes. In this model, the bulk gap closes between $0$ and $\pi$ flux, as shown in Fig.~\ref{fig:Hof}. The gap closing corresponds to the transition between two distinct atomic insulating phases that have different Wannier centers. \par

\section{\label{app:Wilson} Wilson loop and nestfed Wilson loop}
\begin{figure*}[ht]
    \centering
    \includegraphics[width=\linewidth]{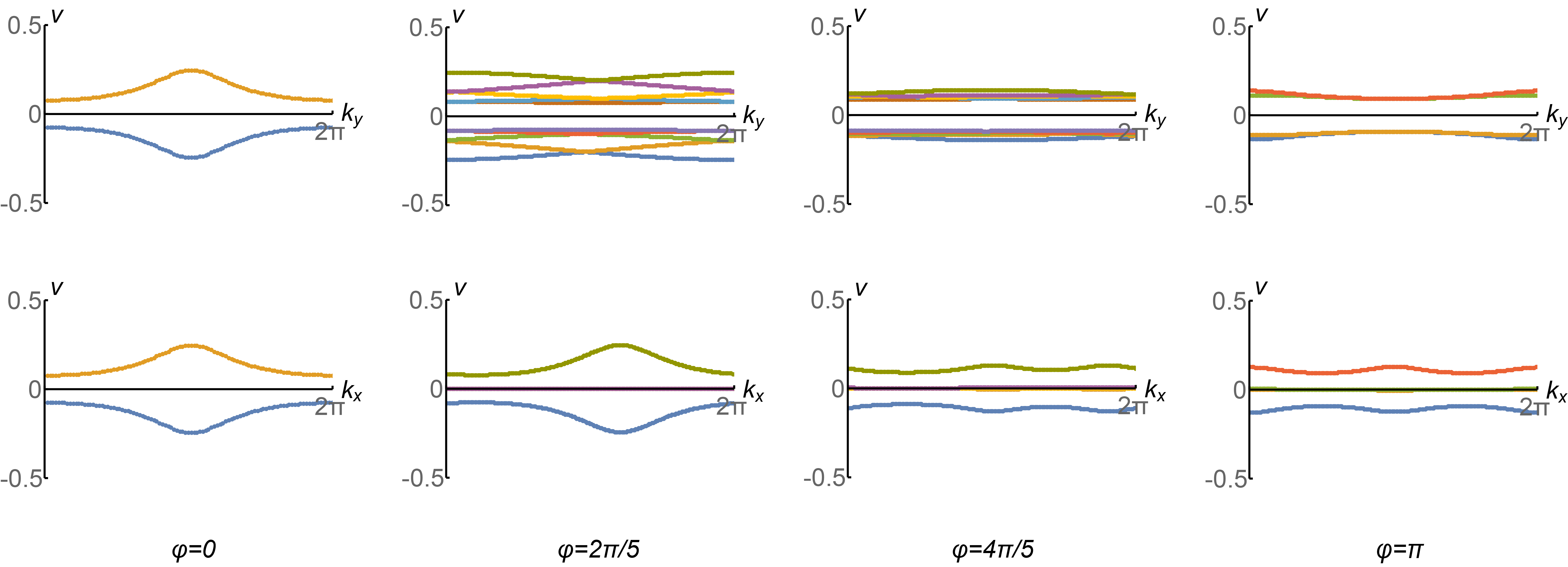}
    \caption{Wilson loop spectrum for the valence bands at half-filling, directed in $x$-direction and in $y$-direction, as defined in Eq.~(\ref{eqn:appH1}). The upper row corresponds to the spectrum of $W_x(k_y)$, while the lower two row corresponds to the spectrum of $W_y(k_x)$. The critical flux is between $2\pi/5$ and $4\pi/5$.}
    \label{fig:W}
\end{figure*}

An alternative topological invariant that characterizes the model in Sec.~\ref{sec:Model} is the quadrupole moment $Q_{xy}$, or equivalently the Wilson loop of Wilson loop (nested Wilson loop)~\cite{benalcazar2017electric}.\par
For discrete tight binding models, the Wilson loop is defined as \cite{alexandradinata2014wilson,bradlyn2019disconnected,benalcazar2017electric}
\begin{equation}
\label{eqn:appH1}
    [W_{\cal C}(\mathbf k_0)]_{mn}=\langle u_{m}(\mathbf k_0)| \prod_{\mathbf k\in{\cal C}} {\cal P}(\mathbf k) |u_{n}(\mathbf k_0)\rangle
\end{equation}
where ${\cal P}(\mathbf k)=\sum_{l=1}^{n_{occ}}|u_{l}(\mathbf k)\rangle \langle u_{l}(\mathbf k)|$ is the projector onto occupied bands, ${\cal C}$ is a loop in the Brillouin zone and $\mathbf k_0$ is the base point. We consider two Wilson loops: $W_x(k_x^0,k_y)$ with the path $k_x:0 \rightarrow 2\pi$ and $W_y(k_x,k_y^0)$ with the path $k_y:0 \rightarrow 2\pi$. The Wilson loop matrices are not gauge invariant since they depend on the gauge choice of Bloch wavefunctions. However, the eigenvalues of the Wilson loop matrices are gauge invariant. Since Wilson loops are unitary, their eigenvalues are unit complex number of the form $e^{2\pi i \nu}$. We plot $\nu_x(k_y)$ for $W_x(k_y)$ and $\nu_y(k_x)$ for $W_y(k_x)$ for the half filling gap for several fluxes in Fig.~\ref{fig:W}. The eigenvalue $\nu_x(k_y)$ has a physical meaning, namely the $x$-coordinate of the hybrid Wannier function $|w(x,k_y)\rangle$. 
The polarization is determined by the trace of the Wilson loop matrix~$\nu$~\cite{benalcazar2017quantized,resta2007theory}
\begin{equation}
    p_x= \int \frac{k_y}{2\pi} \frac{1}{2\pi} \text{Im} \left( ~tr \log W_{x}(k_y) \right),
\end{equation}
which is equal to the sum of Wilson loop eigenvalues.
In our model, polarization vanishes at half filling at any $\phi$.
A similar expression for the quadrupole moment has been studied in Refs.~\onlinecite{wheeler2019many}. However we do not use this approach. Instead we study the nested Wilson loop.
\par
When the Wilson loop spectrum $\nu_x(k_y)$ is gapped, one can compute the nested Wilson loop of the Wilson loop eigenstates that have gaps with other states. The eigenvalues of the nested Wilson loop are denoted as $\nu_y^{\nu_x}$ and $\nu_x^{\nu_y}$. For the selected gapped Wilson loop eigenstates, the sum of the nested Wilson loop eigenvalues determines its quadrupole moment \cite{benalcazar2017quantized}
\begin{equation}
\label{eqn:qxy}
    Q_{xy}=\sum_{i=1}^{n_{occ}^{\nu_x}}\nu_y^{\nu_x} =\sum_{i=1}^{n_{occ}^{\nu_y}}\nu_x^{\nu_y}
\end{equation}

 We choose $q$-by-$1$ unit cell for flux $\phi= \frac{2\pi p}{q}$ to compute the Wilson loop spectrum.
In Fig.~\ref{fig:W} we show the Wilson loop spectrum of the $2q$ valence bands below half filling of the energy spectrum.

The Wilson loop spectrum $\nu_x(k_y)$ has a gap at half filling. Therefore, we study the nested Wilson loop $\nu_y^{\nu_x}$ for the $q$ states below this gap. 
We find that the quadrupole moment Eq.~(\ref{eqn:qxy}) for the chosen bands is $0.5$ when $0\leq \phi< \phi^*$, and $0$ when $\phi^*<\phi<\pi$, where $\phi^*$ is the critical flux where the Hofstadter spectrum shows the bulk gap closing at half filling.

We now turn to the Wilson loop spectrum $\nu_y(k_x)$ which has two gaps that separate one band on the top, one band on the bottom and $2q-2$ bands around zero. Therefore, we study the nested Wilson loop $\nu_x^{\nu_y}$ for the single band on the bottom.
It turns out that the quadrupole moment Eq.~(\ref{eqn:qxy}) for the chosen bands is again $0.5$ when $0\leq \phi< \phi^*$, and $0$ when $\phi^*<\phi<\pi$.

\section{\label{app:Wannier}Magnetic Wannier functions and Balian-Low theorem}
In this appendix, we explain the Balian-Low obstruction of exponentially localized Wannier functions in two dimensional systems in a magnetic field. The key ingredient of this theorem is the projective translation group, which bridges our two dimensional magnetic systems and the quantum phase space of one dimensional quantum mechanics where the theorem was originally introduced.

Two-dimensional system under uniform magnetic field have magnetic translation operators that are projective representations of the translation group. The operators satisfy the following multiplication rule
\begin{equation}
\label{eqn:appI1}
    T(\mathbf a_1) T(\mathbf a_2) =  T(\mathbf a_1+\mathbf a_2) e^{\frac i2 \mathbf B \cdot (\mathbf a_1 \times \mathbf a_2)}
\end{equation}
The magnetic unit cell encloses an integer multiple of $2\pi$ flux, which ensures that lattice translations commute. Then the momentum space and Bloch wavefunctions are defined. The Bloch wavefunctions in momentum space Fourier transform into the magnetic Wannier functions in real space.
\par

The quantum phase space of one dimensional quantum mechanics is labeled by two dependent variables, position $x$ and wave vector $k=p/\hbar=-i\frac{d}{dx}$. Now consider the translation operators in the quantum phase space, $T_x(\Delta_x)=e^{-ik\Delta_x}$, $T_k(\Delta_k)=e^{ix\Delta_k}$. The two translations satisfy commutation relation $T_x(\Delta_x)T_k(\Delta_k)=T_k(\Delta_k)T_x(\Delta_x)e^{i\Delta_x\Delta_k}$. More generally, treating $x$ and $k$ on the same footing as $\mathbf a=(x_1,x_2)\equiv(x,k)$, the general translations in quantum phase space satisfy the same algebra as given by Eq.~(\ref{eqn:appI1})
\begin{equation}
\label{eqn:appI2}
    T(\mathbf a_1) T(\mathbf a_2) =  T(\mathbf a_1+\mathbf a_2) e^{\frac i2 \mathbf a_1 \times \mathbf a_2}
\end{equation}
where $\mathbf a_1 \times \mathbf a_2=-\mathbf a_2 \times \mathbf a_1$ is a scalar in two dimensions.
In one-dimensional quantum mechanics, it was desired to find a set of orthonormal basis functions that are localized in both $x$ and $k$ directions and form a lattice in quantum phase space. The functions are related by discrete translation symmetries
\begin{equation}
    g_{m, n}(x)=e^{ ix m \Delta_k } g(x-n \Delta_x)
\end{equation}
where $g(x)$ is centered at $(0,0)$ and $\Delta_x\Delta_k$ is the size of the unit cell. It was found that the basis is complete if and only if $\Delta_x\Delta_k=2\pi$~\cite{von2018mathematical}. Lattices satisfying this condition are called von Neumann lattices and the basis functions are also called ``Wannier functions.''
In the language of time-frequency signal analysis, this set of basis functions $\{g_{m,n}|m,n\in \mathbb Z\}$ is also called a Gabor system.
\par
In the quantum phase space (or time-frequency analysis), there is a Balian-Low theorem~\cite{benedetto1994differentiation} stating that 
{when $\Delta_x\Delta_k=2\pi$ } for the complete and orthonormal basis of Hilbert space $\{g_{m,n}|m,n\in \mathbb Z\}$, 
\begin{align}
    \text{either~} \int_{-\infty}^{\infty}x^2|g(x)|^2dx &=\infty \nonumber\\
    \text{~or~}  \int_{-\infty}^{\infty}k^2|\widetilde{g}(k)|^2dk &=\infty
\label{eqn:variance}
\end{align}
as a consequence of the algebra in Eq.~(\ref{eqn:appI2})~\cite{battle1988heisenberg}. This theorem forbids the existence of an exponentially localized Gabor system.
\par

Now return to two-dimensional systems in a magnetic field. It is shown that there is also a Balian-Low theorem that forbids the exponentially localized Wannier function for one band purely due to the algebras of the translation group operators~\cite{moscolari2019symmetry}.
This result is analogous to the one that has been understood in condensed matter physics.
\par


For a two dimensional lattice system in a magnetic field, each single gapped band has a nonzero Chern number as one can see from the Streda formula. At a rational flux $\phi=2\pi p/q$, the Streda formula says~\cite{streda1982theory}
\begin{equation}
    \bar{\rho} = C\frac{\phi}{2\pi} + s
\end{equation}
where $s\in \mathbb Z$ for non-interacting systems without symmetry breaking~\cite{bernevig2013topological}.
For a single gapped band $\bar \rho=1/q$. Therefore, we have $Cp=1\mod q$. 
\par

Interestingly, as Thouless showed in Ref.~\onlinecite{thouless1984wannier}, when the Chern number is nonzero, there are no exponentially localized Wannier functions and
the divergence of the variance of either the $x$ or $y$ coordinate is in the form of Eq.~(\ref{eqn:variance}) by replacing $k$ with $y$. 

\end{appendix}

\end{document}